\documentclass[twocolumn]{aastex62}

\usepackage{mathptmx}
\usepackage{verbatim}
\usepackage{graphicx}
\usepackage{amsmath}
\usepackage{latexsym}
\usepackage{multirow}
\usepackage{wasysym}
\usepackage{natbib}
\usepackage{academicons}
\usepackage{float}

\newcommand{\mbh}{$M_{\rm BH}$}
\newcommand{\lledd}{$L/L_{\rm Edd}$}
\newcommand{\et}{et al.\ }
\newcommand{\xray}{\hbox{X-ray}}
\newcommand{\aox}{$\alpha_{\rm ox}$}
\newcommand{\daox}{$\Delta\alpha_{\rm ox}$}
\newcommand{\nh}{$N_{\rm H}$}

\newcommand{\chandra}{{\sl Chandra}}
\newcommand{\hb}{H$\beta$}
\newcommand{\civ}{C~{\sc iv}}
\newcommand{\civdist}{C~{\sc iv}~$\parallel$~Distance}
\newcommand{\mgii}{Mg~{\sc ii}}
\newcommand{\feii}{Fe~{\sc ii}}
\newcommand{\luv}{$L_{2500\,\text{\AA}}$}

\journalinfo{The Astrophysical Journal, ???:??? (??pp), 2025 ??? ??}
\received{2025 Apr 14} 
\accepted{2025 Jul 25}

\shortauthors{MARLAR ET AL.}

\begin{document}
\title{Gemini Near Infrared Spectrograph$-$Distant Quasar Survey: the \chandra\ View}
\author[0000-0002-2969-5470]{Andrea~Marlar}
\affiliation{
	Department of Physics, University of North Texas, Denton, TX 76203, USA; AndreaMarlar@my.unt.edu} 
    
\author[0000-0003-4327-1460]{Ohad~Shemmer}
\affiliation{
	Department of Physics, University of North Texas, Denton, TX 76203, USA; AndreaMarlar@my.unt.edu} 
    
\author[0000-0002-1207-0909]{Michael~S.~Brotherton}
\affiliation{
	Department of Physics and Astronomy, University of Wyoming, Laramie, WY 82071, USA} 
    
\author[0000-0002-1061-1804]{Gordon~T.~Richards}
\affiliation{
	Department of Physics, Drexel University, 3141 Chestnut Street, Philadelphia, PA 19104, USA}
    
\author[0000-0003-0192-1840]{Cooper~Dix}
\affiliation{Department of Astronomy, The University of Texas at Austin, Austin, TX 78712, USA}

\author[0000-0001-8406-4084]{Brandon~M.~Matthews}
\affiliation{
	Department of Physics, University of North Texas, Denton, TX 76203, USA; AndreaMarlar@my.unt.edu} 

\author[0000-0002-0167-2453]{W.N.~Brandt}
\affiliation{
	Department of Astronomy and Astrophysics, 525 Davey Lab, The Pennsylvania State University, University Park, PA, 16802, USA}
\affiliation{
	Institute for Gravitation and the Cosmos, The Pennsylvania State University, University Park, PA, 16802, USA}
\affiliation{
	Department of Physics, 104 Davey Lab, The Pennsylvania State University, University Park, PA, 16802, USA}
    
\author[0000-0002-7092-0326]{R.M.~Plotkin}
\affiliation{Department of Physics, University of Nevada, Reno, NV 89557, USA} 
\affiliation{Nevada Center for Astrophysics, University of Nevada, Las Vegas, NV 89154, USA} 
	
\begin{abstract}
We present \chandra\ observations of 63 sources from the Gemini Near Infrared Spectrograph$-$Distant Quasar Survey (GNIRS-DQS) of which 54 were targeted by snapshot observations in Cycle~24. A total of 55 sources are clearly detected in at least one \xray\ band, and we set stringent upper limits on the \xray\ fluxes of the remaining eight sources. In combination with rest-frame ultraviolet-optical spectroscopic data for these sources, we assess whether \xray s can provide a robust accretion-rate indicator for quasars, particularly at the highest accessible redshifts. We utilize a recently modified \hb-based Eddington luminosity ratio estimator, as well as the \civ\ $\lambda1549$ emission-line parameter space to investigate trends and correlations with the optical-\xray\ spectral slope (\aox) and the effective hard-\xray\ power-law photon index ($\Gamma$). We find that \aox\ does not improve current accretion-rate estimates based on \hb\ or \civ. Instead, within the limitations of our sample, we confirm previous findings that the \civ\ parameter space may be a better indicator of the accretion rate up to $z\sim3.5$. We also find that the average $\Gamma$ values for a small subset of our sources, as well as the average $\Gamma$ value in different groupings of our sources, are consistent with their respective relatively high Eddington luminosity ratios. Deeper \xray\ observations of our \xray-detected sources are needed for measuring $\Gamma$ accurately and testing whether this parameter can serve as a robust, un-biased accretion-rate diagnostic.
\end{abstract}
\section{Introduction}\label{sec:introduction}
A lingering challenge in extragalactic astrophysics is understanding how supermassive black holes (SMBHs) evolve over cosmic time. Reliable estimates of the accretion rate and efficiency of the accretion process in the centers of active galaxies, or quasars, are vital in this investigation. The next generation of multiwavelength observatories will find thousands of quasars at $z > 6$ (e.g., {\sl Athena} and {\sl Rubin}; e.g., Merloni \et 2012; Thomas \et 2020; Tee \et 2023) for which quasar fundamental properties will have to be derived. The most prominent current method for estimating the quasar accretion rate focuses on the Eddington luminosity ratio, \lledd, where $L$ is the quasar bolometric luminosity and $L_{\rm{Edd}}$ is the Eddington luminosity which is proportional to the SMBH mass (\mbh). The most popular method for achieving this relies on measurements of prominent broad-emission line region (BELR) lines such as \hb~$\lambda4861$ or \mgii~$\lambda2800$ from single-epoch spectra, assuming a virialized BELR (e.g., Laor 1998; Vestergaard \& Peterson 2006; Shen \& Liu 2012; Mej{\'\i}a-Restrepo \et 2016; Grier \et 2017; Du \et 2018; Bahk \et 2019; Dalla Bont{\`a} \et 2020). These BELR lines are either not easily accessible at the highest redshifts (i.e., $z > 6$), or their measurements can be highly uncertain due to line weakness with respect to the quasar continuum in the most luminous sources (e.g., Ba{\~n}ados \et 2016; Onoue \et 2019; Reed \et 2019; Wang \et 2021), thus rendering poor constraints on \lledd\ and \mbh\ for the most distant quasars.

One promising indicator of \lledd\ is the parameter space involving the prominent \civ~$\lambda1549$ emission line (e.g., Baskin \& Laor 2004; Shemmer \& Lieber 2015; Rivera \et 2020). Rankine \et (2020), Rivera \et (2022, hereafter R22), and Ha \et (2023, hereafter H23) have shown that a combination of the \civ\ equivalent width, EW(\civ), and the velocity offset of the \civ\ emission-line peak with respect to the systemic redshift (hereafter, \civ\ Blueshift; Mart{\'\i}nez-Aldama \et 2018), known as the `\civdist'\footnote{Defined as the distance along a curve which is a best-fit function to the \civ\ EW-Blueshift parameter space (see R22).}, correlates significantly with \lledd. Furthermore, H23 found the \lledd\ vs. \civdist\ correlation becomes stronger with the addition of a correction to the traditional \hb-based \lledd\ which involves the relative strength of the \feii\ emission blend in the $\lambda4434-\lambda4684$ wavelength range with respect to the \hb\ line (see, Du \& Wang 2019; Maithil \et 2022). 

Other promising \lledd\ indicators rely on quasars' \xray\ emission. The hard-\xray\ power-law photon index ($\Gamma$) has been known to provide an unbiased \hb-based Eddington luminosity ratio estimate in unobscured, moderate-to-high luminosity quasars with an intrinsic uncertainty of $\lesssim0.35$~dex (e.g., Shemmer \et 2006, 2008; Constantin \et 2009; Risaliti \et 2009; Jin \et 2012; Brightman \et 2013; Fanali \et 2013; Kubota \& Done 2018; Liu \et 2021; Maithil \et 2024). A corona of hot electrons surrounding the SMBH at a characteristic distance of $\approx10R_{\rm g}$ is assumed to produce hard-\xray\ emission via Compton upscattering of UV–soft-\xray\ photons from the accretion disk. Thus, a high accretion rate can be manifested by a steepening of the hard-\xray\ power-law spectrum leading to the observed $\Gamma$-\lledd\ correlation (e.g., Haardt \& Maraschi 1991). However, measuring the $\Gamma$ value accurately is not the most practical means for estimating the Eddington luminosity ratio in distant quasars due to the significant observational investment required from current \xray\ missions.

Alternatively, the relative strength of \xray\ emission with respect to that of the optical-UV can trace the quasar spectral energy distribution (SED) which is known to depend on \lledd. The optical-\xray\ spectral slope, \aox\footnote{Defined as \aox$=\log(f_{\rm 2\,keV}/f_{2500\,\text{\AA}})/ \log(\nu_{\rm 2\,keV}/\nu_{2500\,\text{\AA}})$, where $f_{\rm 2\,keV}$ and $f_{2500\,\text{\AA}}$ are the flux densities at frequencies corresponding to $2$\,keV ($\nu_{\rm 2\,keV}$) and $2500$\,\AA\ ($\nu_{2500\,\text{\AA}}$), respectively.}, has previously been shown to correlate with luminosity (e.g., Just \et 2007; Lusso \et 2010; Timlin \et 2020), but shows a weaker correlation with \hb-based \lledd. This weakness is presumably due to the dependance of \aox\ on the black hole mass as well (e.g., Shemmer \et 2008; Grupe \et 2010; Wu \et 2012; Liu \et 2021; Temple \et 2023). 

Marlar \et (2022, hereafter M22) investigated the correlations between \aox, \hb-based \lledd, and EW(\civ) to see if \xray s could improve upon the current accretion-rate indicator using archival data of 53 sources that had all three parameters available. However, almost half of their sources were uniformly selected in a limited region of parameter space (at $z\lesssim0.5$), thus introducing a significant selection bias to the correlation analysis. Mitigating these biases at low redshift would require additional observations of the \civ\ line from the \textit{Hubble Space Telescope} as well as new \xray\ observations.

Therefore, we adopt a more practical approach to obtain new \xray\ observations of luminous quasars at $z\gtrsim1.5$ where \civ\ is available from large optical spectroscopic surveys, such as the Sloan Digital Sky Survey (SDSS; York \et 2000), and \hb\ can be obtained from ground-based near-infrared spectroscopy. Such quasars would allow for an expanded \civ\ and \aox\ parameter space, which is crucial for establishing the desired corrections for the \hb-based \lledd\ estimates. 

In this work, we present a sample of new and archival observations of 63 quasars selected from the \textit{Gemini} Near Infrared Spectrograph-Distant Quasar Survey (GNIRS-DQS; Matthews \et 2021), which were observed with the \chandra\ \xray\ observatory\footnote{https://cxc.cfa.harvard.edu/csc/} (Weisskopf \et 2000). GNIRS-DQS is the largest uniform sample of quasars at `cosmic noon' \hbox{($1.55 \lesssim z \lesssim 3.60$)} with high-quality spectroscopy in the $\sim 0.8 - 2.5$ $\mu$m observed-frame band, including spectral data of, at a minimum, the \hb, \mgii, and \civ\ emission lines. The 260 GNIRS-DQS sources constitute a flux-limited sample of SDSS quasars down to $m_i \sim 19.0$. All GNIRS-DQS sources have Two Micron All Sky Survey (2MASS; Skrutskie \et 2006) detections. Our aim is to overcome the limitations encountered in the archival study of Marlar \et (2022) by expanding the parameter space in a more systematic fashion. The expansion to higher redshifts also provides a more relevant benchmark from which our results could be potentially extrapolated to sources at the highest accessible redshifts ($z\gtrsim6$).

We describe our sample selection, observations, and data reduction in Section 2; in Section 3 we present the results of our analyses. We summarize our findings in Section 4. All \hb-based \lledd\ values presented in this work include the \feii-based correction from Maithil \et (2022). We adopt a flat cosmological model with \hbox{$\Omega_{\Lambda}$ = 0.7}, \hbox{$\Omega_{M}$ = 0.3}, and \hbox{$H_{0}$ = 70 km ${\rm s}^{-1}~{\rm Mpc}^{-1}$} (e.g., Spergel \et 2007).

\section{Target Selection, Observations, and Data Reduction}\label{sec:obs}
Of the 260 GNIRS-DQS sources, we selected the brightest 54 that did not yet have sensitive \xray\ coverage and that comprised a sufficiently large sample of sources for which economical \chandra\ snaphshot observations could be performed. In order to ensure reliable measurements of the intrinsic \aox\ and \civ\ spectral properties, we also required that our sources have not been identified as broad absorption line (BAL) quasars (e.g., Ahmed \et 2024) and are radio quiet\footnote{The radio loudness parameter, $R$, is defined as $f_{\rm 5\,GHz}/f_{4400\,\text{\AA}}$, where $f_{\rm 5\,GHz}$ and $f_{4400\,\text{\AA}}$ are the flux densities at $5$\,GHz and $4400$\,\AA, respectively (Kellermann \et 1989).} ($R<10$; Matthews \et 2023). The first of these criteria ensures that the effects of \xray\ absorption are minimized (e.g., Gallagher \et 2006), and the second criterion is required for minimizing the potential contribution of a jet to the \xray\ continuum emission (e.g., Miller \et 2011), although this contribution is expected to be significant only for extremely radio loud sources (e.g., Zhu \et 2020, 2021). We employed \chandra\ exposure times of \hbox{$3.0 - 7.0$ ks}, per source. These exposures were designed for detecting $\gtrsim85\%$ of our sources with at least $\sim2-3$ photons per source, taking into account the typical \xray\ weakness observed in the general quasar population (e.g., Steffen \et 2006; Gibson \et 2008; Pu \et 2020). Only seven of our sources were not detected,  representing a success-rate of $87\%$, consistent with our prediction. Aiming for a higher detection threshold of, for example, 95\% would have more than doubled the program's exposure time and was deemed unwarranted.

In addition to these 54 sources, we add high-quality archival \chandra\ observations of nine additional GNIRS-DQS sources that were presented in M22. All of these sources were targeted by \chandra\ and the observations preceded the GNIRS-DQS project. The observations represent a variety of science cases and observing strategies.

All 63 GNIRS-DQS sources were observed with the \chandra\ Advanced CCD Imaging Spectrometer (ACIS; Garmire \et 2003). The \chandra\ observation log appears in \hbox{Table~\ref{tab:chandra_log}}. \textit{Column (1)} gives the SDSS quasar name; \textit{Columns (2) $-$ (3)} give the RA and DEC, respectively, taken from the NASA/IPAC Extragalactic Database (NED)\footnote{https://ned.ipac.caltech.edu/}; \textit{Column (4)} gives the systemic redshift ($z_{sys}$) from Matthews \et (2023); \textit{Column (5)} gives the angular distance between the SDSS and \xray\ positions which can be as large as 1.0$''$\footnote{We note that the median offset is 0.4$''$, and even the largest offsets of $\sim0.8-1.0''$ that we find for five of our sources, are well within the limitations of the \chandra\ detectors (see also https://cxc.harvard.edu/cal/ASPECT/celmon/). Additionally, we find that the uncertainty in the \xray\ position as determined by {\sc wavdetect} generally increases as the number of \xray\ counts decreases, an effect which likely partially contributes to the increased offsets for some of our sources. No systematic \chandra\ astrometric offsets were found between our quasars with the largest offsets and other \xray\ sources in their \chandra\ images.}; \textit{Column (6)} gives the Galactic absorption column density in units of $10^{20} {\rm cm}^{-2}$, taken from Dickey $\&$ Lockman (1990) and obtained with the High Energy Astrophysics Science Archive Research Center \nh\ tool\footnote{https://heasarc.gsfc.nasa.gov/cgi-bin/Tools/w3nh/w3nh.pl.}; \textit{Columns (7) $-$ (10)} give the \chandra\ cycle, start date, observation ID, and exposure time, respectively. 

\newpage
\startlongtable
\begin{deluxetable*}{lcccccclcc}
\tabletypesize{\footnotesize}
\tablecolumns{10}
\tablecaption{\chandra\ Observation Log \label{tab:chandra_log}}
\tablehead{
\colhead{} &
\colhead{RA} &
\colhead{DEC} &
\colhead{} &
\colhead{$\Delta_{\rm{Opt-X}}$\tablenotemark{a}} &
\colhead{Galactic \nh\tablenotemark{b}} &
\colhead{} &
\colhead{} &
\colhead{} &
\colhead{Exp. Time\tablenotemark{c}} \\
\colhead{Quasar} &
\colhead{(deg)} &
\colhead{(deg)} &
\colhead{$z_{sys}$} &
\colhead{(arcsec)} &
\colhead{(10$^{20}$\,cm$^{-2}$)} &
\colhead{Cycle} &
\colhead{Obs. Date} & 
\colhead{Obs. ID} &
\colhead{(ks)} \\
\colhead{(1)} &
\colhead{(2)} &
\colhead{(3)} &
\colhead{(4)} &
\colhead{(5)} &
\colhead{(6)} &
\colhead{(7)} &
\colhead{(8)} &
\colhead{(9)} &
\colhead{(10)} 
}
\startdata
\multicolumn{10}{c}{New Observations} \\
\noalign{\smallskip}\hline\noalign{\smallskip}
\object{SDSS~J003416.61$+$002241.1} & 8.569202 & 0.378083 & 1.63 & 0.2 & 2.58 & 24 & 2023 Aug 3 & 26794 & 4.51 \\
\object{SDSS~J004719.71$+$014813.9} & 11.832203 & 1.803716 & 1.59 & 0.7 & 2.15 & 24 & 2023 Nov 17 & 26795 & 3.23 \\
\object{SDSS~J013136.44$+$130331.0} & 22.901861 & 13.058628 & 1.60 & 0.2 & 4.57 & 24 & 2022 Dec 15 & 26796 & 3.03 \\
\object{SDSS~J014128.26$+$070606.1} & 25.367762 & 7.101704 & 2.26 & 0.6 & 3.28 & 24 & 2023 Sep 30 & 26797 & 6.30 \\
\object{SDSS~J035150.97$-$061326.4} & 57.962418 & $-$6.224009 & 2.22 & 0.3 & 6.46 & 24 & 2023 Mar 19 & 26798 & 4.01 \\
\object{SDSS~J072517.52$+$434553.4} & 111.323021 & 43.764845 & 1.60 & 0.6 & 7.67 & 24 & 2023 Oct 3 & 26799 & 3.51 \\
\object{SDSS~J073913.65$+$461858.5} & 114.806899 & 46.316278 & 1.57 & 0.4 & 5.21 & 24 & 2023 Apr 6 & 26800 & 3.03 \\
\object{SDSS~J074941.16$+$262715.9} & 117.421505 & 26.454426 & 1.59 & 0.8 & 3.27 & 24 & 2023 Dec 22 & 26801 & 3.03 \\
\object{SDSS~J075115.43$+$505439.1} & 117.814358 & 50.910879 & 2.31 & 0.5 & 5.69 & 24 & 2023 Mar 14 & 26802 & 3.42 \\
\object{SDSS~J075136.36$+$432732.4} & 117.901509 & 43.459005 & 2.25 & 0.4 & 4.66 & 24 & 2024 Apr 8 & 26803 & 4.21 \\
\object{SDSS~J081019.47$+$095040.9} & 122.581157 & 9.844682 & 2.24 & \nodata\tablenotemark{d} & 2.59 & 24 & 2023 Dec 30 & 26804 & 3.92 \\
\object{SDSS~J081558.35$+$154055.2} & 123.993164 & 15.682026 & 2.24 & 0.6 & 3.61 & 24 & 2024 May 31 & 26805 & 3.91 \\
\object{SDSS~J082603.32$+$342800.6} & 126.513859 & 34.466873 & 2.31 & 0.0 & 4.28 & 24 & 2023 Oct 1 & 26806 & 4.61 \\
\object{SDSS~J085337.36$+$121800.3} & 133.405709 & 12.300094 & 2.20 & 0.6 & 3.62 & 24 & 2024 Apr 21 & 26807 & 4.41 \\
\object{SDSS~J085443.10$+$075223.2} & 133.679597 & 7.873128 & 1.61 & 0.2 & 4.85 & 24 & 2023 Jan 13 & 26808 & 3.04 \\
\object{SDSS~J090247.57$+$304120.7} & 135.698218 & 30.689116 & 1.56 & 0.5 & 1.81 & 24 & 2023 Dec 5 & 26809 & 3.03 \\
\object{SDSS~J090646.98$+$174046.8} & 136.695762 & 17.679675 & 1.58 & 0.4 & 3.36 & 24 & 2024 Apr 8 & 26810 & 3.03 \\
\object{SDSS~J090710.36$+$430000.2} & 136.793203 & 43.000058 & 2.19 & 0.2 & 1.47 & 24 & 2024 Jan 19 & 26811 & 3.22 \\
\object{SDSS~J091941.26$+$253537.7} & 139.921953 & 25.593824 & 2.27 & 0.3 & 3.14 & 24 & 2024 Feb 8 & 26812 & 5.31 \\
\object{SDSS~J092216.04$+$160526.4} & 140.566849 & 16.090700 & 2.37 & 0.6 & 2.82 & 24 & 2023 Feb 5 & 26813 & 6.97 \\
\object{SDSS~J092523.24$+$214119.8} & 141.346845 & 21.688842 & 2.36 & 0.3 & 3.24 & 24 & 2023 Dec 9 & 26814 & 6.48 \\
\object{SDSS~J092555.05$+$490338.2} & 141.479370 & 49.060631 & 2.34 & 0.6 & 1.25 & 24 & 2023 Apr 7 & 26815 & 5.00 \\
\object{SDSS~J093533.88$+$235720.5} & 143.891180 & 23.955698 & 2.30 & 0.2 & 2.34 & 24 & 2023 Dec 14 & 26816 & 5.36 \\
\object{SDSS~J094637.83$-$012411.5} & 146.657638 & $-$1.403220 & 2.22 & \nodata\tablenotemark{d} & 3.01 & 24 & 2024 Apr 17 & 26817 & 4.22 \\
\object{SDSS~J094648.59$+$171827.7} & 146.702512 & 17.307724 & 2.30 & 0.4 & 2.44 & 24 & 2023 Dec 13 & 26818 & 5.01 \\
\object{SDSS~J095330.36$+$353223.1} & 148.376519 & 35.539758 & 2.39 & 0.2 & 0.97 & 24 & 2023 Dec 14 & 26819 & 4.92 \\
\object{SDSS~J095555.68$+$351652.6} & 148.981917 & 35.281364 & 1.62 & 0.6 & 0.99 & 24 & 2024 Apr 18 & 26820 & 4.12 \\
\object{SDSS~J095823.07$+$371218.3} & 149.596156 & 37.205091 & 2.28 & \nodata\tablenotemark{d} & 1.19 & 24 & 2024 Apr 20 & 26821 & 3.92 \\
\object{SDSS~J100212.63$+$520800.2} & 150.552653 & 52.133414 & 1.62 & 0.8 & 0.97 & 24 & 2023 May 7 & 26822 & 3.63 \\
\object{SDSS~J101106.74$+$114759.4} & 152.778113 & 11.799869 & 2.25 & 0.2 & 3.45 & 24 & 2023 Feb 10 & 26823 & 5.61 \\
\object{SDSS~J101429.57$+$481938.4} & 153.623187 & 48.327338 & 1.57 & 0.1 & 0.82 & 24 & 2023 Dec 14 & 26824 & 2.89 \\
\object{SDSS~J102731.49$+$541809.7} & 156.881250 & 54.302724 & 1.59 & 0.2 & 0.96 & 24 & 2023 May 27 & 26825 & 3.52 \\
\object{SDSS~J103209.78$+$385630.5} & 158.040782 & 38.941832 & 1.59 & 0.2 & 1.67 & 24 & 2024 Jan 21 & 26826 & 3.73 \\
\object{SDSS~J103236.98$+$230554.1} & 158.154101 & 23.098391 & 2.38 & \nodata\tablenotemark{d} & 1.53 & 24 & 2024 Mar 22 & 26827 & 5.51 \\
\object{SDSS~J104336.73$+$494707.6} & 160.903066 & 49.785482 & 2.20 & 0.8 & 1.22 & 24 & 2023 Dec 3 & 26828 & 7.32 \\
\object{SDSS~J105045.72$+$544719.2} & 162.690507 & 54.788695 & 2.17 & 0.4 & 0.91 & 24 & 2024 Sep 19 & 26829 & 5.50 \\
\object{SDSS~J105926.43$+$062227.4} & 164.860139 & 6.374300 & 2.20 & 0.1 & 2.48 & 24 & 2024 Jun 20 & 26830 & 3.03 \\
\object{SDSS~J110735.58$+$642008.6} & 166.898269 & 64.335771 & 2.31 & 0.2 & 0.96 & 24 & 2022 Nov 30 & 26831 & 5.20 \\
\object{SDSS~J110810.87$+$014140.7} & 167.045298 & 1.694650 & 1.62 & \nodata\tablenotemark{d} & 3.45 & 24 & 2023 Mar 19 & 26832 & 3.63 \\
\object{SDSS~J111850.02$+$351311.7} & 169.708427 & 35.219911 & 2.18 & 0.5 & 1.73 & 24 & 2023 Dec 5 & 26833 & 6.93 \\
\object{SDSS~J114212.25$+$233250.5} & 175.551051 & 23.547369 & 1.59 & 1.0
& 2.27 & 24 & 2024 Jul 4 & 26834 & 3.31 \\
\object{SDSS~J114350.30$+$362911.3} & 175.959617 & 36.486499 & 2.35 & 0.2 & 1.60 & 24 & 2024 May 21 & 26835 & 3.63 \\
\object{SDSS~J114907.15$+$004104.3} & 177.279794 & 0.684565 & 2.31 & 0.7 & 2.32 & 24 & 2023 Mar 25 & 26836 & 3.13 \\
\object{SDSS~J121314.03$+$080703.6} & 183.308446 & 8.117679 & 2.40 & 0.5 & 1.29 & 24 & 2023 Mar 13 & 26837 & 4.48 \\
\object{SDSS~J121810.98$+$241410.9} & 184.545799 & 24.236359 & 2.38 & 0.6 & 2.13 & 24 & 2024 Jan 15 & 26838 & 3.62 \\
\object{SDSS~J125150.45$+$114340.7} & 192.960216 & 11.727984 & 2.21 & 0.7 & 2.41 & 24 & 2023 Mar 18 & 26839 & 3.46 \\
\object{SDSS~J125159.90$+$500203.6} & 192.999620 & 50.034341 & 2.38 & \nodata\tablenotemark{d} & 1.16 & 24 & 2024 Jan 6 & 26840 & 4.67 \\
\object{SDSS~J134341.99$+$255652.9} & 205.924969 & 25.948030 & 1.60 & 0.3 & 1.01 & 24 & 2022 Dec 3 & 26841 & 3.03 \\
\object{SDSS~J135908.35$+$305830.8} & 209.784836 & 30.975238 & 2.30 & 0.3 & 1.09 & 24 & 2023 May 4 & 26842 & 3.72 \\
\object{SDSS~J144624.29$+$173128.8} & 221.601221 & 17.524675 & 2.21 & \nodata\tablenotemark{d} & 1.74 & 24 & 2023 Jan 13 & 26843 & 3.22 \\
\object{SDSS~J144706.81$+$212839.2} & 221.778378 & 21.4775805 & 3.22 & 0.3 & 2.09 & 24 & 2023 Jan 9 & 26844 & 5.98 \\
\object{SDSS~J144948.62$+$123047.4} & 222.452599 & 12.513189 & 1.59 & 0.5 & 1.55 & 24 & 2023 Apr 10 & 26845 & 3.42 \\
\object{SDSS~J151727.68$+$133358.6} & 229.365355 & 13.566271 & 2.24 & 0.6 & 2.99 & 24 & 2023 Jan 13 & 26846 & 4.32 \\
\object{SDSS~J214901.21$-$073141.6} & 327.255031 & $-$7.528257 & 2.21 & 0.8 & 3.45 & 24 & 2023 Apr 23 & 26847 & 3.02 \\
\noalign{\smallskip}\hline\noalign{\smallskip}
\multicolumn{10}{c}{Archival Observations from M22} \\
\noalign{\smallskip}\hline\noalign{\smallskip}
\object{SDSS~J080117.79$+$521034.5} & 120.324137 & 52.176277 & 3.26 & 0.6 & 4.66 & 15 & 2014 Dec 11 & 17081 & 43.50 \\
\object{SDSS~J084846.11$+$611234.6} & 132.192113 & 61.209644 & 2.26 & 0.1 & 4.43 & 13 & 2011 Dec 22 & 13353 & 1.54 \\
\object{SDSS~J094602.31$+$274407.0} & 146.509648 & 27.735303 & 2.49 & 0.1 & 1.77 & 11 & 2010 Jan 16 & 11489 & 4.98 \\
\object{SDSS~J094646.94$+$392719.0} & 146.695581 & 39.455285 & 2.23 & 0.3 & 1.57 & 12 & 2011 Feb 27 & 12857 & 27.30 \\
\object{SDSS~J095852.19$+$120245.0} & 149.717487 & 12.045853 & 3.31 & 0.1 & 3.22 & 13 & 2012 Apr 22 & 13354 & 1.56 \\
\object{SDSS~J102907.09$+$651024.6} & 157.279438 & 65.173497 & 2.17 & 0.2 & 1.20 & 9 & 2008 Jun 17 & 9228 & 10.64 \\
\object{SDSS~J111119.10$+$133603.8} & 167.829613 & 13.601082 & 3.48 & 0.2 & 1.57 & 16 & 2015 Jan 26 & 17082 & 43.06 \\
\object{SDSS~J141028.14$+$135950.2} & 212.617255 & 13.997281 & 2.22 & 0.1 & 1.42 & 10 & 2009 Nov 28 & 10741 & 4.03 \\
\object{SDSS~J141951.84$+$470901.3} & 214.965988 & 47.150379 & 2.31 & 0.0 & 1.52 & 3 & 2002 Jun 2 & 3076 & 7.66
\enddata
\tablenotetext{a}{Angular separation between the optical and \xray\ sources.}
\tablenotetext{b}{Obtained from Dickey \& Lockman (1990).}
\tablenotetext{c}{The exposure time has been corrected for detector dead time.}
\tablenotetext{d}{{\sc wavdetect} did not detect an \xray\ source at this location.}
\end{deluxetable*}

Source counts in three different bands were extracted using \chandra\ Interactive Analysis of Observations ({\sc ciao})\footnote{http://cxc.cfa.harvard.edu/ciao/} v4.14 tools. The \xray\ counts for all sources were obtained using {\sc wavdetect} (Freeman \et 2002) with wavelet transforms of scale sizes $1$, $1.4$, $2$, $2.8$, and $4$ pixels, a false-positive probability threshold of $10^{-3}$, and confirmed by visual inspection. Errors on the \xray\ counts correspond to the $1\sigma$ level, and were computed according to Tables~1 and 2 of Gehrels (1986) using Poisson statistics. Upper limits were computed according to Kraft \et (1991) and represent the 95\% confidence level; upper limits of $3.0$, $4.8$, $6.4$, $8.0$, and $9.4$ indicate that $0$, $1$, $2$, $3$, and $4$ \xray\ counts, respectively, have been found within an extraction region of radius $1$\arcsec\ centered on the source's SDSS position (considering the background within this source-extraction region to be negligible).

The sources' \xray\ spectra were extracted with the {\sc ciao} {\sc specextract} task using circular regions of $1''$ radius centered on the \xray\ centroid of each source; the background regions were determined using annuli of different sizes, centered on each source, to avoid contamination from nearby \xray\ sources. We individually fit each of our 63 sources in the \hbox{$>2$ keV} rest-frame band using {\sc xspec} v12.14.1 (Arnaud 1996) and a power-law model with a Galactic absorption component (i.e., {\sc phabs*pow}); the C-statistic (Cash 1979) was used throughout.

For eight of our sources that had $<3$ counts in the fitting range, the spectral fitting failed to provide a best-fit photon index. For all other sources that had $<100$ counts per source, the spectral fitting resulted in highly uncertain photon indices that are deemed unusable for any practical purposes (however, the majority of these photon indices are consistent with $\Gamma=2.0$). Therefore, we derive Galactic absorption-corrected flux densities at the rest-frame energy of $2$ keV ($f_{2~\rm keV}$) utilizing {\sc xspec} and fixing the photon index to $\Gamma=2.0$. We estimated the respective errors on those flux densities using two additional {\sc xspec} runs. In the first run, the $\Gamma$ values were fixed to $1.6$, and in the second run, $\Gamma$ values were fixed to $2.4$. This range of $\Gamma$ values is observed in the majority of luminous quasars (see e.g., Vignali \et 2003, Page \et 2005, Risaliti \et 2009). For the eight sources with $<3$ counts, we utilized the \chandra\ {\sc PIMMS} tool\footnote{https://cxc.harvard.edu/toolkit/pimms.jsp} to derive these values using the count rate in the soft band and a fixed photon index value of $\Gamma=2.0$. 

We also attempted to add an intrinsic absorption component to the above {\sc xspec} model (i.e., {\sc phabs*zphabs*pow}) for all sources with $3<$ counts $<100$, but given the small number of counts, {\sc xspec} could not provide meaningful values or upper limits on the photon index and intrinsic neutral absorption column density in any of these sources.

Table~\ref{tab:chandra_uv-opt} presents the basic \xray\ measurements and UV-optical data used for our analyses. \textit{Column (1)} gives the SDSS quasar name; \textit{Columns (2) $-$ (4)} give the \xray\ counts in the soft (observed-frame \hbox{$0.5-2$ keV}), hard (observed-frame \hbox{$2-8$ keV}), and full (observed-frame \hbox{$0.5-8$ keV}) bands, respectively; \textit{Column (5)} gives the count rate in the soft band; \textit{Column (6)} gives the band ratio of hard- to soft-band counts, calculated with the {\sc Bayesian Estimation of Hardness Ratios} ({\sc BEHR}; Park \et 2006); \textit{Column (7)} gives $f_{2~\rm keV}$ values and their errors as described above; \textit{Column (8)} gives the flux density at rest-frame wavelength of $2500$\,\text{\AA} and corresponding $1\sigma$ error, measured from the SDSS spectra of the sources (for four of the sources at $z>3$, these values were extrapolated linearly using the mean flux densities measured in the rest-frame $1825-1835$\,\text{\AA} and $1960-1970$\,\text{\AA} intervals given that the continuum at rest-frame $2500$\,\text{\AA} is not covered by the sources' SDSS spectra); \textit{Column (9)} gives the monochromatic luminosity at a rest-frame wavelength of $2500$\,\text{\AA}; \textit{Column (10)} gives the \aox\ parameter and its error; \textit{Column (11)} gives the $\Delta$\aox\ parameter (and its error), defined as the difference between the measured \aox\ from Column (10) and the predicted \aox, based on the \hbox{\aox-$L_{\nu}(2500\,\text{\AA})$} relation in quasars (given as Equation [3] of Timlin \et 2020); \textit{Column (12)} gives the monochromatic luminosity at a rest-frame wavelength of $5100\,\text{\AA}$ [$\nu L_{\nu}(5100\,\text{\AA})$], taken from Matthews \et (2023); \textit{Column (13)} gives the FWHM of the broad \hb\ line, taken from Matthews \et (2023); \textit{Column (14)} gives the \feii-corrected Eddington luminosity ratio, taken from H23; \textit{Columns (15) $-$ (17)} give the rest-frame \civ\ EW, \civ\ Blueshift, and \civdist\ values taken from H23.

\newpage
\startlongtable
\begin{longrotatetable}
\begin{deluxetable*}{lcccccccccccccccc}
\tabletypesize{\tiny}
\tablecolumns{17}
\tablecaption{Basic X-ray and UV-Optical Data \label{tab:chandra_uv-opt}}
\tablehead{ 
\colhead{} &
\colhead{} &
\colhead{} &
\colhead{} &
\colhead{} &
\colhead{} &
\colhead{} &
\colhead{} &
\colhead{} &
\colhead{} &
\colhead{} &
\colhead{} &
\colhead{} &
\colhead{} &
\multicolumn{3}{c}{\civ} \\
\cline{15-17}
\colhead{} &
\colhead{} &
\multicolumn{1}{c}{Counts} &
\colhead{} &
\colhead{} &
\colhead{} &
\colhead{} &
\colhead{} &
\colhead{$\log L_{\nu}(2500\,\text{\AA})$} &
\colhead{} &
\colhead{} &
\colhead{$\log \nu L_{\nu}(5100\,\text{\AA})$\tablenotemark{d}} &
\colhead{FWHM \hb\tablenotemark{d}} &
\colhead{} &
\colhead{EW} &
\colhead{Blueshift} &
\colhead{$\parallel$ Distance} \\
\cline{2-4} 
\colhead{Quasar} &
\colhead{$0.5-2$~keV} & 
\colhead{$2-8$~keV} &
\colhead{$0.5-8$~keV} & 
\colhead{Count Rate\tablenotemark{a}} &
\colhead{Band Ratio} &
\colhead{$f_{2~\rm keV}$\tablenotemark{b}} &
\colhead{$f_{2500\,\text{\AA}}$\tablenotemark{c}} &
\colhead{(erg\,s$^{-1}$\,Hz$^{-1}$)} &
\colhead{\aox} &
\colhead{\daox} &
\colhead{(erg\,s$^{-1}$)} &
\colhead{(km\,s$^{-1}$)} &
\colhead{\lledd} &
\colhead{(\text{\AA})} &
\colhead{(km\,s$^{-1}$)} &
\colhead{} \\
\colhead{(1)} &
\colhead{(2)} &
\colhead{(3)} &
\colhead{(4)} &
\colhead{(5)} &
\colhead{(6)} &
\colhead{(7)} &
\colhead{(8)} &
\colhead{(9)} &
\colhead{(10)} &
\colhead{(11)} &
\colhead{(12)} &
\colhead{(13)} &
\colhead{(14)} &
\colhead{(15)} &
\colhead{(16)} &
\colhead{(17)}
}
\startdata
\multicolumn{17}{c}{New Observations} \\
\noalign{\smallskip}\hline\noalign{\smallskip}
\object{SDSS~J003416.61$+$002241.1}	& 5.0$^{+3.4}_{-2.2}$ & 7.8$^{+3.9}_{-2.7}$ & 12.8$^{+4.7}_{-3.5}$ & 1.1$^{+0.8}_{-0.5}$ & 1.53$^{+5.48}_{-1.19}$ & $1.7^{+0.6}_{-0.5}$ & $3.0\pm 0.1$ & $31.30\pm 0.01$ & $-1.63\pm 0.05$ & $+0.03\pm 0.05$ & 46.24 & 5527 & 0.23 & 28.5 & 597 & 0.611 \\
\object{SDSS~J004719.71$+$014813.9}	& 15.8$^{+5.1}_{-3.9}$ & 11.9$^{+4.6}_{-3.4}$ & 29.5$^{+6.5}_{-5.4}$ & 4.9$^{+1.6}_{-1.2}$ & 0.72$^{+0.62}_{-0.12}$ & $4.3^{+1.3}_{-1.0}$ & $3.4\pm 0.1$ & $31.33\pm 0.01$ & $-1.50^{+0.05}_{-0.04}$ & $+0.17^{+0.05}_{-0.04}$ & 46.13 & 5605 & 0.20 & 48.9 & 456 & 0.505 \\
\object{SDSS~J013136.44$+$130331.0} & 4.0$^{+3.2}_{-1.9}$ & 1.9$^{+2.6}_{-1.3}$ & 6.8$^{+3.8}_{-2.5}$ & 1.3$^{+1.0}_{-0.6}$ & 0.48$^{+0.80}_{-0.38}$\tablenotemark{e} & $1.0^{+0.3}_{-0.2}$ & $3.8\pm 0.1$ & $31.39\pm 0.01$ & $-1.76\pm 0.05$ & $-0.08\pm 0.05$ & 46.45 & 2294 & 1.63 & 2.8 & 2320 & 1.030 \\
\object{SDSS~J014128.26$+$070606.1} & 7.0$^{+3.8}_{-2.6}$ & $< 4.8$ & 8.9$^{+4.1}_{-2.9}$ & 1.1$^{+0.6}_{-0.4}$ & $<0.65$ & $0.8^{+0.4}_{-0.2}$ & $1.6\pm 0.1$ & $31.27\pm 0.03$ & $-1.64\pm 0.06$ & $+0.01\pm 0.06$ & 46.25 & 3312 & 0.64 & 42.7 & 518 & 0.537 \\
\object{SDSS~J035150.97$-$061326.4} & 8.8$^{+4.1}_{-2.9}$ & 11.7$^{+4.5}_{-3.4}$ & 20.6$^{+5.6}_{-4.5}$ & 2.2$^{+1.0}_{-0.7}$ & 1.28$^{+1.57}_{-0.24}$ & $2.5^{+0.8}_{-0.6}$ & $3.9\pm 0.3$ & $31.66\pm 0.03$ & $-1.61\pm 0.05$ & $+0.11\pm 0.05$ & 46.51 & 2616 & 1.33 & 13.8 & 500 & 0.728 \\
\object{SDSS~J072517.52$+$434553.4} & 11.9$^{+4.6}_{-3.4}$ & 10.8$^{+4.4}_{-3.2}$ & 22.7$^{+5.8}_{-4.7}$ & 3.4$^{+1.3}_{-1.0}$ & 0.87$^{+0.96}_{-0.15}$ & $3.1^{+1.3}_{-0.9}$ & $4.4\pm 0.1$ & $31.44\pm 0.01$ & $-1.59\pm 0.06$ & $+0.09\pm 0.06$ & 46.37 & 1705 & 2.73 & 19.1 & 429 & 0.664 \\
\object{SDSS~J073913.65$+$461858.5} & 9.9$^{+4.3}_{-3.1}$ & 9.8$^{+4.3}_{-3.1}$ & 19.7$^{+5.5}_{-4.4}$ & 3.3$^{+1.4}_{-1.0}$ & 0.95$^{+1.23}_{-0.17}$ & $3.1^{+1.3}_{-0.9}$ & $4.5\pm 0.1$ & $31.45\pm 0.01$ & $-1.60\pm 0.06$ & $+0.09\pm 0.06$ & 46.33 & 4060 & 0.46 & 15.9 & 1770 & 0.822 \\
\object{SDSS~J074941.16$+$262715.9} & 9.8$^{+4.3}_{-3.1}$ & 6.9$^{+3.8}_{-2.6}$ & 16.7$^{+5.2}_{-4.0}$ & 3.2$^{+1.4}_{-1.0}$ & 0.65$^{+1.01}_{-0.11}$ & $2.3^{+0.7}_{-0.5}$ & $2.2\pm 0.1$ & $31.16\pm 0.02$ & $-1.53\pm 0.04$ & $+0.10\pm 0.04$ & 46.37 & 3592 & 0.61 & 27.3 & 1023 & 0.662 \\
\object{SDSS~J075115.43$+$505439.1} & 7.9$^{+3.9}_{-2.8}$ & 5.8$^{+3.6}_{-2.3}$ & 13.7$^{+4.8}_{-3.7}$ & 2.3$^{+1.2}_{-0.8}$ & 0.69$^{+1.39}_{-0.12}$ & $2.4^{+0.7}_{-0.5}$ & $6.4\pm 0.1$ & $31.90\pm 0.01$ & $-1.70\pm 0.04$ & $+0.08\pm 0.04$ & 46.59 & 3077 & 1.05 & 6.6 & 5953 & 1.353 \\
\object{SDSS~J075136.36$+$432732.4} & 3.9$^{+3.2}_{-1.9}$ & 3.8$^{+3.1}_{-1.9}$ & 7.7$^{+3.9}_{-2.7}$ & 0.9$^{+0.8}_{-0.4}$ & 0.90$^{+6.35}_{-0.14}$ & $1.8^{+0.8}_{-0.5}$ & $3.5\pm 0.0$ & $31.62\pm 0.01$ & $-1.65\pm 0.06$ & $+0.07\pm 0.06$ & 46.42 & 3736 & 0.60 & 33.5 & 1996 & 0.724 \\
\object{SDSS~J081019.47$+$095040.9} & $<4.8$ & $<4.8$ & $<6.4$ & $<1.2$ & $<0.92$ & $<0.8$ & $3.1\pm 0.1$ & $31.55\pm 0.01$ & $<-1.76$ & $<-0.05$ & 46.42 & 3297 & 0.77 & 25.8 & 2175 & 0.789 \\
\object{SDSS~J081558.35$+$154055.2} & 12.9$^{+4.7}_{-3.5}$ & 6.8$^{+3.7}_{-2.5}$ & 19.6$^{+5.5}_{-4.4}$ & 3.3$^{+1.2}_{-0.9}$ & 0.50$^{+0.68}_{-0.08}$ & $2.7^{+1.0}_{-0.7}$ & $3.9\pm 0.1$ & $31.66\pm 0.01$ & $-1.60\pm 0.05$ & $+0.13\pm 0.05$ & 46.53 & 4622 & 0.44 & 28.9 & 549 & 0.604 \\
\object{SDSS~J082603.32$+$342800.6} & 10.0$^{+4.3}_{-3.1}$ & 7.0$^{+3.8}_{-2.6}$ & 16.9$^{+5.2}_{-4.1}$ & 2.2$^{+0.9}_{-0.7}$ & 0.65$^{+1.01}_{-0.11}$ & $2.5^{+0.8}_{-0.6}$ & $3.6\pm 0.1$ & $31.65\pm 0.01$ & $-1.60\pm 0.05$ & $+0.13\pm 0.05$ & 46.40 & 5763 & 0.25 & 26.0 & 936 & 0.662 \\
\object{SDSS~J085337.36$+$121800.3} & 3.9$^{+3.2}_{-1.9}$ & $<6.4$ & 6.0$^{+3.6}_{-2.4}$ & 0.9$^{+0.7}_{-0.4}$ & $<1.41$ & $1.2^{+0.4}_{-0.3}$ & $5.1\pm 0.1$ & $31.76\pm 0.01$ & $-1.77\pm 0.04$ & $-0.03\pm 0.04$ & 46.56 & 4502 & 0.47 & 7.7 & 791 & 0.825 \\
\object{SDSS~J085443.10$+$075223.2} & 4.8$^{+3.4}_{-2.1}$ & 3.0$^{+2.9}_{-1.6}$ & 7.7$^{+3.9}_{-2.7}$ & 1.6$^{+1.1}_{-0.7}$ & 0.51$^{+3.24}_{-0.06}$ & $1.4^{+0.6}_{-0.4}$ & $3.3\pm 0.0$ & $31.33\pm 0.00$ & $-1.67\pm 0.06$ & $-0.01\pm 0.06$ & 46.32 & 3138 & 0.77 & 29.1 & 1466 & 0.696 \\
\object{SDSS~J090247.57$+$304120.7} & 14.9$^{+5.0}_{-3.8}$ & 12.5$^{+4.6}_{-3.5}$ & 27.5$^{+6.3}_{-5.2}$ & 4.9$^{+1.6}_{-1.3}$ & 0.84$^{+0.66}_{-0.15}$ & $3.8^{+1.5}_{-1.0}$ & $4.0\pm 0.3$ & $31.39\pm 0.03$ & $-1.54\pm 0.05$ & $+0.13\pm 0.05$ & 46.52 & 5370 & 0.32 & 47.7 & 229 & 0.490 \\
\object{SDSS~J090646.98$+$174046.8} & 4.8$^{+3.4}_{-2.1}$ & 5.9$^{+3.6}_{-2.4}$ & 10.7$^{+4.4}_{-3.2}$ & 1.6$^{+1.1}_{-0.7}$ & 1.12$^{+3.77}_{-0.22}$ & $2.7^{+0.9}_{-0.6}$ & $5.8\pm 0.3$ & $31.56\pm 0.02$ & $-1.66^{+0.05}_{-0.04}$ & $+0.05^{+0.05}_{-0.04}$ & 46.39 & 3174 & 0.80 & 16.7 & 2369 & 0.875 \\
\object{SDSS~J090710.36$+$430000.2} & 4.0$^{+3.2}_{-1.9}$ & 6.9$^{+3.8}_{-2.6}$ & 10.9$^{+4.4}_{-3.3}$ & 1.2$^{+1.0}_{-0.6}$ & 1.66$^{+6.47}_{-0.35}$ & $2.2^{+0.7}_{-0.5}$ & $7.9\pm 0.1$ & $31.95\pm 0.00$ & $-1.75\pm 0.05$ & $+0.03\pm 0.05$ & 46.62 & 3136 & 1.04 & 24.4 & 1230 & 0.703 \\
\object{SDSS~J091941.26$+$253537.7} & 4.8$^{+3.4}_{-2.1}$ & 11.7$^{+4.5}_{-3.4}$ & 17.3$^{+5.3}_{-4.1}$ & 0.9$^{+0.6}_{-0.4}$ & 2.35$^{+5.14}_{-0.55}$ & $2.0^{+0.7}_{-0.5}$ & $3.1\pm 0.1$ & $31.57\pm 0.01$ & $-1.61\pm 0.05$ & $+0.10\pm 0.05$ & 46.28 & 6110 & 0.19 & 36.7 & 159 & 0.530 \\
\object{SDSS~J092216.04$+$160526.4} & 22.4$^{+5.8}_{-4.7}$ & 14.8$^{+4.9}_{-3.8}$ & 37.1$^{+7.2}_{-6.1}$ & 3.2$^{+0.8}_{-0.7}$ & 0.66$^{+0.46}_{-0.10}$ & $2.6^{+1.0}_{-0.7}$ & $4.0\pm 0.1$ & $31.72\pm 0.01$ & $-1.61^{+0.05}_{-0.06}$ & $+0.13^{+0.05}_{-0.06}$ & 46.37 & 3026 & 0.87 & 43.7 & 53 & 0.493 \\
\object{SDSS~J092523.24$+$214119.8} & 28.0$^{+6.4}_{-5.3}$ & 21.7$^{+5.7}_{-4.6}$ & 50.7$^{+8.2}_{-7.1}$ & 4.3$^{+1.0}_{-0.8}$ & 0.77$^{+0.39}_{-0.12}$ & $4.6^{+1.5}_{-1.1}$ & $2.4\pm 0.1$ & $31.48\pm 0.01$ & $-1.42\pm 0.05$ & $+0.27\pm 0.05$ & 46.22 & 3133 & 0.69 & 44.1 & 287 & 0.510 \\
\object{SDSS~J092555.05$+$490338.2} & 2.9$^{+2.9}_{-1.6}$ & 1.9$^{+2.6}_{-1.3}$ & 6.6$^{+3.7}_{-2.5}$ & 0.6$^{+0.6}_{-0.3}$ & 0.68$^{+1.25}_{-0.56}$\tablenotemark{e} & $1.1^{+0.3}_{-0.2}$ & $1.9\pm 0.1$ & $31.37\pm 0.02$ & $-1.63\pm 0.04$ & $+0.04\pm 0.04$ & 46.34 & 6598 & 0.18 & 32.2 & 548 & 0.587 \\
\object{SDSS~J093533.88$+$235720.5} & 7.0$^{+3.8}_{-2.6}$ & 5.8$^{+3.6}_{-2.3}$ & 12.8$^{+4.7}_{-3.5}$ & 1.3$^{+0.7}_{-0.5}$ & 0.79$^{+1.76}_{-0.14}$ & $1.6^{+0.5}_{-0.4}$ & $4.3\pm 0.4$ & $31.73\pm 0.04$ & $-1.70\pm 0.05$ & $+0.04\pm 0.05$ & 46.34 & 6736 & 0.17 & 44.1 & 820 & 0.561 \\
\object{SDSS~J094637.83$-$012411.5} & $<4.8$ & $<3.0$ & $<4.8$ & $<1.1$ & $<0.51$ & $<0.7$ & $1.9\pm 0.1$ & $31.35\pm 0.02$ & $<-1.70$ & $<-0.04$ & 46.52 & 3366 & 0.81 & 19.4 & 92 & 0.625 \\
\object{SDSS~J094648.59$+$171827.7} & 15.4$^{+5.0}_{-3.9}$ & 6.7$^{+3.7}_{-2.5}$ & 22.0$^{+5.8}_{-4.6}$ & 3.1$^{+1.0}_{-0.8}$ & 0.43$^{+0.57}_{-0.07}$ & $2.5^{+1.0}_{-0.7}$ & $3.0\pm 0.0$ & $31.56\pm 0.01$ & $-1.56\pm 0.06$ & $+0.15\pm 0.06$ & 46.43 & 4627 & 0.39 & 44.5 & 346 & 0.513 \\
\object{SDSS~J095330.36$+$353223.1} & 9.0$^{+4.1}_{-2.9}$ & $<8.0$ & 11.9$^{+4.6}_{-3.4}$ & 1.8$^{+0.8}_{-0.6}$ & $<0.83$ & $1.4^{+0.5}_{-0.4}$ & $3.8\pm 0.2$ & $31.70\pm 0.02$ & $-1.70\pm 0.05$ & $+0.04\pm 0.05$ & 46.39 & 4235 & 0.45 & 13.5 & 2488 & 0.914 \\
\object{SDSS~J095555.68$+$351652.6} & 24.4$^{+6.0}_{-4.9}$ & 18.8$^{+5.4}_{-4.3}$ & 43.1$^{+7.6}_{-6.5}$ & 5.9$^{+1.5}_{-1.2}$ & 0.78$^{+0.43}_{-0.12}$ & $4.1^{+1.4}_{-1.1}$ & $2.7\pm 0.0$ & $31.25\pm 0.01$ & $-1.47\pm 0.05$ & $+0.18\pm 0.05$ & 46.26 & 6595 & 0.16 & 29.2 & $-149$ & 0.538 \\
\object{SDSS~J095823.07$+$371218.3} & $<4.8$ & $<3.0$ & $<4.8$ & $<1.2$ & $<0.51$ & $<0.8$ & $2.5\pm 0.0$ & $31.48\pm 0.00$ & $<-1.72$ & $<-0.03$ & 46.47 & 3566 & 0.69 & 22.5 & 1898 & 0.783 \\
\object{SDSS~J100212.63$+$520800.2} & 6.9$^{+3.8}_{-2.6}$ & 5.8$^{+3.6}_{-2.3}$ & 12.7$^{+4.7}_{-3.5}$ & 1.9$^{+1.0}_{-0.7}$ & 0.79$^{+1.76}_{-0.14}$ & $1.2^{+0.5}_{-0.3}$ & $2.4\pm 0.0$ & $31.20\pm 0.01$ & $-1.65\pm 0.06$ & $-0.02\pm 0.06$ & 46.21 & 3123 & 0.69 & 17.7 & 889 & 0.724 \\
\object{SDSS~J101106.74$+$114759.4} & 7.9$^{+3.9}_{-2.7}$ & 9.9$^{+4.3}_{-3.1}$ & 17.7$^{+5.3}_{-4.2}$ & 1.4$^{+0.7}_{-0.5}$ & 1.19$^{+1.76}_{-0.22}$ & $1.8^{+0.7}_{-0.5}$ & $2.4\pm 0.0$ & $31.46\pm 0.01$ & $-1.58\pm 0.05$ & $+0.10\pm 0.05$ & 46.35 & 2275 & 1.50 & 30.4 & 575 & 0.598 \\
\object{SDSS~J101429.57$+$481938.4} & 3.0$^{+2.9}_{-1.6}$ & 5.0$^{+3.4}_{-2.2}$ & 7.9$^{+3.9}_{-2.8}$ & 1.0$^{+1.0}_{-0.6}$ & 1.53$^{+13.8}_{-0.14}$ & $1.5^{+0.4}_{-0.3}$ & $4.3\pm 0.0$ & $31.42\pm 0.00$ & $-1.71\pm 0.04$ & $-0.03\pm 0.04$ & 46.36 & 2841 & 0.97 & 19.0 & 2147 & 0.834 \\
\object{SDSS~J102731.49$+$541809.7} & 9.7$^{+4.2}_{-3.1}$ & 7.9$^{+3.9}_{-2.7}$ & 17.6$^{+5.3}_{-4.2}$ & 2.8$^{+1.2}_{-0.9}$ & 0.75$^{+1.09}_{-0.13}$ & $2.5^{+0.8}_{-0.6}$ & $3.9\pm 0.1$ & $31.40\pm 0.01$ & $-1.61\pm 0.05$ & $+0.07\pm 0.05$ & 46.31 & 6892 & 0.16 & 21.2 & 382 & 0.640 \\
\object{SDSS~J103209.78$+$385630.5} & 5.0$^{+3.4}_{-2.1}$ & 3.0$^{+2.9}_{-1.6}$ & 7.9$^{+3.9}_{-2.7}$ & 1.3$^{+0.9}_{-0.6}$ & 0.51$^{+3.24}_{-0.06}$ & $0.9^{+0.3}_{-0.2}$ & $4.9\pm 0.1$ & $31.49\pm 0.01$ & $-1.81\pm 0.05$ & $-0.12\pm 0.05$ & 46.28 & 4281 & 0.40 & 15.6 & 658 & 0.723 \\
\object{SDSS~J103236.98$+$230554.1} & $<6.4$ & $<4.8$ & $<8.0$ & $<1.2$ & $<0.76$ & $<0.8$ & $3.1\pm 0.1$ & $31.61\pm 0.01$ & $<-1.76$ & $<-0.04$ & 46.36 & 3267 & 0.74 & 18.3 & $-33$ & 0.621 \\
\object{SDSS~J104336.73$+$494707.6} & 24.0$^{+6.0}_{-4.9}$ & 18.4$^{+5.4}_{-4.2}$ & 43.2$^{+7.6}_{-6.5}$ & 3.3$^{+0.8}_{-0.7}$ & 0.73$^{+0.43}_{-0.11}$ & $2.5^{+0.8}_{-0.6}$ & $4.5\pm 0.1$ & $31.70\pm 0.01$ & $-1.63\pm 0.05$ & $+0.10\pm 0.05$ & 46.52 & 7025 & 0.18 & 29.7 & 1669 & 0.713 \\
\object{SDSS~J105045.72$+$544719.2} & 10.9$^{+4.4}_{-3.2}$ & 15.5$^{+5.0}_{-3.9}$ & 26.5$^{+6.2}_{-5.1}$ & 2.0$^{+0.8}_{-0.6}$ & 1.32$^{+1.22}_{-0.25}$ & $2.4^{+0.9}_{-0.7}$ & $6.7\pm 0.0$ & $31.87\pm 0.00$ & $-1.70\pm 0.05$ & $+0.07\pm 0.05$ & 46.66 & 4781 & 0.46 & 32.8 & 914 & 0.621 \\
\object{SDSS~J105926.43$+$062227.4} & $<9.4$ & 9.8$^{+4.3}_{-3.1}$ & 13.7$^{+4.8}_{-3.7}$ & $<3.1$ & $<1.06$ & $2.8^{+1.0}_{-0.7}$ & $8.1\pm 0.2$ & $31.96\pm 0.01$ & $-1.71\pm 0.05$ & $+0.08\pm 0.05$ & 46.72 & 6361 & 0.28 & 48.3 & 297 & 0.493 \\
\object{SDSS~J110735.58$+$642008.6} & 5.9$^{+3.6}_{-2.4}$ & 6.0$^{+3.6}_{-2.4}$ & 11.9$^{+4.6}_{-3.4}$ & 1.1$^{+0.7}_{-0.5}$ & 0.93$^{+2.37}_{-0.18}$ & $1.6^{+0.5}_{-0.4}$ & $6.5\pm 0.0$ & $31.91\pm 0.00$ & $-1.77\pm 0.05$ & $+0.00\pm 0.05$ & 46.40 & 2937 & 0.95 & 24.3 & 1887 & 0.769 \\
\object{SDSS~J110810.87$+$014140.7} & $<4.8$ & $<3.0$ & $<4.8$ & $<1.3$ & $<0.51$ & $<0.7$ & $3.9\pm 0.2$ & $31.40\pm 0.02$ & $<-1.82$ & $<-0.14$ & 46.33 & 4784 & 0.33 & 22.2 & $-123$ & 0.581 \\
\object{SDSS~J111850.02$+$351311.7} & 11.7$^{+4.5}_{-3.4}$ & 9.9$^{+4.3}_{-3.1}$ & 21.7$^{+5.7}_{-4.6}$ & 1.7$^{+0.7}_{-0.5}$ & 0.79$^{+0.91}_{-0.14}$ & $1.9^{+0.7}_{-0.5}$ & $4.0\pm 0.1$ & $31.65\pm 0.01$ & $-1.66\pm 0.05$ & $+0.07\pm 0.05$ & 46.51 & 3664 & 0.68 & 22.3 & 2266 & 0.823 \\
\object{SDSS~J114212.25$+$233250.5} & 2.9$^{+2.9}_{-1.6}$ & $<4.8$ & 4.8$^{+3.4}_{-2.1}$ & 0.9$^{+0.9}_{-0.5}$ & $<1.53$ & $0.6^{+0.3}_{-0.2}$ & $8.1\pm 0.1$ & $31.71\pm 0.00$ & $-1.96^{+0.06}_{-0.05}$ & $-0.22^{+0.06}_{-0.05}$ & 46.35 & 3518 & 0.63 & 19.3 & 2427 & 0.862 \\
\object{SDSS~J114350.30$+$362911.3} & 8.8$^{+4.1}_{-2.9}$ & 5.9$^{+3.6}_{-2.4}$ & 16.7$^{+5.2}_{-4.0}$ & 2.4$^{+1.1}_{-0.8}$ & 0.61$^{+1.16}_{-0.10}$ & $3.0^{+1.0}_{-0.7}$ & $5.6\pm 0.2$ & $31.86\pm 0.01$ & $-1.64\pm 0.05$ & $+0.13\pm 0.05$ & 46.44 & 4090 & 0.51 & 15.3 & 424 & 0.703 \\
\object{SDSS~J114907.15$+$004104.3} & $<4.8$ & 3.0$^{+2.9}_{-1.6}$ & 3.9$^{+3.2}_{-1.9}$ & $<1.5$ & $<0.51$ & $<0.5$ & $2.7\pm 0.1$ & $31.53\pm 0.01$ & $<-1.82$ & $<-0.12$ & 46.50 & 5920 & 0.26 & 23.9 & 1931 & 0.776 \\
\object{SDSS~J121314.03$+$080703.6} & 17.6$^{+5.3}_{-4.2}$ & 10.9$^{+4.4}_{-3.2}$ & 28.4$^{+6.4}_{-5.3}$ & 3.9$^{+1.2}_{-0.9}$ & 0.58$^{+0.52}_{-0.10}$ & $3.7^{+1.4}_{-1.0}$ & $4.6\pm 0.1$ & $31.78\pm 0.00$ & $-1.57\pm 0.05$ & $+0.18\pm 0.05$ & 46.44 & 3761 & 0.60 & 15.0 & 2950 & 0.955 \\
\object{SDSS~J121810.98$+$241410.9} & 11.6$^{+4.5}_{-3.4}$ & 6.8$^{+3.7}_{-2.5}$ & 18.4$^{+5.4}_{-4.2}$ & 3.2$^{+1.2}_{-0.9}$ & 0.54$^{+0.79}_{-0.09}$ & $2.7^{+0.9}_{-0.7}$ & $4.3\pm 0.3$ & $31.75\pm 0.03$ & $-1.61\pm 0.05$ & $+0.13\pm 0.05$ & 46.73 & 3981 & 0.72 & 30.8 & 1221 & 0.662 \\
\object{SDSS~J125150.45$+$114340.7} & 8.8$^{+4.1}_{-2.9}$ & 9.8$^{+4.3}_{-3.1}$ & 18.7$^{+5.4}_{-4.3}$ & 2.6$^{+1.2}_{-0.8}$ & 1.06$^{+1.41}_{-0.21}$ & $2.8^{+1.0}_{-0.8}$ & $3.0\pm 0.0$ & $31.54\pm 0.01$ & $-1.55\pm 0.05$ & $+0.16\pm 0.05$ & 46.49 & 4635 & 0.42 & 47.0 & 2031 & 0.671 \\
\object{SDSS~J125159.90$+$500203.6} & $<6.4$ & $<4.8$ & $<8.0$ & $<1.4$ & $<0.76$ & $<1.0$ & $4.0\pm 0.1$ & $31.72\pm 0.01$ & $<-1.77$ & $<-0.03$ & 46.51 & 3365 & 0.81 & 28.7 & 1113 & 0.663 \\
\object{SDSS~J134341.99$+$255652.9} & 2.0$^{+2.6}_{-1.3}$ & 3.0$^{+2.9}_{-1.6}$ & 4.8$^{+3.4}_{-2.1}$ & 0.6$^{+0.9}_{-0.4}$ & 1.24$^{+7.24}_{-0.03}$ & $0.7^{+0.3}_{-0.2}$ & $7.6\pm 0.1$ & $31.69\pm 0.01$ & $-1.94\pm 0.06$ & $-0.21\pm 0.06$ & 46.60 & 12197 & 0.07 & 16.7 & $-4$ & 0.641 \\
\object{SDSS~J135908.35$+$305830.8} & 7.0$^{+3.8}_{-2.6}$ & 2.9$^{+2.9}_{-1.6}$ & 9.8$^{+4.3}_{-3.1}$ & 1.9$^{+1.0}_{-0.7}$ & 0.36$^{+1.45}_{-0.04}$ & $1.9^{+0.9}_{-0.6}$ & $4.1\pm 0.2$ & $31.70\pm 0.02$ & $-1.66^{+0.07}_{-0.06}$ & $+0.07^{+0.07}_{-0.06}$ & 46.50 & 5795 & 0.27 & 27.9 & 3593 & 0.935 \\
\object{SDSS~J144624.29$+$173128.8} & $<4.8$ & $<4.8$ & $<6.4$ & $<1.5$ & $<0.92$ & $<1.0$ & $4.3\pm 0.0$ & $31.69\pm 0.00$ & $<-1.78$ & $<-0.04$ & 46.52 & 2938 & 1.07 & 14.9 & 1139 & 0.773 \\
\object{SDSS~J144706.81$+$212839.2} & 4.9$^{+3.4}_{-2.1}$ & 6.9$^{+3.8}_{-2.6}$ & 11.8$^{+4.5}_{-3.4}$ & 0.8$^{+0.6}_{-0.4}$ & 1.32$^{+3.67}_{-0.28}$ & $1.6^{+0.6}_{-0.4}$ & $1.1\pm 0.2$ & $31.38\pm 0.06$ & $-1.48\pm 0.05$ & $+0.19\pm 0.05$ & 46.76 & 6113 & 0.31 & 14.1 & 1354 & 0.800 \\
\object{SDSS~J144948.62$+$123047.4} & 16.8$^{+5.2}_{-4.1}$ & 20.7$^{+5.6}_{-4.5}$ & 37.5$^{+7.2}_{-6.1}$ & 4.9$^{+1.5}_{-1.2}$ & 1.22$^{+0.72}_{-0.22}$ & $5.2^{+1.9}_{-1.4}$ & $3.1\pm 0.2$ & $31.29\pm 0.02$ & $-1.45\pm 0.05$ & $+0.21\pm 0.05$ & 46.34 & 6505 & 0.18 & 66.2 & 471 & 0.405 \\
\object{SDSS~J151727.68$+$133358.6} & 6.7$^{+3.7}_{-2.5}$ & 6.0$^{+3.6}_{-2.4}$ & 15.6$^{+5.0}_{-3.9}$ & 1.5$^{+0.9}_{-0.6}$ & 0.79$^{+1.76}_{-1.11}$ & $1.6^{+0.6}_{-0.4}$ & $3.7\pm 0.1$ & $31.64\pm 0.01$ & $-1.67\pm 0.05$ & $+0.05\pm 0.05$ & 46.45 & 4680 & 0.39 & 26.8 & 1421 & 0.706 \\
\object{SDSS~J214901.21$-$073141.6} & 6.9$^{+3.8}_{-2.6}$ & $<6.4$ & 8.8$^{+4.1}_{-2.9}$ & 2.3$^{+1.2}_{-0.8}$ & $<0.79$ & $6.0^{+3.6}_{-2.1}$ & $2.6\pm 0.3$ & $31.47\pm 0.05$ & $-1.39^{+0.08}_{-0.07}$ & $+0.30^{+0.08}_{-0.07}$ & 46.44 & 3181 & 0.84 & 21.1 & 1232 & 0.728 \\
\noalign{\smallskip}\hline\noalign{\smallskip}
\multicolumn{17}{c}{Archival Observations from M22} \\
\noalign{\smallskip}\hline\noalign{\smallskip}
\object{SDSS~J080117.79$+$521034.5} & 116.0$^{+11.8}_{-10.8}$ & 49.5$^{+8.1}_{-7.0}$ & 168.3$^{+14.0}_{-13.0}$ & 2.7$^{+0.3}_{-0.2}$ & 0.43$^{+0.10}_{-0.05}$ & $1.4^{+0.7}_{-0.5}$ & $6.4\pm 0.2$ & $32.15\pm 0.01$ & $-1.79\pm 0.07$ & $+0.04\pm 0.07$ & 46.92 & 5361 & 1.63 & 19.3 & 3267 & 0.959 \\
\object{SDSS~J084846.11$+$611234.6} & 30.3$^{+6.6}_{-5.5}$ & 8.8$^{+4.1}_{-2.9}$ & 40.0$^{+7.4}_{-6.3}$ & 19.7$^{+4.3}_{-3.6}$ & 0.28$^{+0.25}_{-0.04}$ & $7.7^{+4.6}_{-2.9}$ & $7.7\pm 0.0$ & $31.97\pm 0.00$ & $-1.54\pm 0.08$ & $+0.25\pm 0.08$ & 46.82 & 4180 & 1.48 & 29.2 & 139 & 0.563 \\
\object{SDSS~J094602.31$+$274407.0} & 4.0$^{+3.2}_{-1.9}$ & $<4.8$ & 5.0$^{+3.4}_{-2.1}$ & 0.8$^{+0.6}_{-0.4}$ & 1.15$^{+6.67}_{-0.20}$ & $0.3^{+0.2}_{-0.1}$ & $7.1\pm 0.0$ & $32.00\pm 0.00$ & $-2.08\pm 0.08$ & $-0.28\pm 0.08$ & 46.75 & 3833 & 2.89 & 5.9 & 8477 & 1.359 \\
\object{SDSS~J094646.94$+$392719.0} & 13.9$^{+4.8}_{-3.7}$ & 6.8$^{+3.7}_{-2.5}$ & 20.6$^{+5.6}_{-4.5}$ & 0.5$^{+0.2}_{-0.1}$ & 0.46$^{+0.62}_{-0.07}$ & $0.2\pm 0.1$ & $3.5\pm 0.2$ & $31.61\pm 0.02$ & $-2.00^{+0.07}_{-0.08}$ & $-0.28^{+0.07}_{-0.08}$ & 46.37 & 5214 & 1.08 & 29.2 & 3901 & 0.989 \\
\object{SDSS~J095852.19$+$120245.0} & 15.8$^{+5.1}_{-3.9}$ & 2.0$^{+2.6}_{-1.3}$ & 20.8$^{+5.6}_{-4.5}$ & 10.2$^{+3.3}_{-2.5}$ & 0.10$^{+0.39}_{-0.00}$ & $5.3^{+3.3}_{-2.1}$ & $1.6\pm 0.2$ & $31.54\pm 0.05$ & $-1.33\pm 0.08$ & $+0.37\pm 0.08$ & 46.76 & 4417 & 1.07 & 16.8 & 707 & 0.715 \\
\object{SDSS~J102907.09$+$651024.6} & 114.1$^{+11.7}_{-10.7}$ & 24.6$^{+6.0}_{-4.9}$ & 139.4$^{+12.8}_{-11.8}$ & 10.7$^{+1.1}_{-1.0}$ & 0.22$^{+0.07}_{-0.03}$ & $3.2^{+1.8}_{-1.2}$ & $10.0\pm 0.5$ & $32.05\pm 0.02$ & $-1.73\pm 0.08$ & $+0.08\pm 0.08$ & 46.66 & 4545 & 0.90 & 23.8 & 1764 & 0.760 \\
\object{SDSS~J111119.10$+$133603.8} & 134.5$^{+12.6}_{-11.6}$ & 45.2$^{+7.8}_{-6.7}$ & 179.5$^{+14.4}_{-13.4}$ & 3.1$\pm 0.3$ & 0.33$^{+0.08}_{-0.04}$ & $1.9^{+1.0}_{-0.7}$ & $2.9\pm 0.2$ & $31.86\pm 0.03$ & $-1.61\pm 0.07$ & $+0.16\pm 0.07$ & 46.94 & 6936 & 0.70 & 19.5 & 988 & 0.718 \\
\object{SDSS~J141028.14$+$135950.2} & 49.2$^{+8.1}_{-7.0}$ & 14.8$^{+4.9}_{-3.8}$ & 63.8$^{+9.0}_{-8.0}$ & 12.2$^{+2.0}_{-1.7}$ & 0.30$^{+0.16}_{-0.04}$ & $3.9^{+2.3}_{-1.4}$ & $4.5\pm 0.0$ & $31.72\pm 0.00$ & $-1.56\pm 0.08$ & $+0.18\pm 0.08$ & 46.67 & 6555 & 0.87 & 38.7 & 1353 & 0.637 \\
\object{SDSS~J141951.84$+$470901.3} & 125.1$^{+12.2}_{-11.2}$ & 28.7$^{+6.4}_{-5.3}$ & 154.4$^{+13.5}_{-12.4}$ & 16.3$^{+1.6}_{-1.5}$ & 0.23$^{+0.07}_{-0.03}$ & $4.8^{+2.9}_{-1.9}$ & $6.6\pm 0.2$ & $31.92\pm 0.01$ & $-1.59\pm 0.08$ & $+0.19\pm 0.08$ & 46.76 & 4938 & 1.19 & 24.9 & 3100 & 0.909
\enddata
\tablenotetext{a}{Count rate computed in the soft band (observed-frame \hbox{$0.5-2$~keV}) in units of $10^{-3}$~counts~s$^{-1}$.}
\tablenotetext{b}{Galactic absorption-corrected flux density at rest-frame $2$~keV in units of $10^{-31}$~erg~cm$^{-2}$~s$^{-1}$~Hz$^{-1}$ assuming a power-law model with $\Gamma=2.0$.}
\tablenotetext{c}{Flux density at rest-frame $2500$\,\text{\AA} with units of $10^{-27}$\,erg\,cm$^{-2}$\,s$^{-1}$\,Hz$^{-1}$.}
\tablenotetext{d}{Taken from Matthews \et (2023).}
\tablenotetext{e}{For the two faintest \xray\ sources in our sample, the band ratio was derived from the ratio of the hard-band counts to the soft-band counts.}
\end{deluxetable*}
\end{longrotatetable}

\section{Results and Discussion}\label{sec:results}

\subsection{Optical-to-X-ray Spectral Slope (\aox)}
The main goal of this work is to test if basic \xray\ data can provide improved quasar \lledd\ estimates with respect to those obtained solely from optical-UV \lledd\ indicators such as in the \hb\ or \civ\ spectroscopic parameter space. To this end, we present new \xray\ data that allow us to derive \aox\ values, or upper limits, for 63 GNIRS-DQS quasars which are luminous sources at $1.5 \lesssim z \lesssim 3.5$. The \xray\ data for nine of these sources were presented in M22.

\subsubsection{\aox\ vs. \luv}\label{sec:aox_luv}
Our first step is to test whether our sources follow the well-known \aox-\luv\ anti-correlation (e.g., Steffen \et 2006; Just \et 2007; Lusso \et 2010; Timlin \et 2020). We obtain a Spearman-rank correlation coefficient $r_{\rm S} = -0.430$ and a chance probability value of $p = 0.001$ indicating that a significant anti-correlation is observed between these two parameters for 55 \xray-detected sources, likely due to their relatively narrow luminosity range. Adding \aox\ and \luv\ values for the remaining 44 SDSS, non-GNIRS-DQS quasars from M22 to the correlation results in \hbox{$r_{\rm S} = -0.354$} and a chance probability value of $p = 3.286\times 10^{-4}$ indicating a strong anti-correlation for 99 sources across three orders of magnitude in UV luminosity.\footnote{For additional details regarding the number of sources used for each sub-sample, see Appendix~\ref{sec:A}.}

\begin{figure}
\hspace{-0.5cm}
\epsscale{1.2}
\plotone{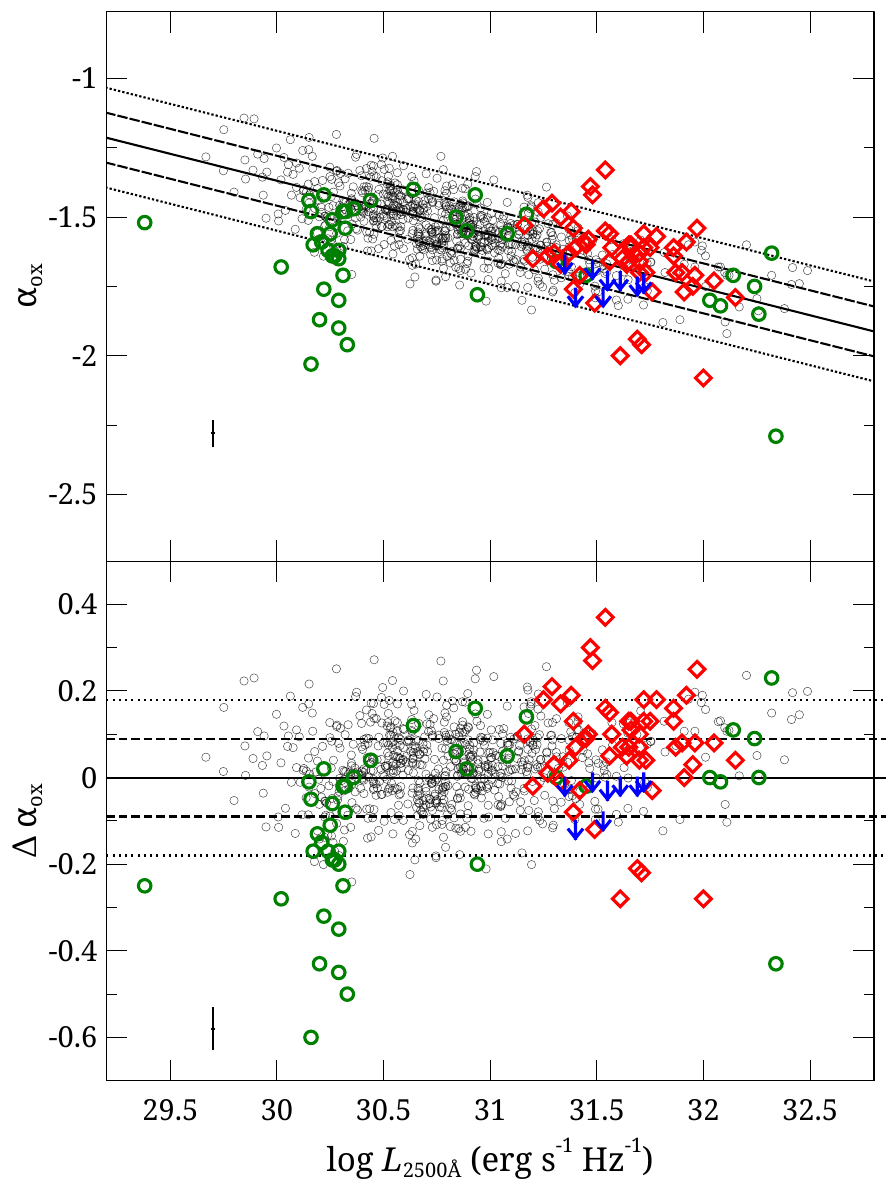}
\caption{Optical-to-X-ray spectral slope (\aox) (top) and the luminosity corrected \aox\ value, \daox\ (bottom) versus \luv. Red diamonds represent \xray-detected sources from this work, with blue arrows representing \xray-undetected sources; green and black circles represent sources from M22 and R22, respectively; median error bars are shown in the lower left-hand corner of each panel, for clarity. The solid line in the top panel represents the best fit relation from Eq. [3] of Timlin \et (2020); the dashed and dotted lines show the 1$\sigma$ and 2$\sigma$ deviations from this relation, respectively. The solid line in the bottom panel represents \daox$=0$ with the dashed and dotted lines representing the 1$\sigma$ and 2$\sigma$ deviations from the Timlin \et (2020) relation (as in the top panel). With the exception of $\sim5$\% of the sources lying above or below the 2$\sigma$ boundaries, our new observations as well as those of M22 and R22 are consistent with the Timlin \et (2020) relation.}
\label{fig:rivera}
\end{figure}

For context, we also compare our results with those obtained with the ``Good'' sample of R22, which utilizes \aox\ and \luv\ values for 779 SDSS quasars spanning three orders of magnitude in UV luminosity; most of this sample was constructed from samples presented in Lusso \et (2020) and Timlin \et (2020). A Spearman-rank correlation results in $r_{\rm S} = -0.624$ and a chance probability value of $p << 0.001$ for a total of 858 sources\footnote{We have removed 20 sources from the R22 sample that have data in M22; see Appendix~\ref{sec:B} for additional details.}, including the M22 sample and the sample presented in this work. The top panel of Figure~\ref{fig:rivera} shows the dependence of \aox\ on \luv\ for the sample presented in this work, as well as the M22 and R22 samples. Our results are consistent with the well-known \aox-\luv\ anti-correlation. We find that this anti-correlation becomes stronger with the increase in sample size and in the probed luminosity range.

The bottom panel of Figure~\ref{fig:rivera} shows the dependence of \daox\ on \luv\ for our sample, as well as the samples of M22 and R22. Although the \daox\ parameter is constructed to remove the dependence of \xray\ emission strength on UV luminosity, it may display a correlation with \luv. For our sample of 55 GNIRS-DQS sources, the Spearman-rank correlation coefficient and chance probability are \hbox{$r_{\rm S} = -0.040$}, and \hbox{$p = 0.771$}, respectively, becoming \hbox{$r_{\rm S} = 0.478$}, \hbox{$p << 0.001$} when 44 sources from the M22 sample are added, and resulting in $r_{\rm S} = 0.203$, $p << 0.001$ when 759 sources from R22 are added to the analysis. Formally, we observe a significant, positive correlation between \daox\ and \luv\ for 859 quasars across three orders of magnitude in UV luminosity. We note, however, that this is likely a consequence of the inability of the Timlin \et (2020) relation to completely remove the luminosity dependence from a sample that is subject to diverse selection criteria across multiple sub-samples, particularly that of M22 which possesses an overabundance of \xray\ weak sources (see Figure~\ref{fig:rivera}).

\subsubsection{\aox\ vs. \lledd}\label{sec:aox_lledd}
\begin{figure}
\hspace{-0.5cm}
\epsscale{1.2}
\plotone{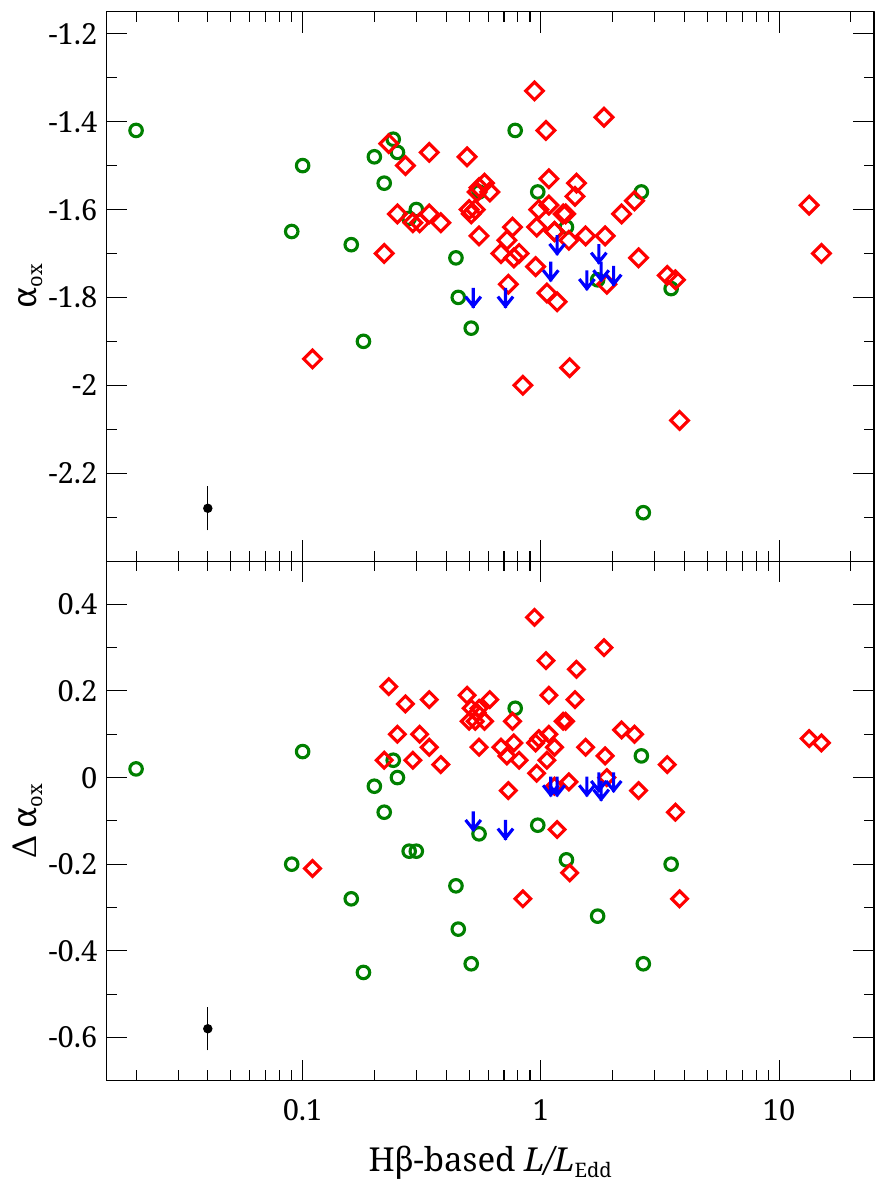}
\caption{Same as Figure~\ref{fig:rivera} but for \aox\ (top) and \daox\ (bottom) versus \hb-based \lledd.} 
\label{fig:corr}
\end{figure}

Figure~\ref{fig:corr} shows the dependence of \aox\ and \daox\ on \hb-based \lledd\ for the sources in this work as well as 22 non-GNIRS-DQS sources\footnote{We have removed 22 sources from the M22 sample that lack publicly available Fe~{\sc ii} data; see Figure~\ref{fig:venn_diagram} and Appendix~\ref{sec:C} for additional details.} from M22. We find no significant Spearman-rank correlations between \hb-based \lledd\ and either \aox\ or \daox\ ($r_{\rm S} = -0.245$, $p = 0.071$, and \hbox{$r_{\rm S} = -0.217$}, $p = 0.112$, respectively) for the 55 GNIRS-DQS \xray-detected sources in this work, and also when the 22 M22 sources are added (\hbox{$r_{\rm S} = -0.275$}, $p = 0.016$ and $r_{\rm S} = 0.057$, $p = 0.624$, respectively). These results are generally consistent with those of Shemmer \et (2008) who found that the anti-correlation between \aox\ and \hb-based \lledd, while significant, is milder with respect to, and likely driven by, the \aox-\luv\ anti-correlation. These results confirm that the \aox\ parameter, by itself, cannot serve as a robust \lledd\ indicator; it is expected to depend on \mbh\ as well (see, e.g., Vasudevan \& Fabian 2007; Grupe \et 2010; Wu \et 2012; Liu \et 2021). However, a considerably larger sample is required to explore the dependence of \aox\ on \hb-based \lledd\ and \mbh\ separately.

\subsubsection{\aox\ vs. \civ}\label{sec:aox_civ}
\begin{figure}
\hspace{-0.5cm}
\epsscale{1.2}
\plotone{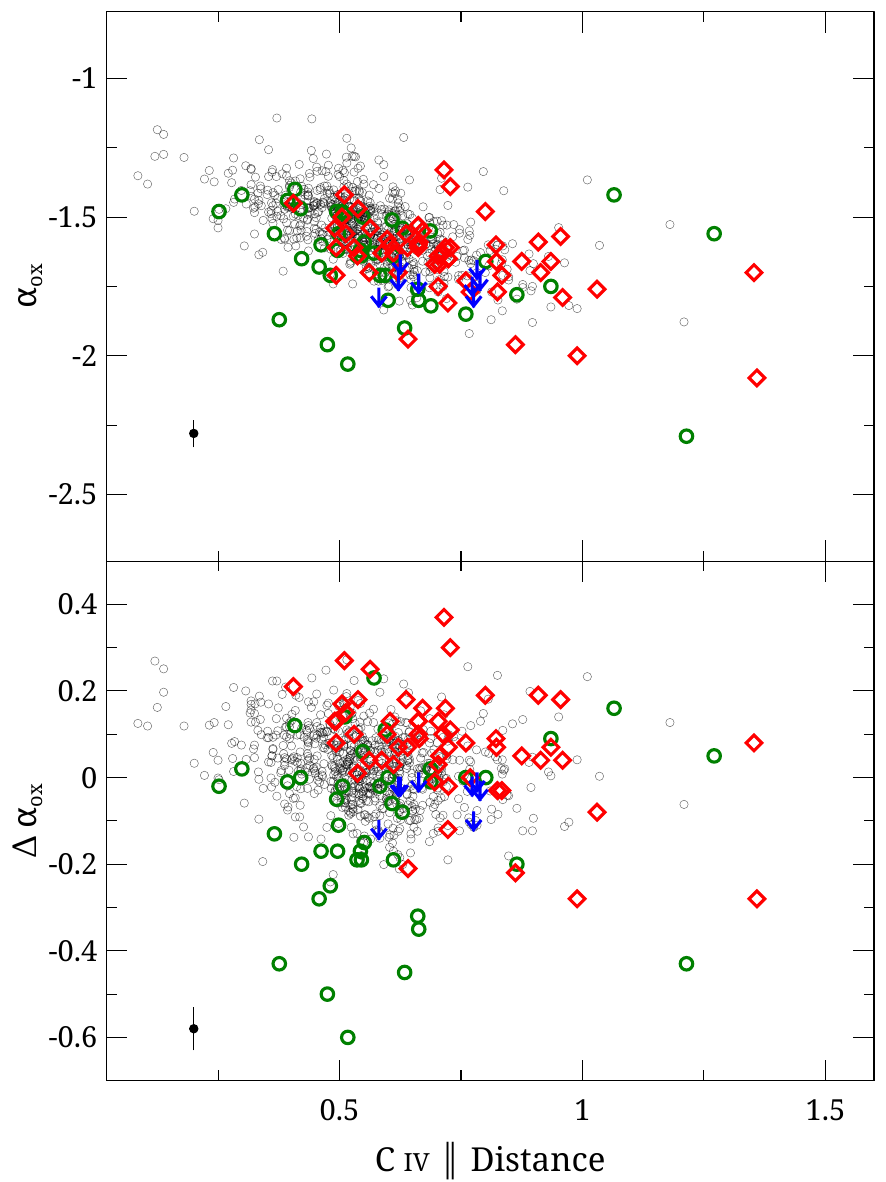}
\caption{Same as Figure~\ref{fig:rivera} but for \aox\ (top) and \daox\ (bottom) versus the \civdist\ parameter. Both \aox\ and \daox\ are significantly anti-correlated with \civdist\ for the combination of all three samples, consistent with the R22 findings.} 
\label{fig:rivera2}
\end{figure}

Figure~\ref{fig:rivera2} shows the dependence of \aox\ (top) and \daox\ (bottom) on the \civdist\ parameter, respectively, for the sample presented in this work, 42 sources\footnote{We have removed 2 sources from the M22 sample that do not have publicly available \civdist\ data; see Figure~\ref{fig:venn_diagram} and Appendix~\ref{sec:C} for additional details.} from the M22 sample, as well as the R22 sample for comparison. For the dependence of \aox\ on \civdist, we find a significant anti-correlation ($r_{\rm S} = -0.489$, \hbox{$p = 1.531\times10^{-4}$}) for the 55 GNIRS-DQS \xray-detected sources in this work. This anti-correlation remains significant when adding 42 sources from M22 ($r_{\rm S} = -0.372$, \hbox{$p = 1.777\times 10^{-4}$).} When adding 759 sources from R22, the correlation results in  \hbox{$r_{\rm S} = -0.606$,} and $p << 0.001$ for a total of 856 sources.\footnote{We note that \civdist\ values in R22 were derived using a different procedure (i.e., Coatman \et 2016) than those in this work (i.e., H23).} Such a strong anti-correlation would seem to support the connection between the BELR and the \xray-producing corona.

The bottom panel of Figure~\ref{fig:rivera2} shows the dependence of \daox\ on the \civdist\ parameter. As with \aox\ above, we ran correlations for only the GNIRS-DQS sources, the addition of the 42 M22 sources, and the addition of the 759 R22 sources to that. The resulting correlation coefficients and their chance probabilities are $r_{\rm S} = -0.387$ and $p = 0.004$, $r_{\rm S} = 0.105$ and $p = 0.308$, $r_{\rm S} = -0.204$ and $p << 0.001$, respectively. We therefore find that \aox\ and \daox\ are significantly anti-correlated with the \civdist\ parameter only when a sufficiently large sample of sources spanning wide ranges in parameter space is considered. These results are consistent with the R22 findings where larger \civdist\ typically indicates more negative values of both \aox\ and \daox. These trends seem to indicate that more \xray\ weak quasars occupy the more extreme regions in the \civ\ parameter space. Specifically, sources that are \xray\ weaker, whether in relative or absolute terms, do not have a sufficient amount of highly ionizing photons reaching the BELR to produce strong \civ\ emission; the larger \civ\ blueshifts also indicate the lower levels of ionization (e.g., Richards \et 2011; Luo \et 2015; Ni \et 2018; Giustini \& Proga 2019).

\subsubsection{\civ\ vs. \lledd}\label{sec:civ_lledd}
\begin{figure*}
\epsscale{1.0}
\plotone{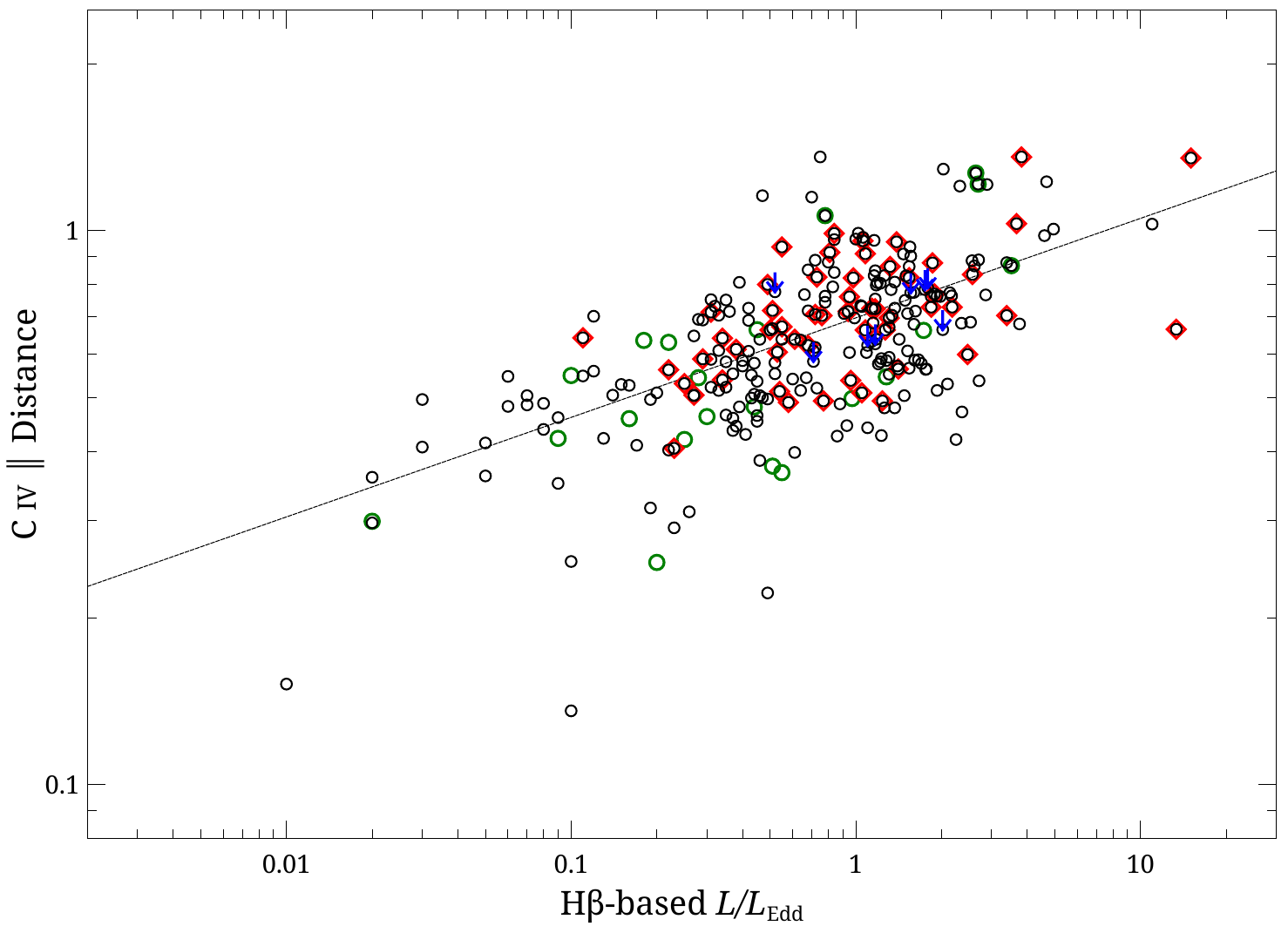}
\caption{The \civdist\ parameter versus \hb-based \lledd. Black (green) circles represent 248 (21) sources from H23 (M22). Red diamonds (blue arrows) represent \xray-detected (\xray-undetected) sources from this work, which constitute a subsample of H23 sources, and the solid line represents the best fit relation from H23.} 
\label{fig:ha}
\end{figure*}

The EW of the \civ\ emission line has been observed to be significantly anti-correlated with \hb-based \lledd\ for sources with EW(\civ) $\gtrsim10$\AA\ (Shemmer \& Lieber 2015; H23). Additionally, H23 found that the combination of EW(\civ) and \civ\ blueshift, i.e., the \civdist\ parameter, has a strong correlation with \hb-based \lledd\ across wide ranges of these parameters. Figure~\ref{fig:ha} shows this correlation for the entire H23 sample of 248 sources which includes the 63 sources from this work (overplotted with red and blue symbols, representing \xray\ detected and undetected sources, respectively). Also plotted are 17 sources from M22 that have \hb-based \lledd\ and \civdist\ values and that are not matched with the H23 sample.\footnote{See Appendix~\ref{sec:C}}

A Spearman-rank correlation between the two parameters in our GNIRS-DQS sample of 63 sources shows that both parameters are strongly correlated \hbox{($r_{\rm S} = 0.470$}, \hbox{$p = 1.037\times 10^{-4}$)}. Adding the 17 M22 sources to the correlation results in $r_{\rm S} = 0.585$, $p << 0.001$, and for a sample consisting of 248 sources from H23 and 17 sources from M22, the Spearman-rank correlation coefficient and chance probability result in $r_{\rm S} = 0.584$ and $p << 0.001$. These results are consistent with the correlation obtained for these parameters in H23 for 248 sources: $r_{\rm S} = 0.566$ and $p << 0.001$.

\begin{figure}
\hspace{-0.5cm}
\epsscale{1.2}
\plotone{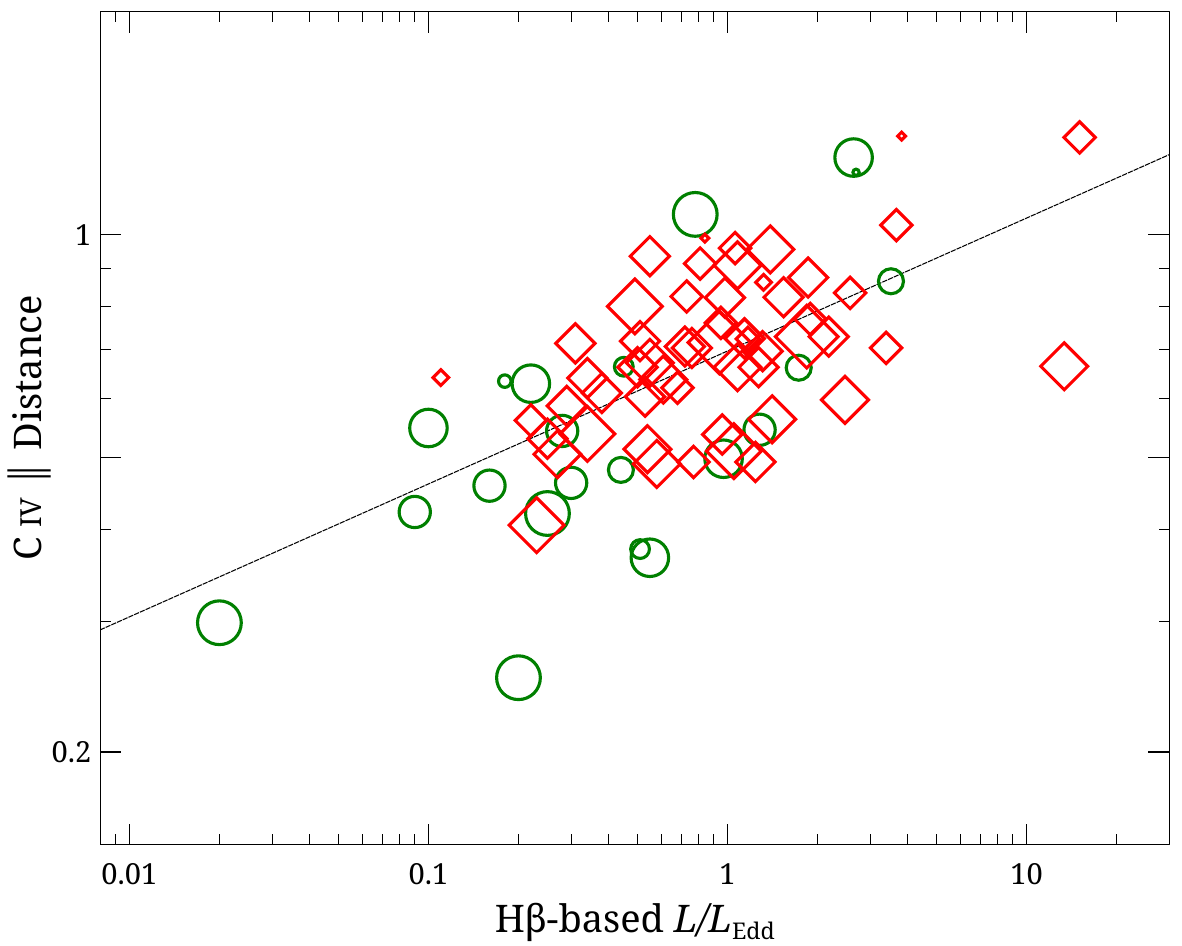}
\caption{Same as Figure~\ref{fig:ha}, but limited to the 55 \xray-detected sources from this work (diamonds) as well as 21 sources from M22 (circles). The solid line represents the best fit relation from H23. Larger symbol sizes correspond to sources with larger (i.e., less negative) \aox\ values. The \aox\ information does not contribute any additional trends to the correlation.} 
\label{fig:ledd_civ}
\end{figure}

To test whether \xray\ information can strengthen this relation, Figure~\ref{fig:ledd_civ} shows the dependence of \civdist\ versus \hb-based \lledd\ but only for the 76 \xray-detected sources from Figure~\ref{fig:ha} (four of these are H23 sources with \xray\ detections in M22; see Appendix~\ref{sec:C} for more details). Symbol sizes in Figure~\ref{fig:ledd_civ} scale with the \aox\ value of each source. We do not find any clear trend related to \aox. At low \lledd\ (and low \civdist) values, we find mostly \xray\ strong sources, but also a few \xray\ weak sources. Similarly, at the higher end of the \hb\ and \civ\ parameter space, we find a wide range of \xray\ strength with no clear trend as a function of \aox. Quantitatively, the mean \aox\ values above and below the median \lledd\ value for our sample of 76 sources differ by only 0.05 (or, a factor of $\sim35$\% in \xray\ strength). The mean \aox\ values above and below the median \civdist\ value for the sample differ by 0.11, indicating a difference of a factor of $\sim2$ in \xray\ strength.

Furthermore, we compared correlations involving \lledd\ only for the 55 \xray-detected GNIRS-DQS sources. A correlation between \aox\ and \lledd\ resulted in $r_{\rm S} = -0.245$ and $p = 0.071$, while the correlation between \civdist\ and \lledd\ resulted in $r_{\rm S} = 0.503$ and $p << 0.001$. Thus, a significant correlation is found only between \civdist\ and \lledd\ for these sources.

These results are consistent with our findings above, where \aox\ appears to be more strongly dependent on the \civdist\ parameter than on \hb-based \lledd. Our results also suggest that the \civ\ parameter space provides a better \lledd\ indicator, compared with the \hb\ parameter space, and that the \aox\ parameter does not contribute to improving correlations between the Eddington luminosity ratio and its diagnostics in the optical-UV band.

\begin{deluxetable}{lcccc}
\tabletypesize{\footnotesize}
\tablecolumns{4}
\tablecaption{Photon Indices \label{tab:gamma}}
\tablehead{
\colhead{} &
\colhead{Counts} &
\multicolumn{1}{r}{$\Gamma$} &
\colhead{} &
\colhead{$N_{\rm H}$} \\
\cline{3-4}
\colhead{Quasar} &
\colhead{$0.5-8$~keV} &
\colhead{Effective} &
\colhead{Measured} &
\colhead{$10^{22}$~cm$^{-2}$}
}
\startdata
\object{SDSS J004719.71+014813.9} & 29.5 & 2.03$^{+0.22}_{-0.72}$ & \nodata & \nodata \\ 
\object{SDSS J092216.04+160526.4} & 37.1 & 2.13$^{+0.20}_{-0.61}$ & \nodata & \nodata \\
\object{SDSS J092523.24+214119.8} & 50.7 & 1.95$^{+0.20}_{-0.46}$ & \nodata & \nodata \\
\object{SDSS J095555.68+351652.6} & 43.1 & 1.93$^{+0.20}_{-0.51}$ & \nodata & \nodata \\
\object{SDSS J104336.73+494707.6} & 43.2 & 1.99$^{+0.20}_{-0.53}$ & \nodata & \nodata \\
\object{SDSS J121314.03+080703.6} & 28.4 & 2.27$^{+0.21}_{-0.75}$ & \nodata & \nodata \\
\object{SDSS J144948.62+123047.5} & 37.5 & 1.42$^{+0.22}_{-0.52}$ & \nodata & \nodata \\
\object{SDSS J080117.79+521034.5} & 168.3 & 1.63$^{+0.12}_{-0.19}$ & 1.84$^{+0.27}_{-0.26}$ & $<1.64$ \\
\object{SDSS J084846.11+611234.6} & 40.0 & 1.93$^{+0.15}_{-0.58}$ & \nodata & \nodata \\
\object{SDSS J102907.09+651024.6} & 139.4 & 2.06$^{+0.12}_{-0.26}$ & 2.10$^{+0.41}_{-0.29}$ & $<1.85$ \\
\object{SDSS J111119.10+133603.8} & 179.5 & 1.95$^{+0.11}_{-0.21}$ & 2.06$^{+0.37}_{-0.35}$ & $<8.13$ \\
\object{SDSS J141028.14+135950.2} & 63.8 & 1.78$^{+0.14}_{-0.39}$ & \nodata & \nodata \\
\object{SDSS J141951.84+470901.3} & 154.4 & 1.73$^{+0.11}_{-0.22}$ & 2.03$^{+0.27}_{-0.26}$ & $<0.82$
\enddata
\end{deluxetable}

\subsection{Hard X-ray Photon Index ($\Gamma$)}
Our results suggest that the \aox\ parameter cannot provide a reliable \lledd\ indicator, and it does not improve upon optical-UV \lledd\ indicators, likely due, in part, to its dependence on both \lledd\ and \mbh. The photon index ($\Gamma$) of the hard-\xray\ power law spectrum, typically measured above a rest-frame energy of $\sim2$~keV, has been observed to provide a reliable \xray\ indicator of \lledd\ in `ordinary' type~1 quasars  (e.g., Shemmer \et 2006, 2008; Constantin \et 2009; Risaliti \et 2009; Brightman \et 2013; Fanali \et 2013; Liu \et 2021). However, our snapshot observations of the GNIRS-DQS sample were not designed to provide accurate measurements of $\Gamma$ values in these sources. Instead, we derive effective $\Gamma$ values from band ratios for 13 of our sources (six of which are archival from M22) that have $\gtrsim30$ full-band counts (see Table~\ref{tab:chandra_uv-opt}) using the \chandra\ {\sc  PIMMS} tool.\footnote{This was achieved by an iterative process, where the $\Gamma$ value was adjusted until counts in both the soft and hard bands were consistent with the band ratio.} The effective $\Gamma$ values are given in Table~\ref{tab:gamma}. We find no clear trends and no significant Spearman-rank correlations between the effective $\Gamma$ values and the respective \hb-based \lledd\ or \civdist\ values of these 13 sources, which is expected due to the small number of sources with sufficient counts for this analysis, and the mean uncertainties on the $\Gamma$ parameter that are too large to reveal any meaningful trend with \lledd.

\begin{figure*}
\hspace*{-0.4cm}
\includegraphics[width=.5\linewidth]{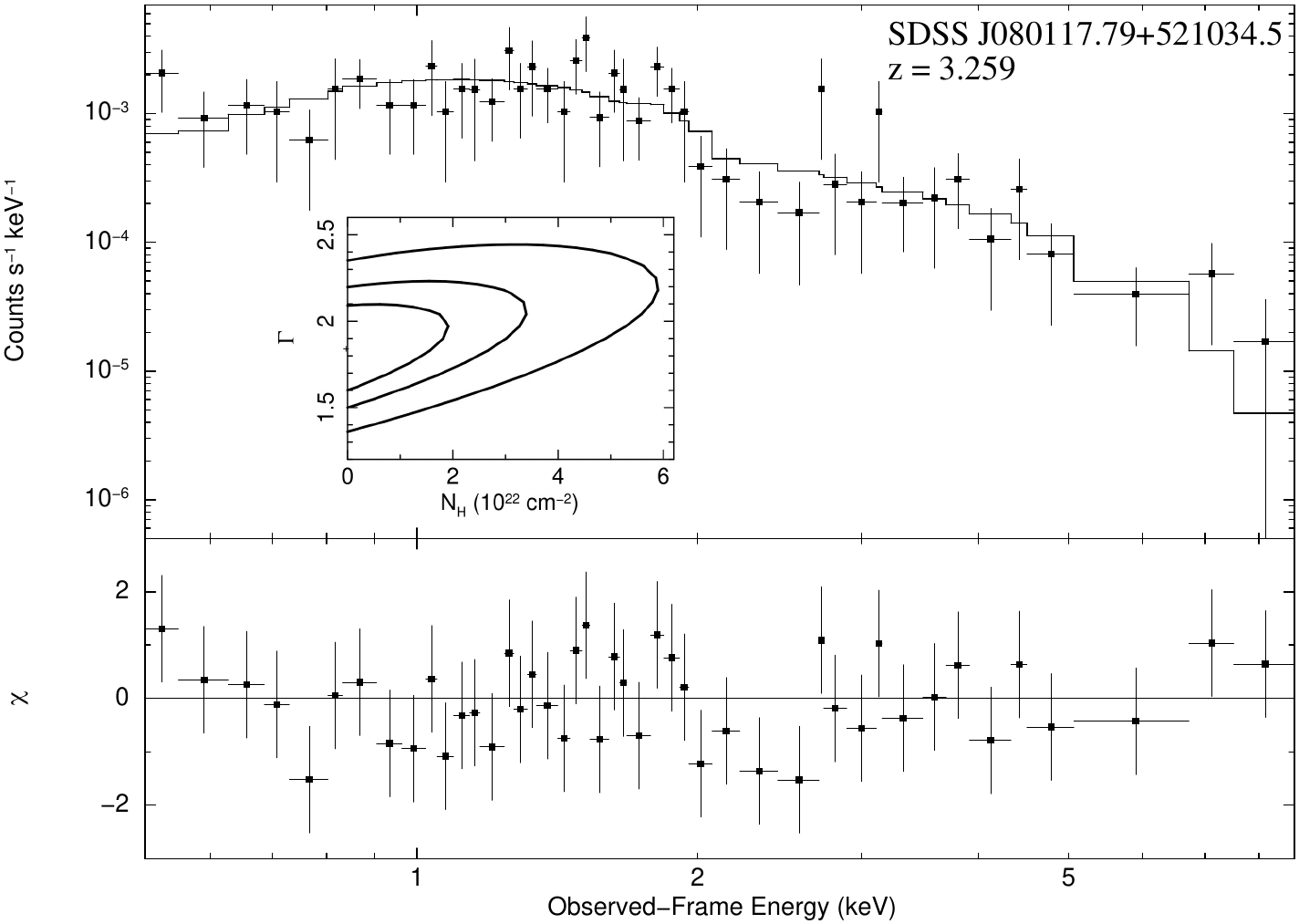}\quad\includegraphics[width=.5\linewidth]{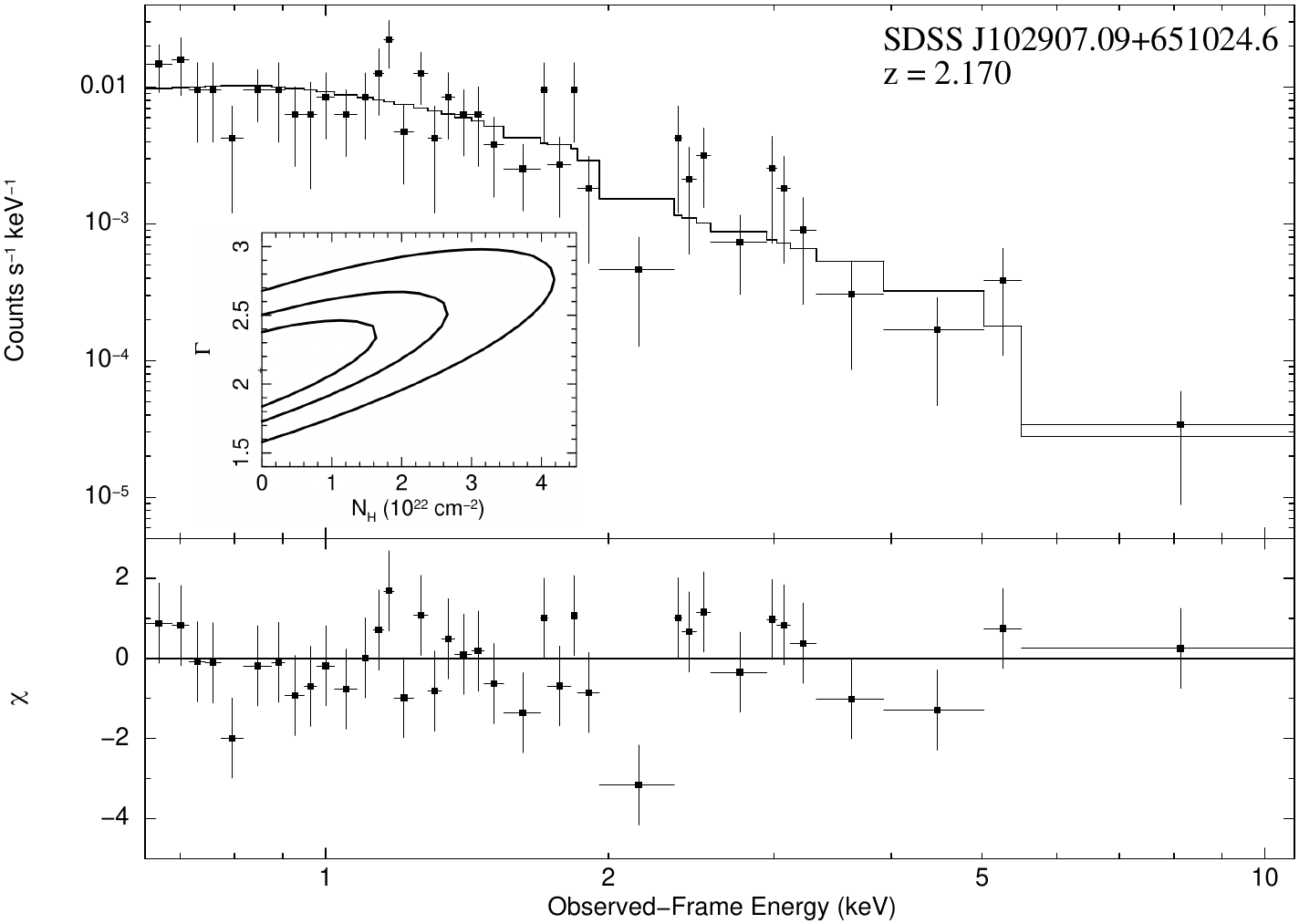}
\\[\baselineskip]
\hspace*{-0.4cm}
\includegraphics[width=.5\linewidth]{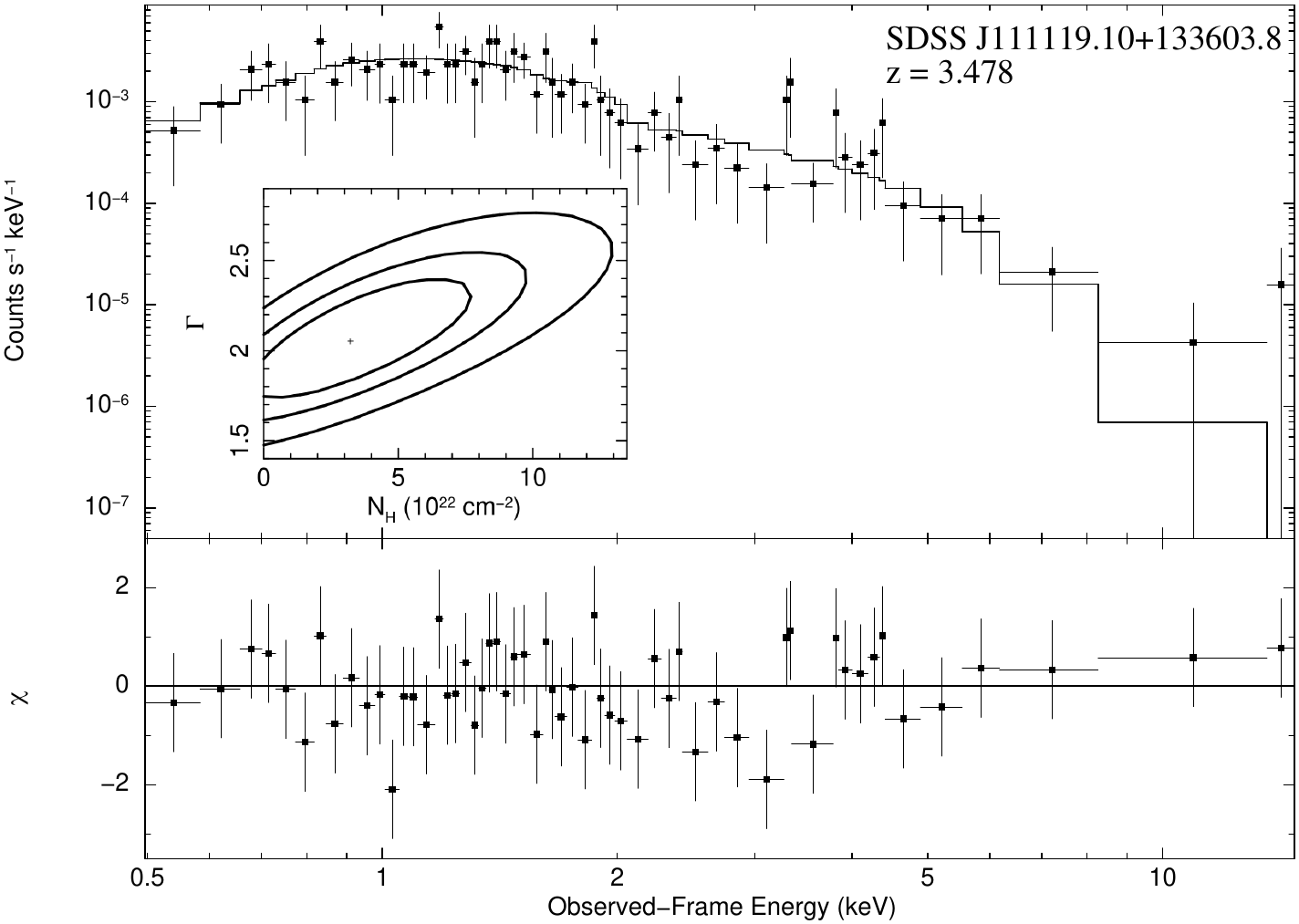}\quad\includegraphics[width=.5\linewidth]{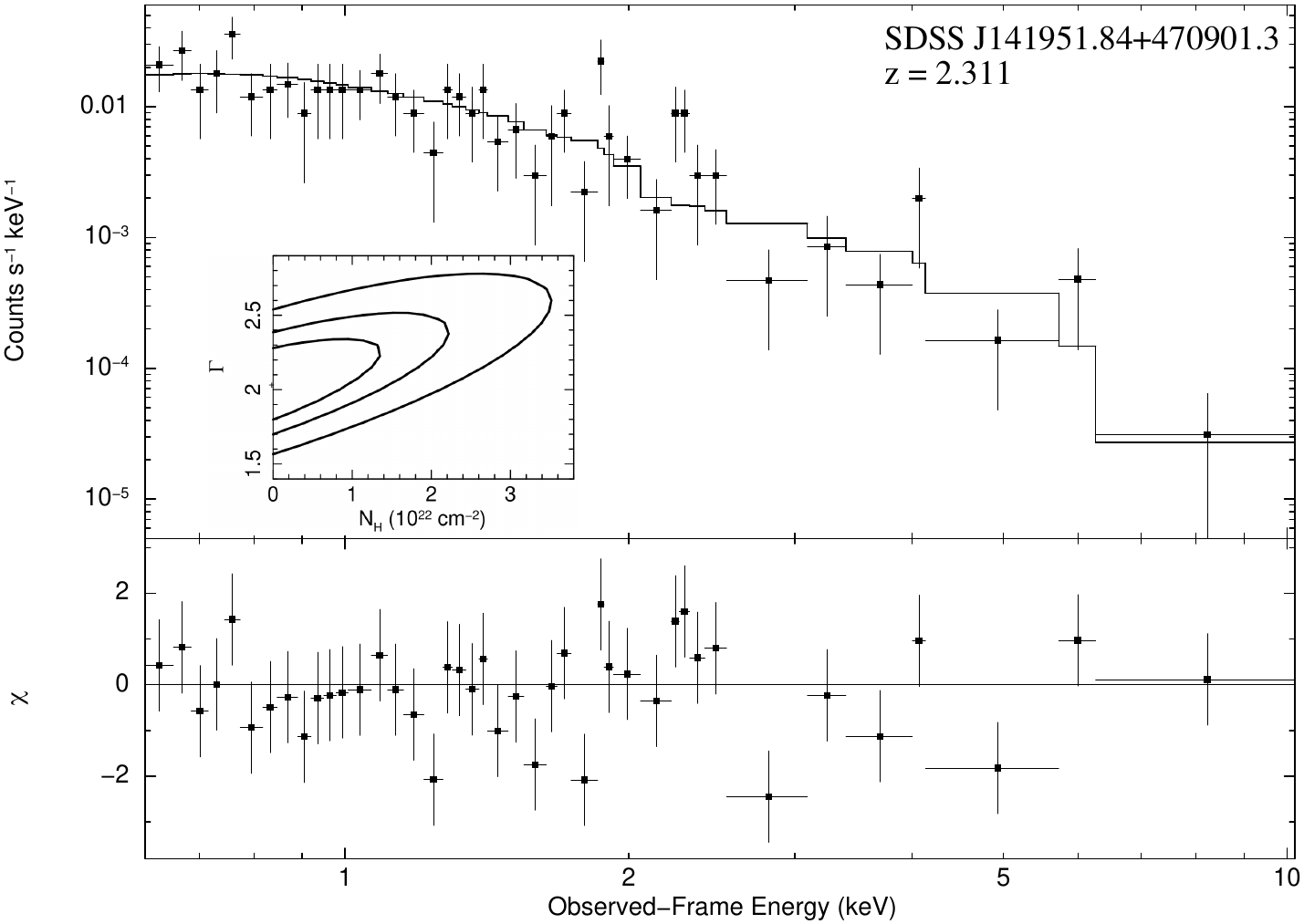}
\caption{Data, best-fit spectra, and residuals of the four GNIRS-DQS sources with $> 100$ full-band counts. The data were fitted using a Galactic absorption power-law model with an added intrinsic neutral absorption component above a rest-frame energy of 2 keV. The $\chi$ residuals are in units of $\sigma$ with error bars of size 1. Insets show the 68\%, 90\%, and 99\% confidence regions for the photon index vs. intrinsic neutral absorption column density. Data were binned with two counts per data point for presentation purposes.}
\label{fig:spectra}
\end{figure*}

For the four GNIRS-DQS sources with $>100$ full-band counts (all of which are from M22) we measured $\Gamma$ and intrinsic absorption column density (\nh) values at $>2$ keV rest-frame energies using the {\sc xspec} {\sc phabs*zphabs*pow} model and the chi-square statistic. The results of these spectral fits appear in Table~\ref{tab:gamma}. We note that the $\Gamma$ values obtained through these fits are consistent, within the errors, with the respective values obtained from the band ratios. We also note that, for each of these sources, the spectral fits resulted in an upper limit on the \nh\ value (see Table~\ref{tab:gamma}). Figure~\ref{fig:spectra} shows the best-fit spectra and residuals for these sources; the respective insets show the 68\%, 90\%, and 99\% confidence regions for the photon index versus intrinsic neutral absorption column density. The measured $\Gamma$ values for these sources, with $\Delta \Gamma \sim0.3 - 0.4$, are consistent with these sources having relatively high \hb-based \lledd\ values in the range $0.70 - 1.63$ (Table~\ref{tab:chandra_uv-opt}; e.g., Shemmer \et 2008).

\begin{figure*}
\hspace*{-0.4cm}
\includegraphics[width=.5\linewidth]{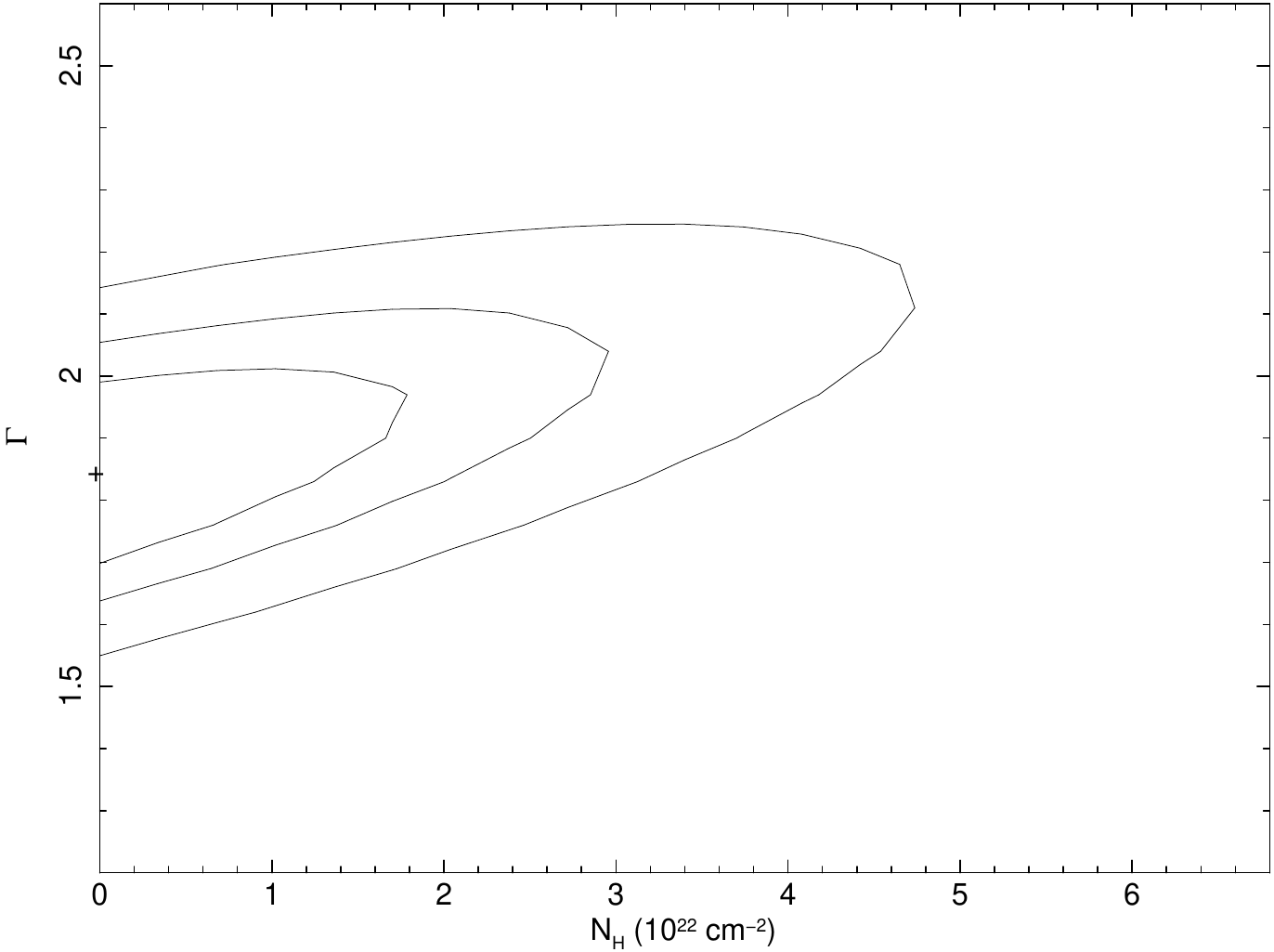}\quad\includegraphics[width=.5\linewidth]{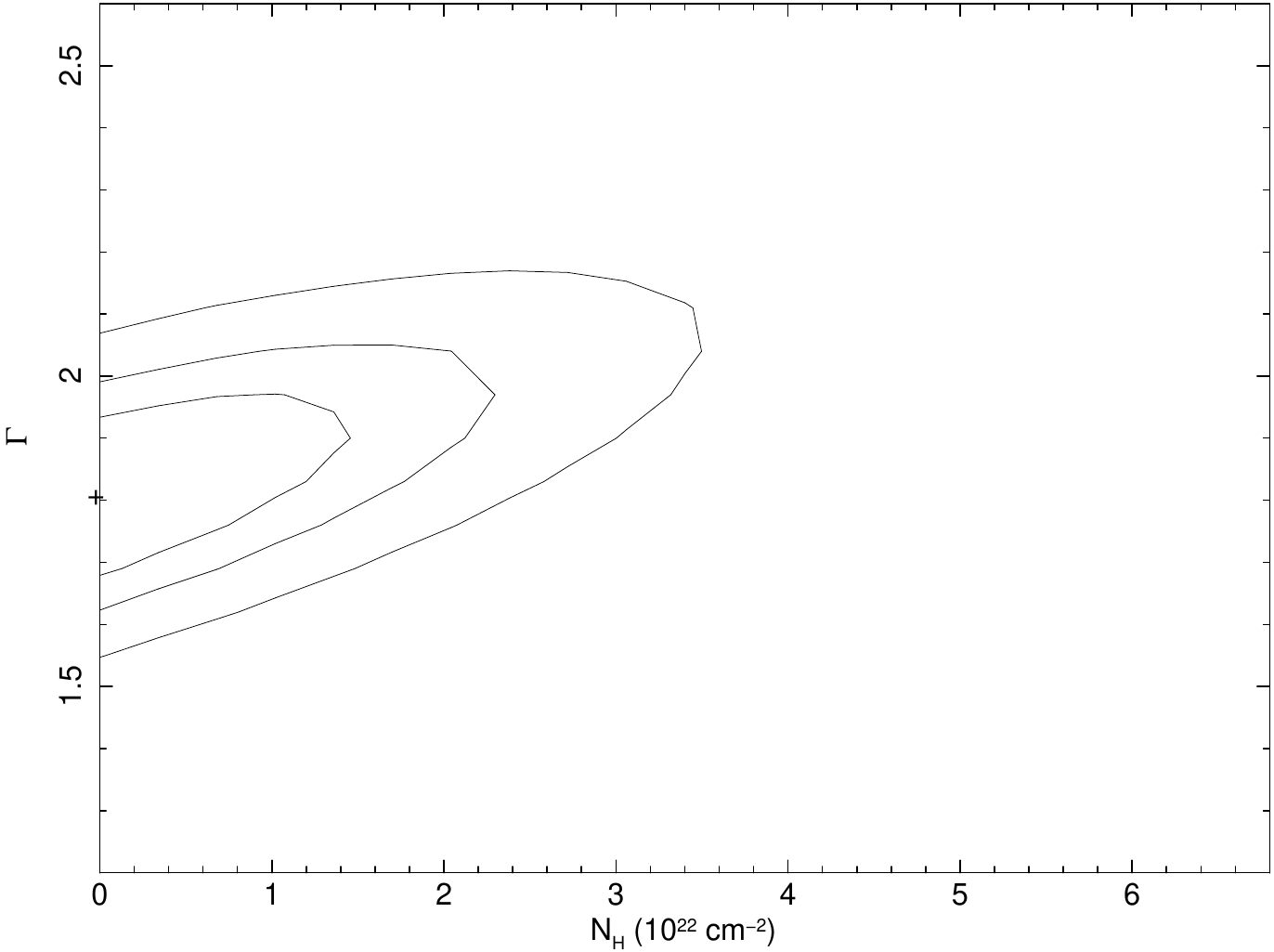}
\\[\baselineskip]
\hspace*{-0.4cm}
\includegraphics[width=.5\linewidth]{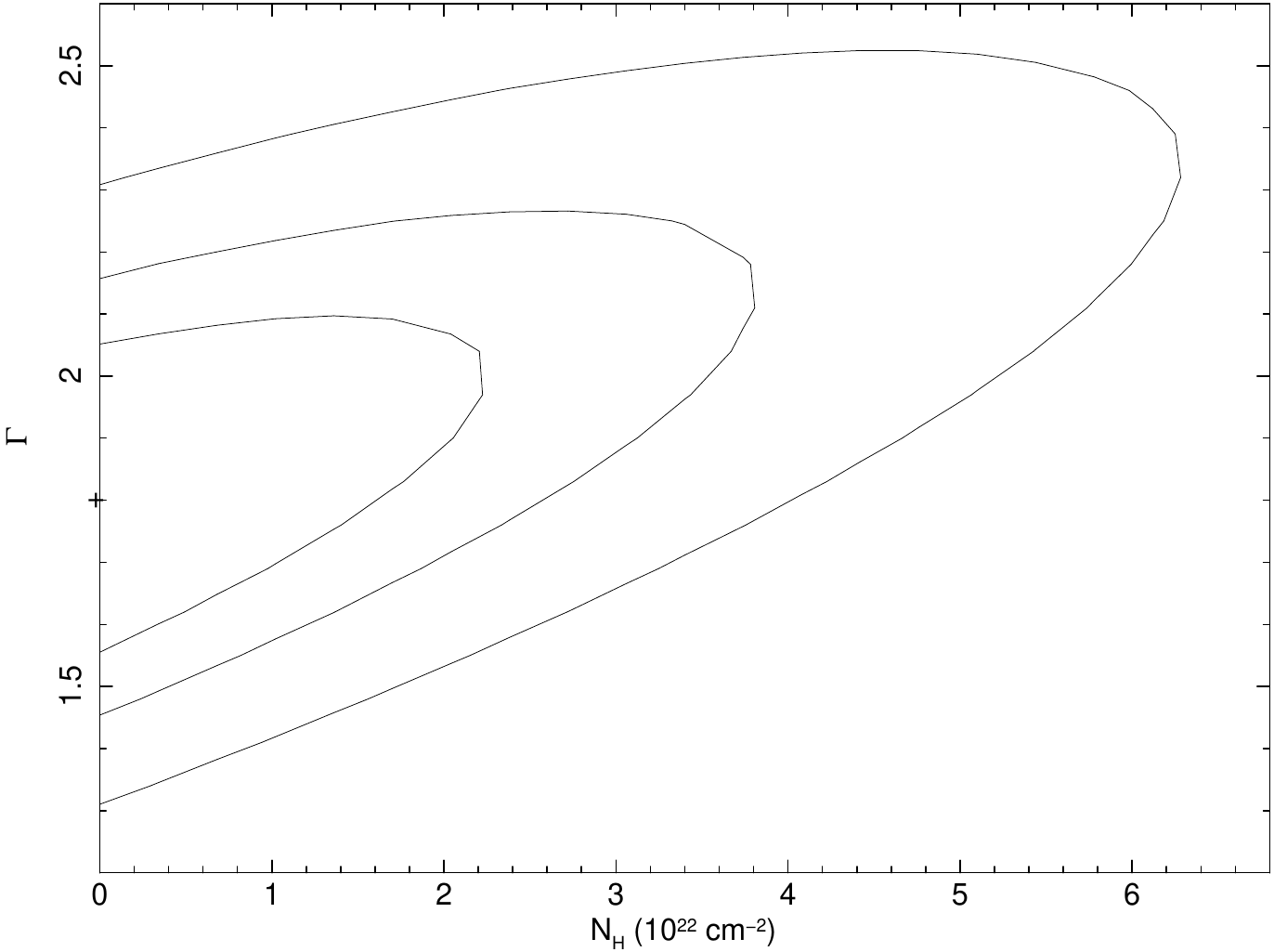}\quad\includegraphics[width=.5\linewidth]{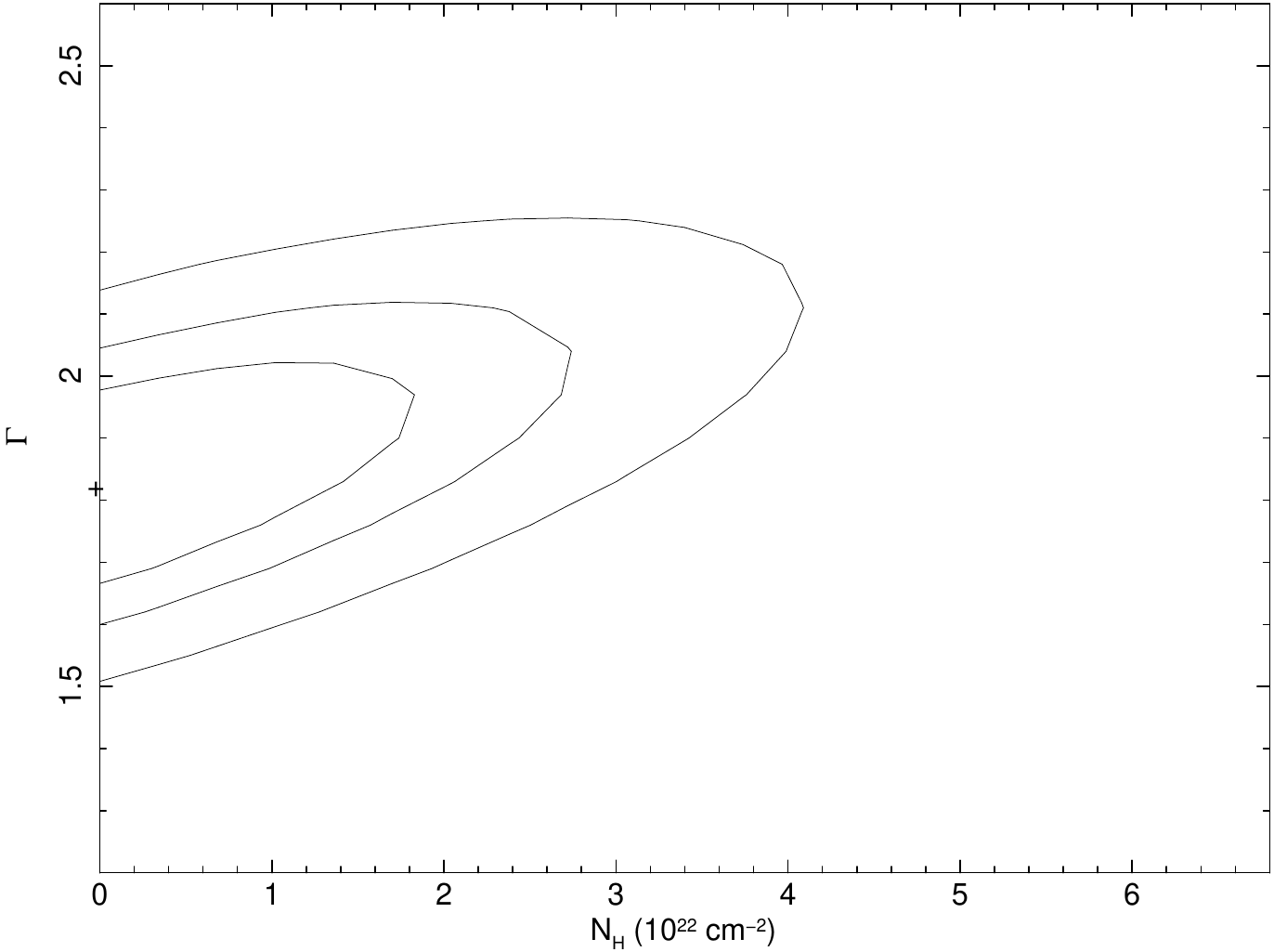}
\caption{Contours representing confidence regions of 68\%, 90\%, and 99\% for $\Gamma$ versus $N_{\rm H}$ for four sub-samples of GNIRS-DQS sources, as given in Table~\ref{tab:stack_z}, that were jointly fit with a Galactic absorption corrected power law and an intrinsic absorption model.  Clockwise from top-left: 50 targeted sources, all 55 sources for which the joint fit was feasible, 36 sources at $z\sim2$, and 17 sources at $z\sim1.5$.} 
\label{fig:cont}
\end{figure*}

Given the limited photon statistics for the majority of our GNIRS-DQS sources, we obtain average $\Gamma$ values for these sources by jointly fitting their \chandra\ data. We performed this procedure in three separate runs, each of which excluded the four sources that have $>100$ counts each. In the first run, we stacked the data of all 55 sources that had a meaningful number of counts, 17 sources at $z\sim1.5$, 36 sources at $z\sim2$, and all 50 targeted sources. We followed the procedure described in Section 2 and used {\sc xspec} to fit the unbinned spectra of each sub-sample jointly with the {\sc phabs*zphabs*pow} model, fixing the Galactic absorption component of each individual source; the chi-square statistic was used throughout the fits. Table~\ref{tab:stack_z} presents the joint fits results. Figure~\ref{fig:cont} shows the 68\%, 90\%, and 99\% confidence regions for the photon index versus intrinsic neutral absorption column density of each sub-sample.

Overall, the average $\Gamma$ values for all four sub-samples are {$\gtrsim1.8$}, consistent with the relatively high \hb-based \lledd\ values of the GNIRS-DQS sources. We do not detect significant differences between the average $\Gamma$ values of the targeted sources with respect to the entire sample (which includes five archival sources), or between sources at $z\sim1.5$ and those at $z\sim2$. We also note that, for each sub-sample, the spectral fits resulted in an upper limit on the value of the neutral absorption column density (see Table~\ref{tab:stack_z}).

\begin{figure*}
\gridline{\hspace*{-1.7cm}\fig{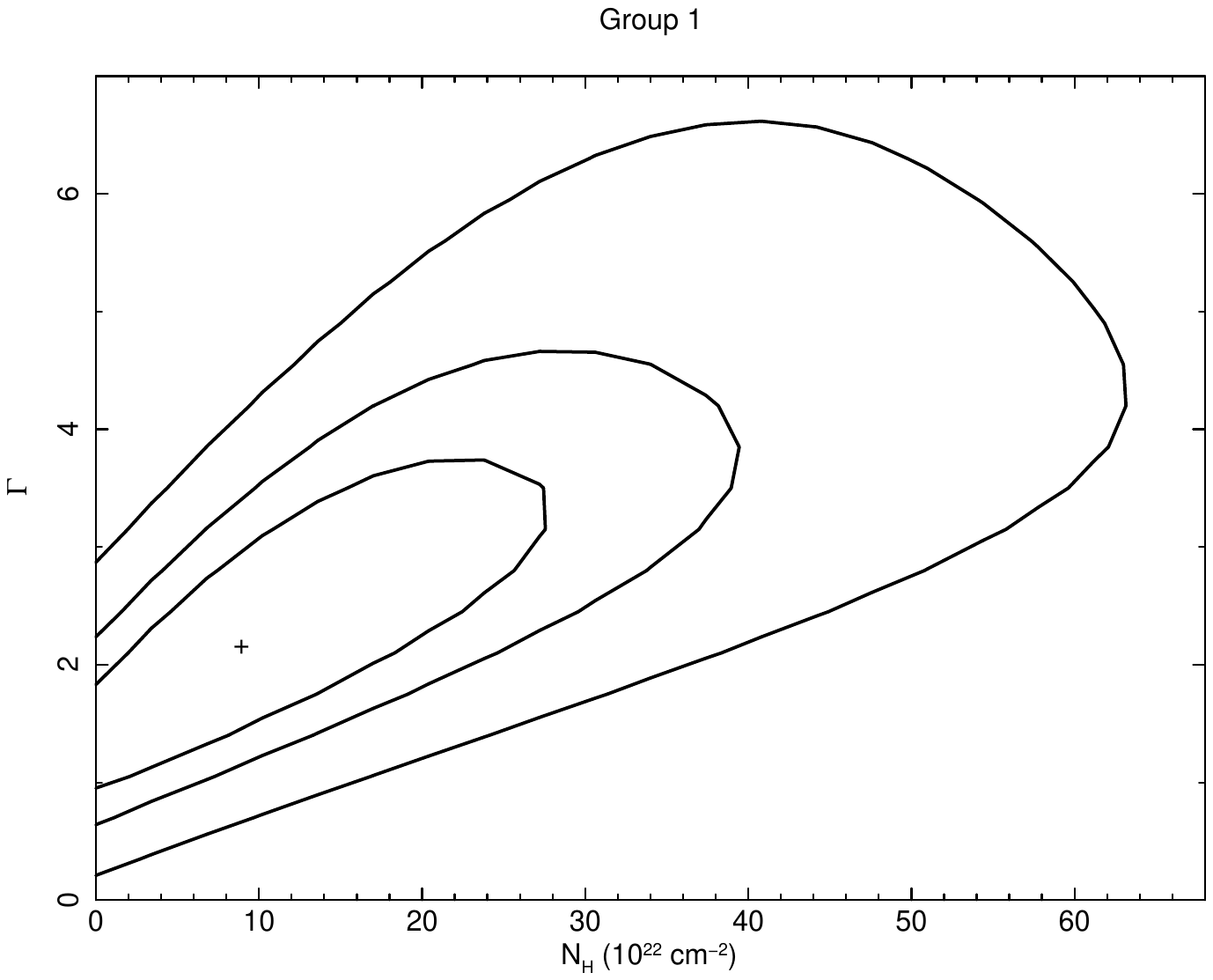}{0.45\textwidth}{}
          \hspace*{-1.2cm}\fig{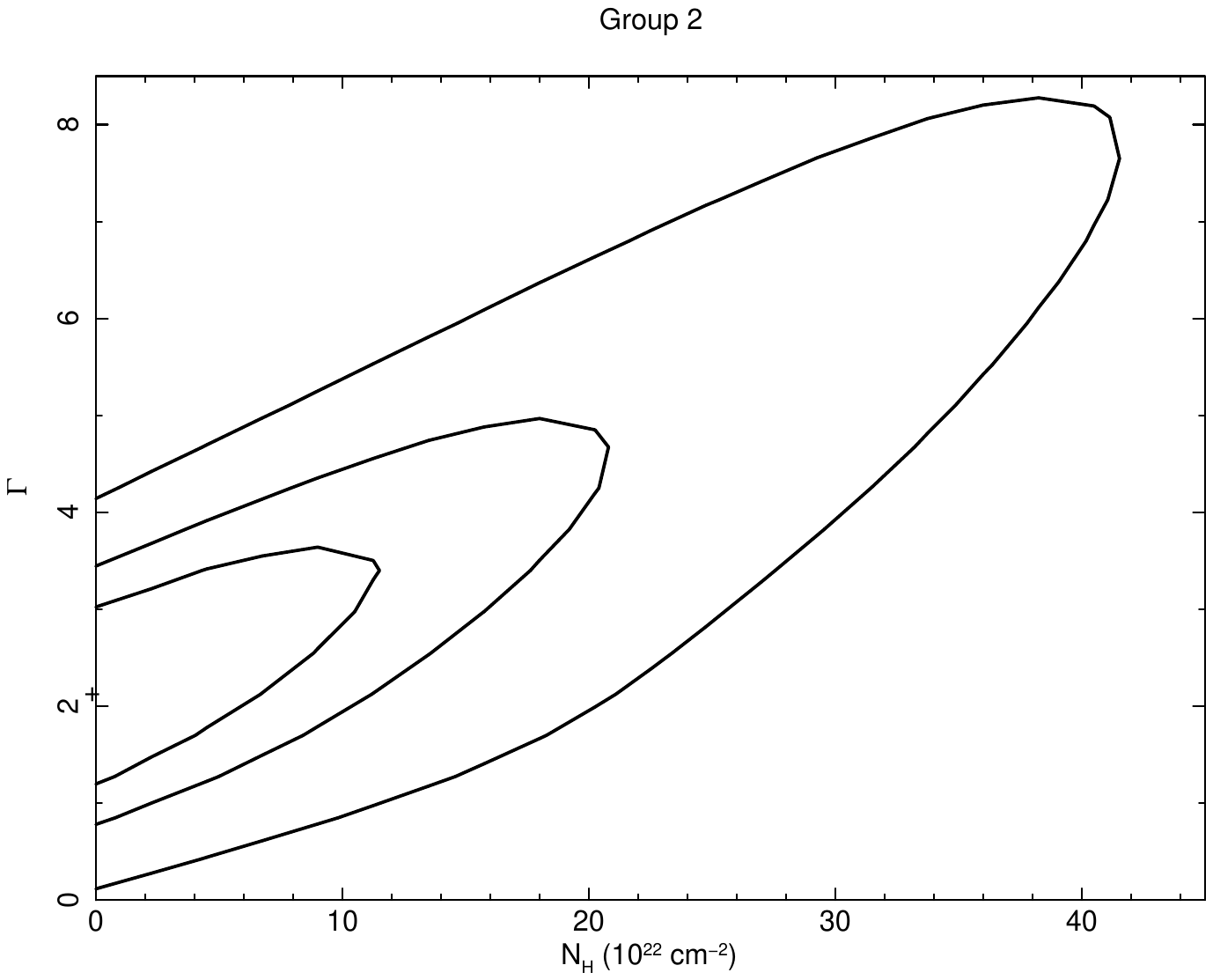}{0.45\textwidth}{}\hspace*{-1.2cm}\fig{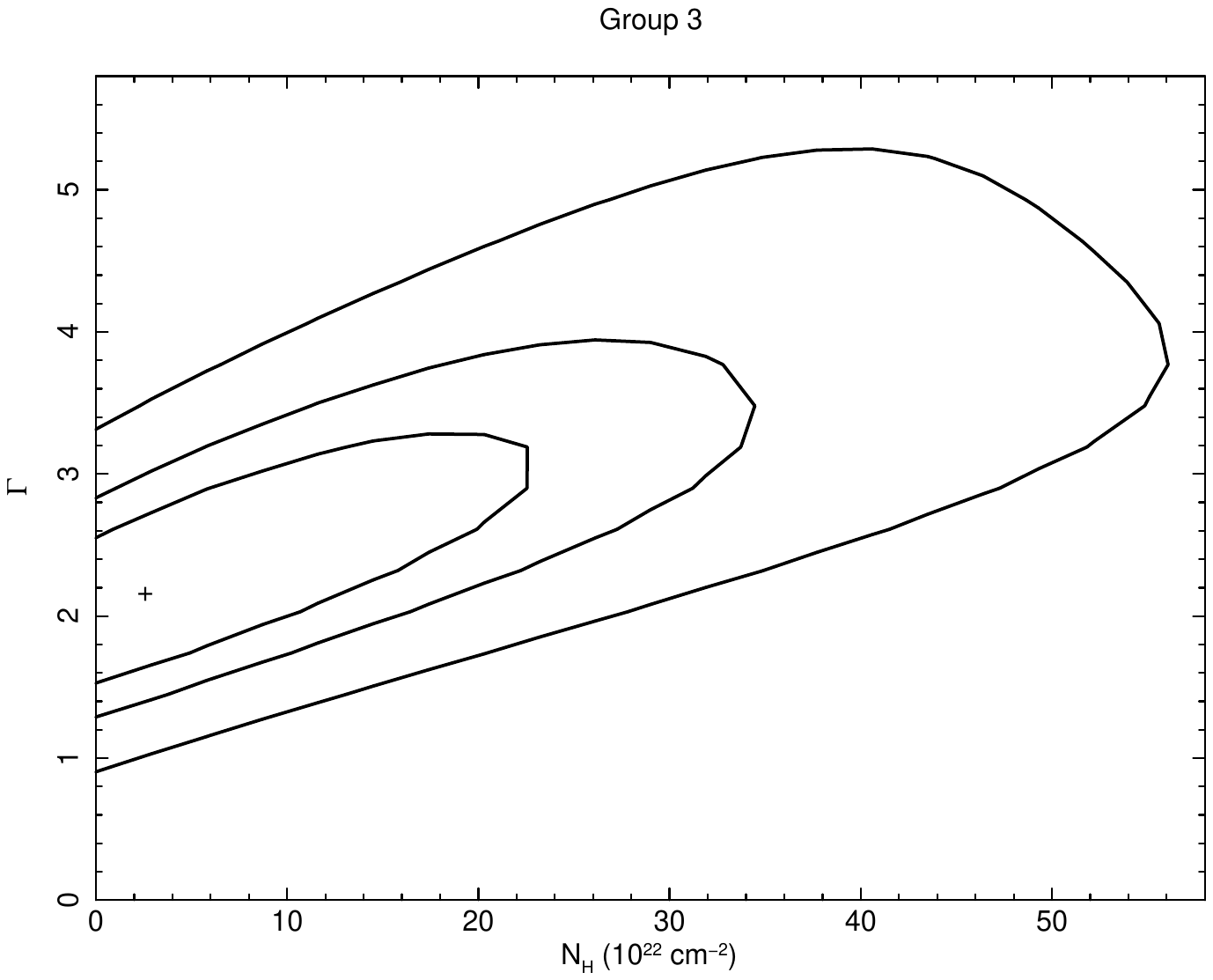}{0.45\textwidth}{}}
          \vspace*{-1.5cm}
\gridline{\hspace*{-1.7cm}\fig{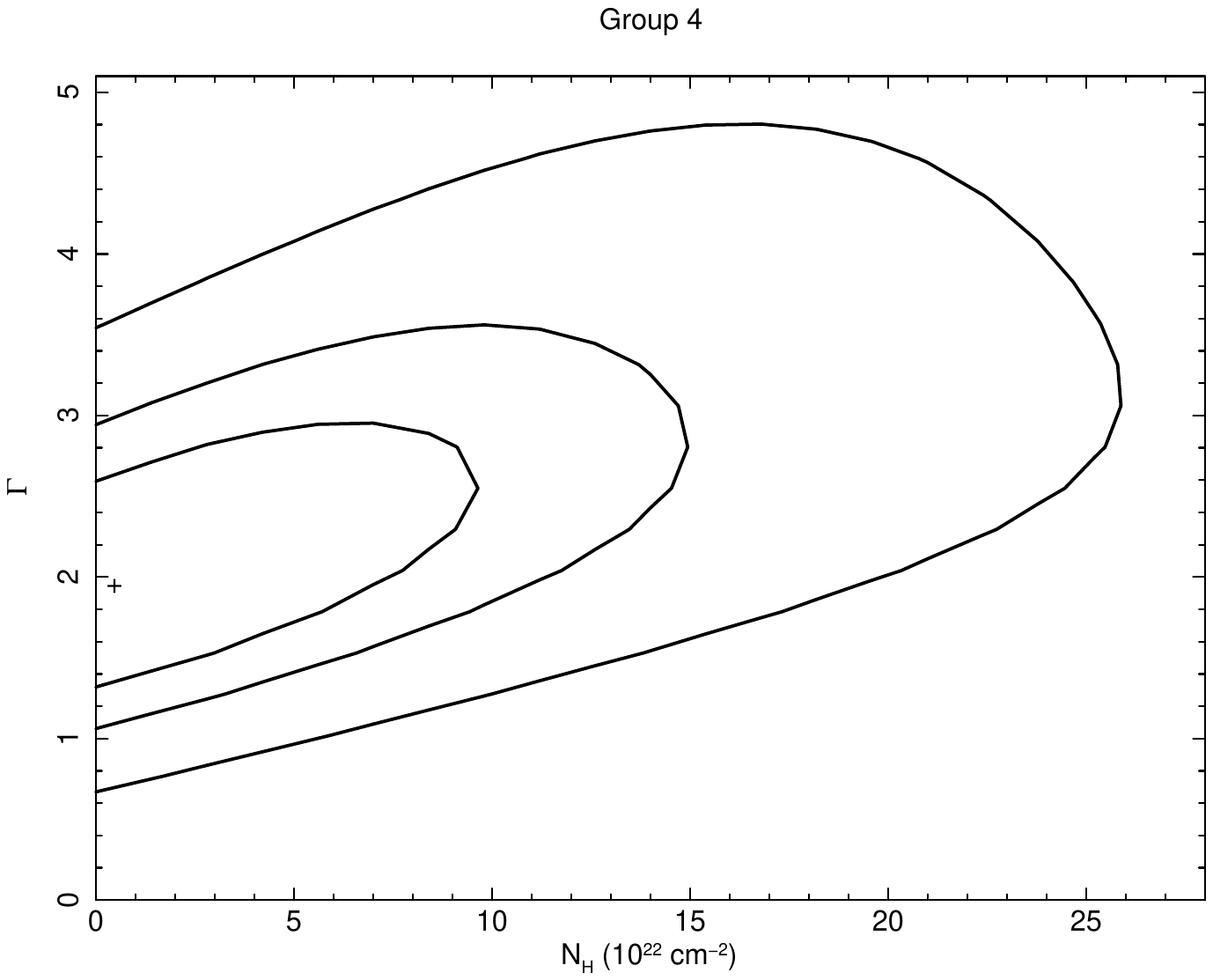}{0.45\textwidth}{}
          \hspace*{-1.2cm}\fig{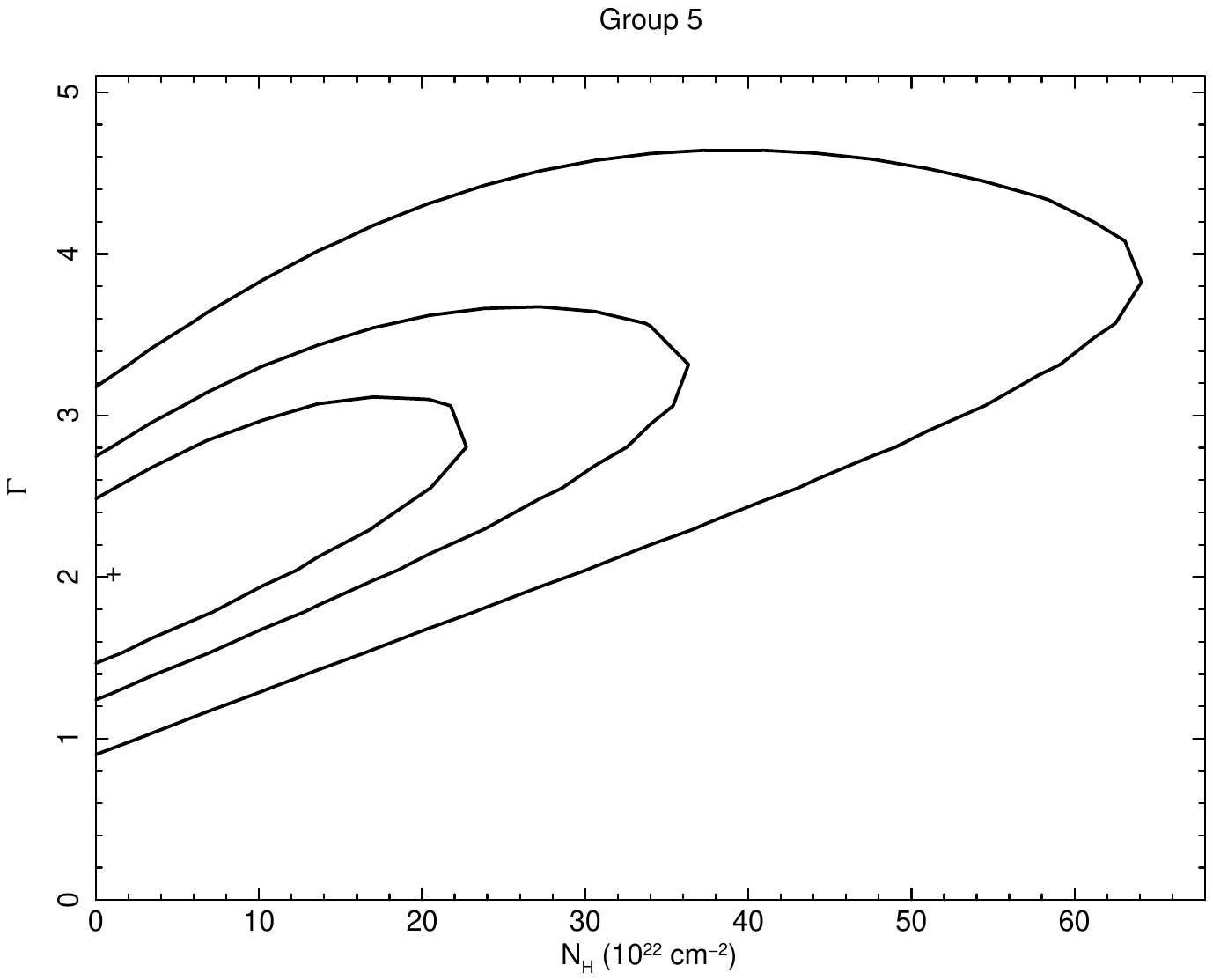}{0.45\textwidth}{}\hspace*{-1.2cm}\fig{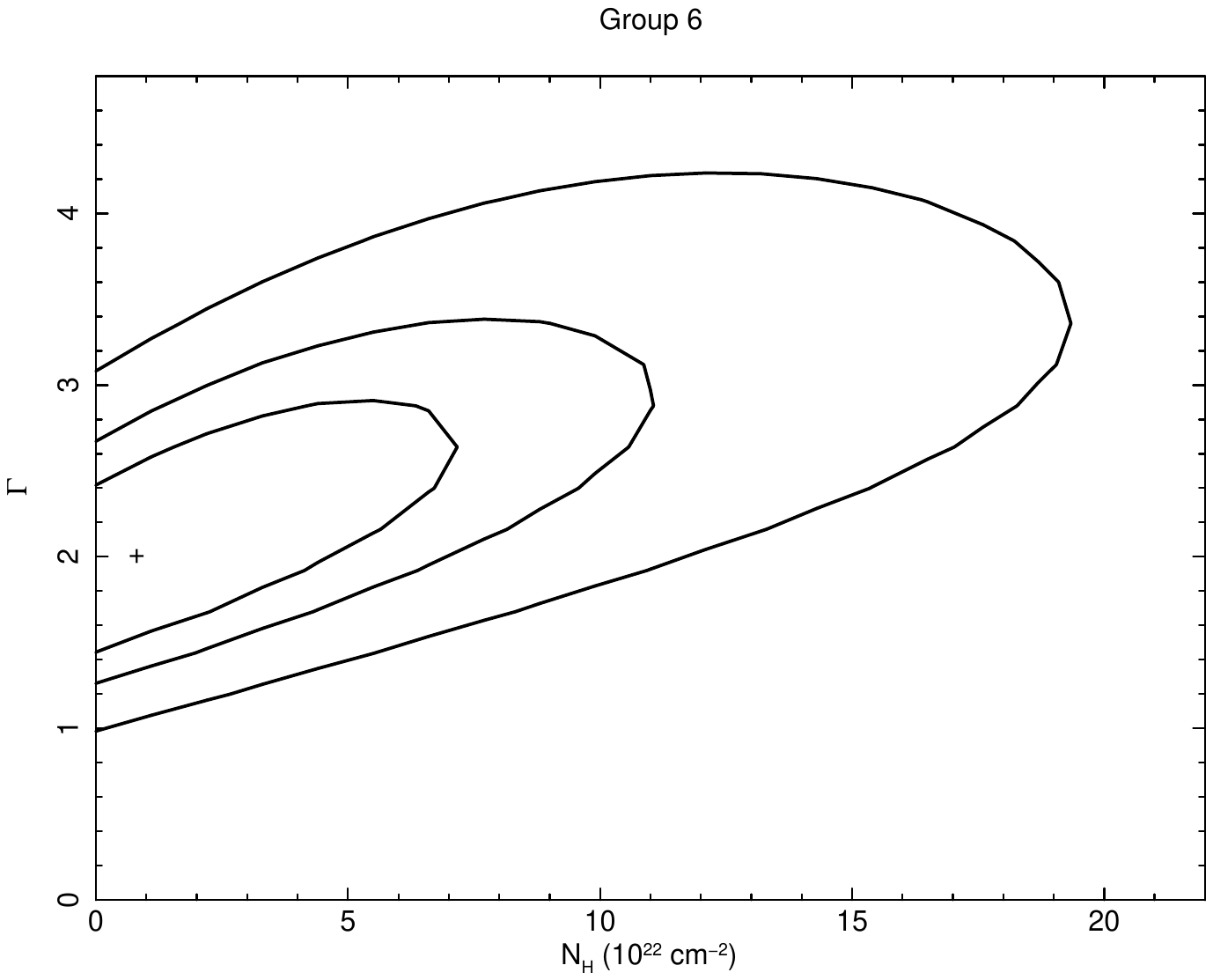}{0.45\textwidth}{}}
          \vspace*{-1.5cm}
\gridline{\fig{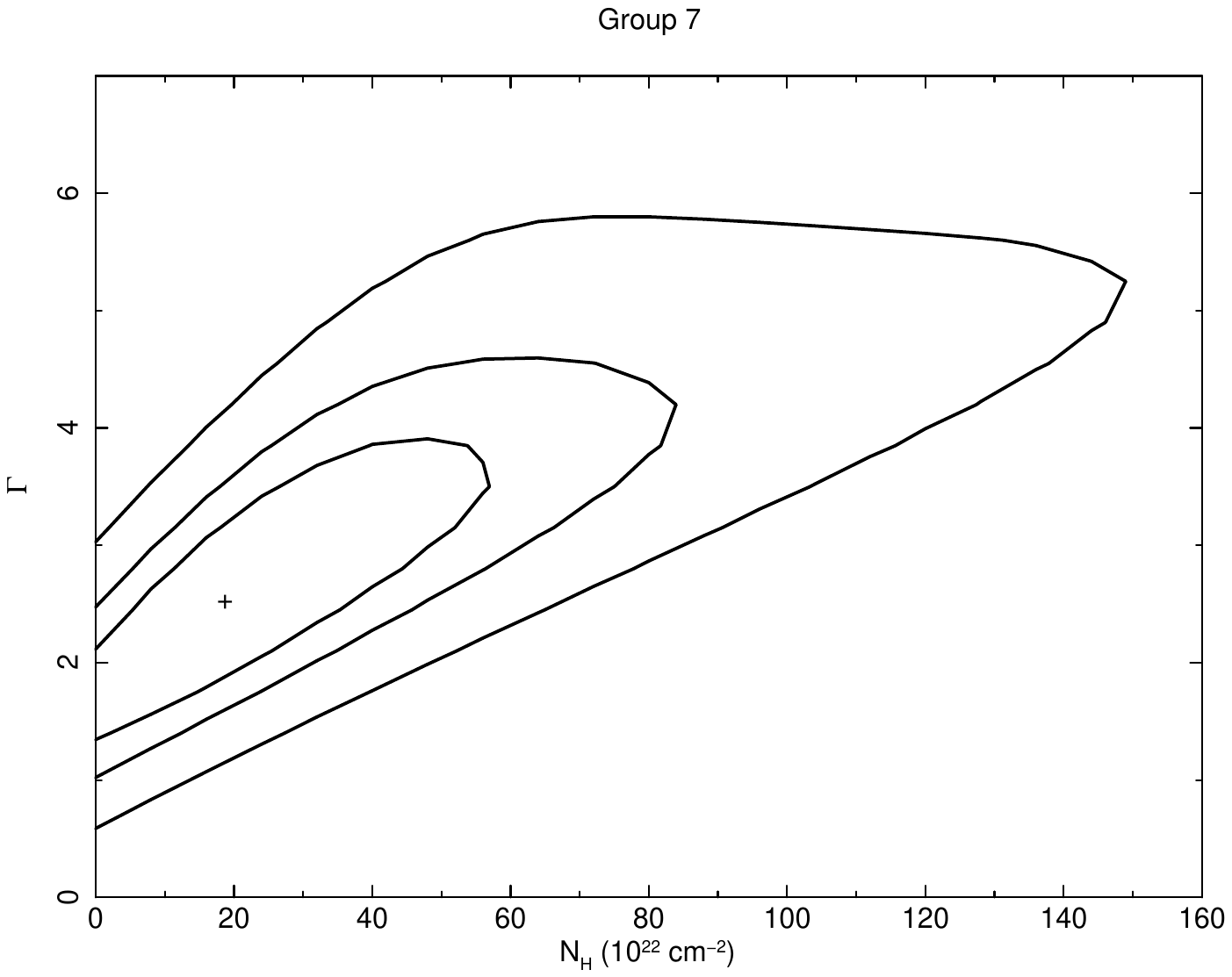}{0.45\textwidth}{}
          \hspace*{-1.2cm}\fig{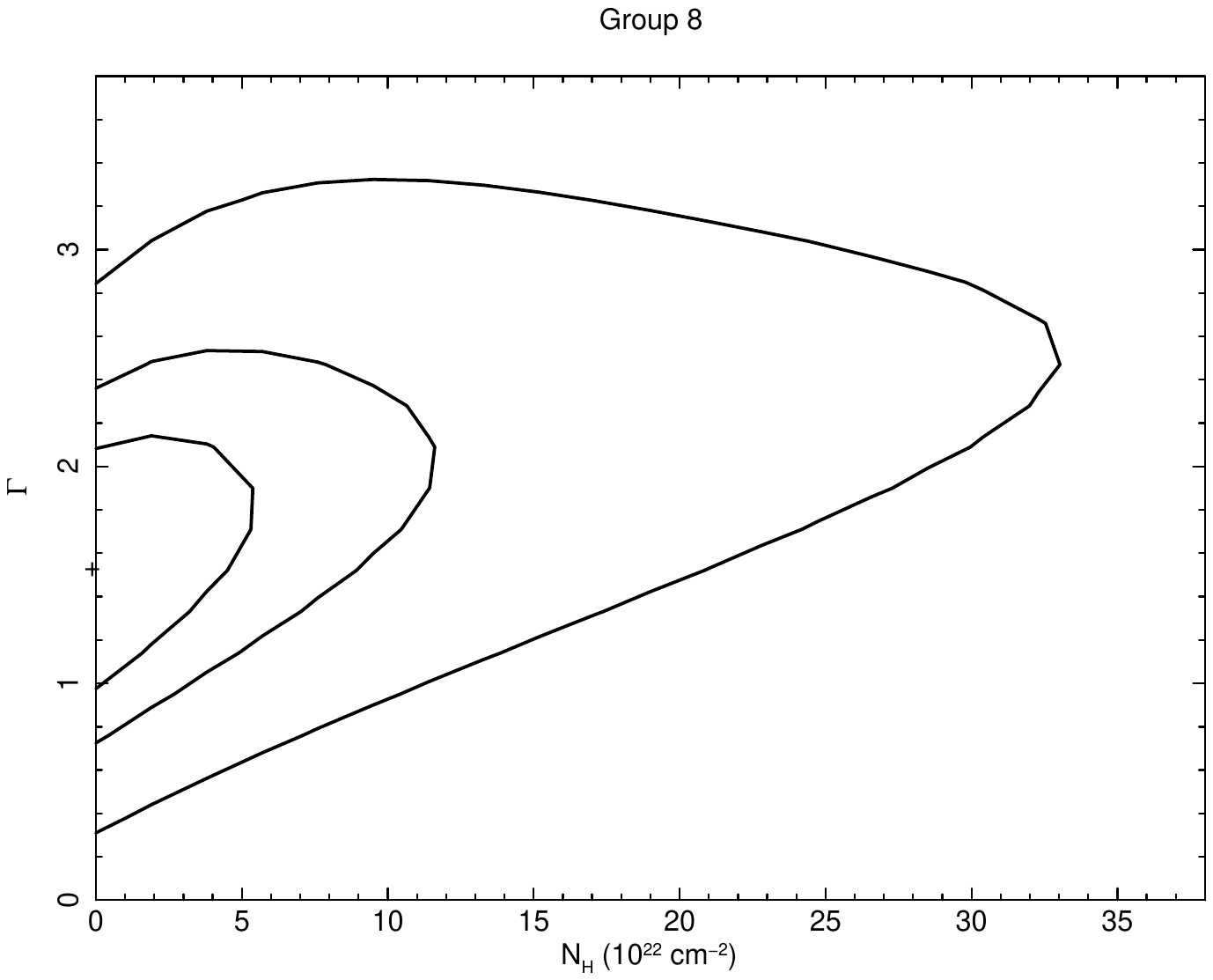}{0.45\textwidth}{}}
\caption{Same as Figure~\ref{fig:cont} but for the \luv\ bins given in Table~\ref{tab:stack_luv}.}
\label{fig:luv_cont}
\end{figure*}

\begin{figure*}
\gridline{\hspace*{-1.7cm}\fig{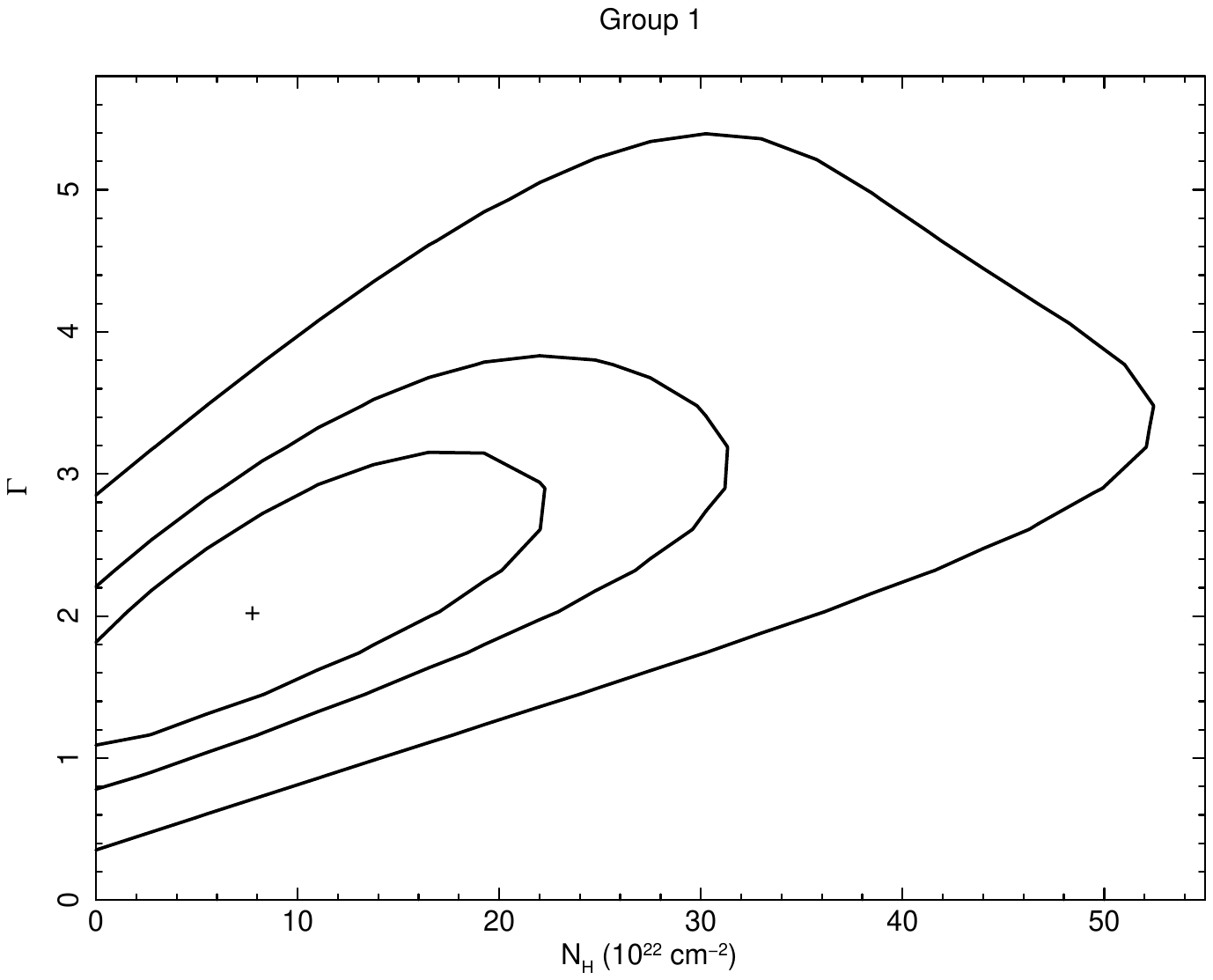}{0.45\textwidth}{}
          \hspace*{-1.2cm}\fig{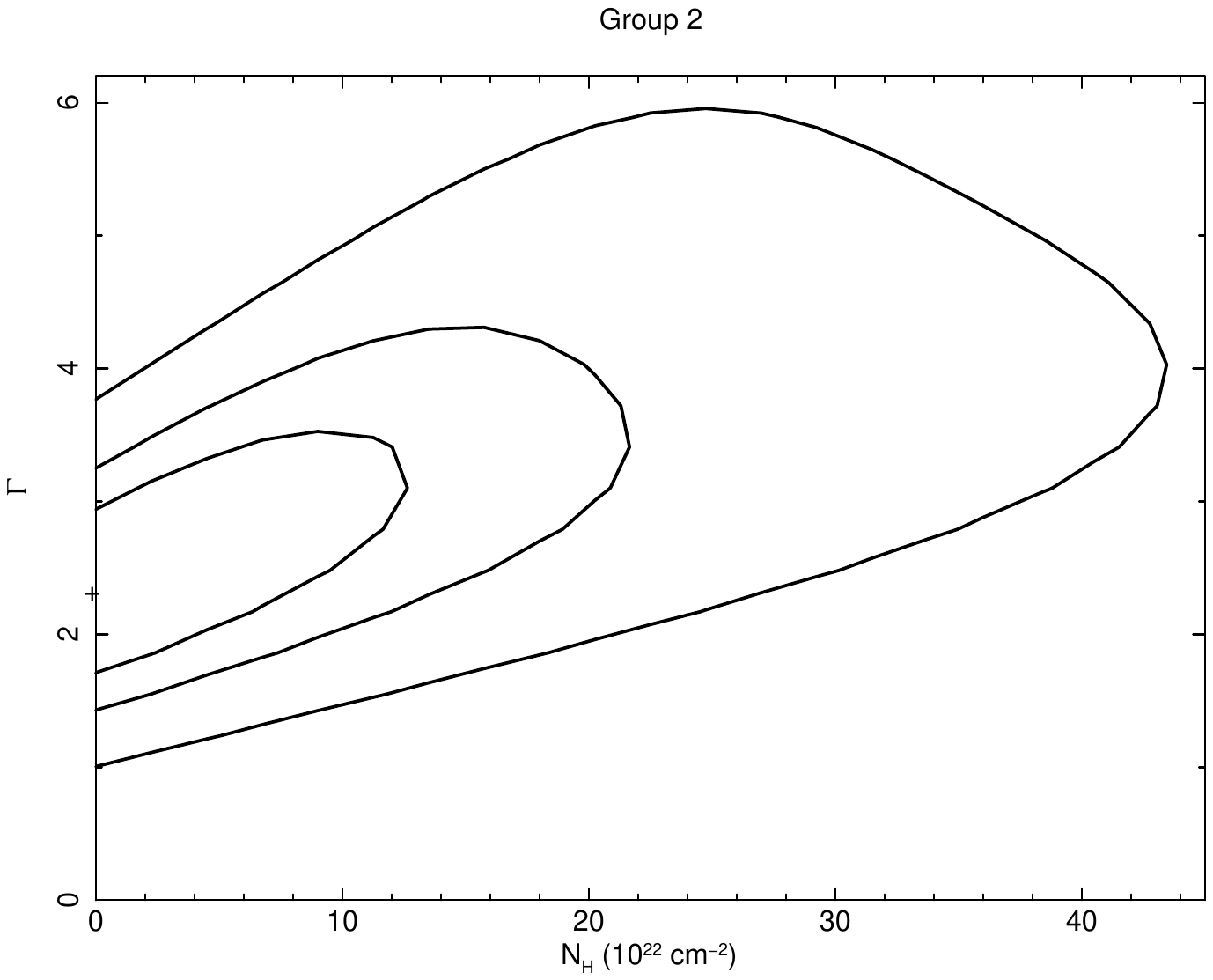}{0.45\textwidth}{}\hspace*{-1.2cm}\fig{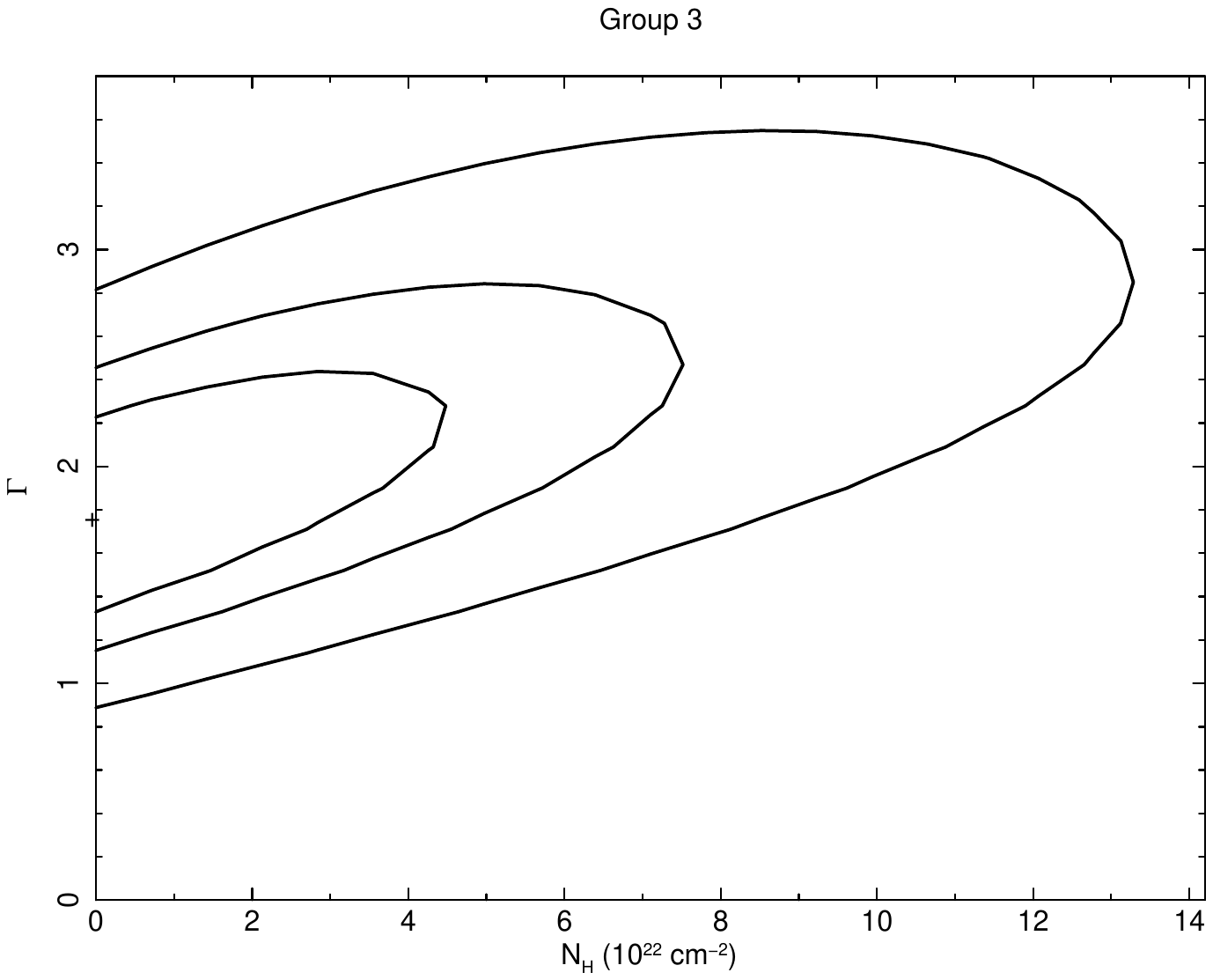}{0.45\textwidth}{}}
          \vspace*{-1.5cm}
\gridline{\fig{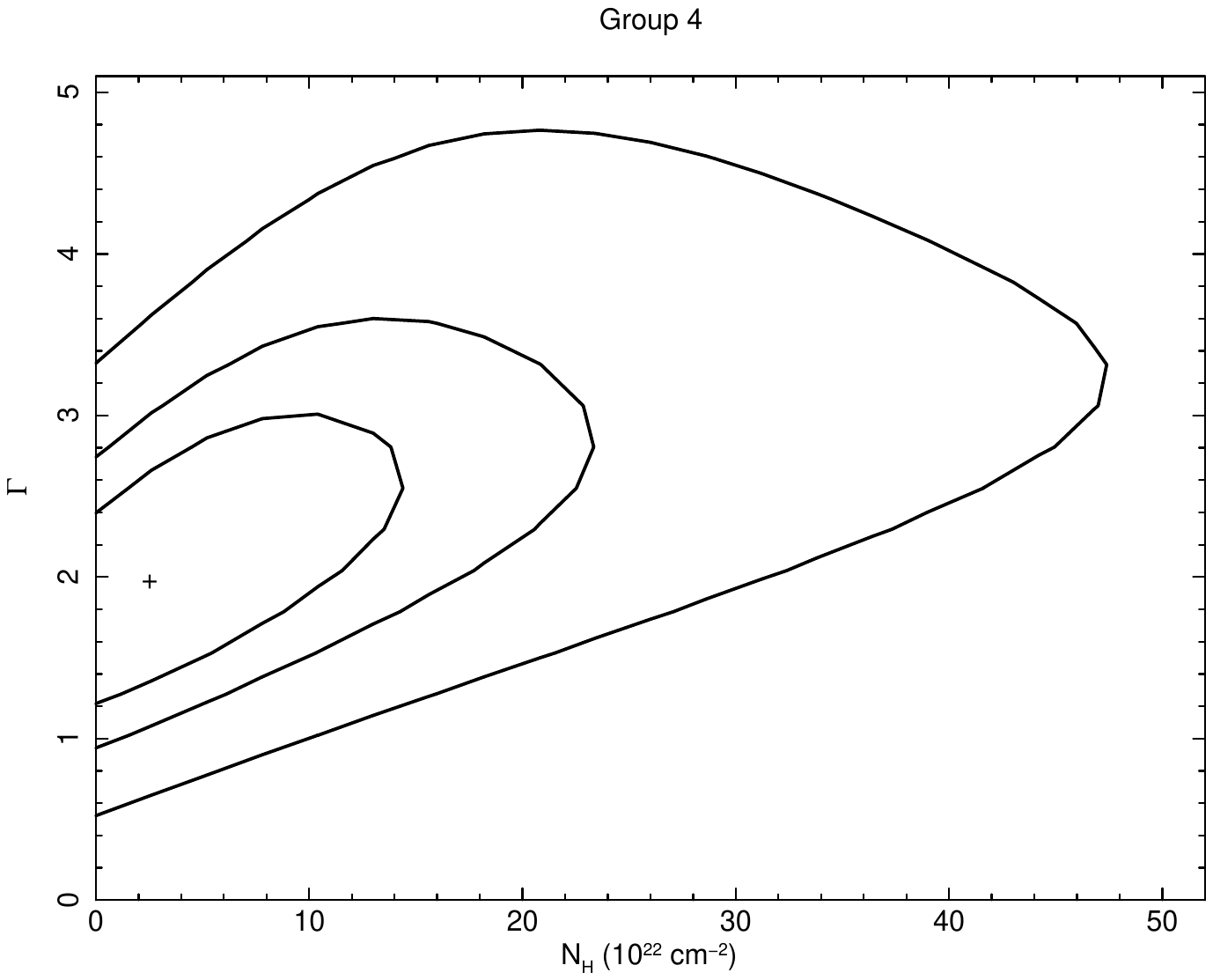}{0.45\textwidth}{}
          \hspace*{-1.5cm}\fig{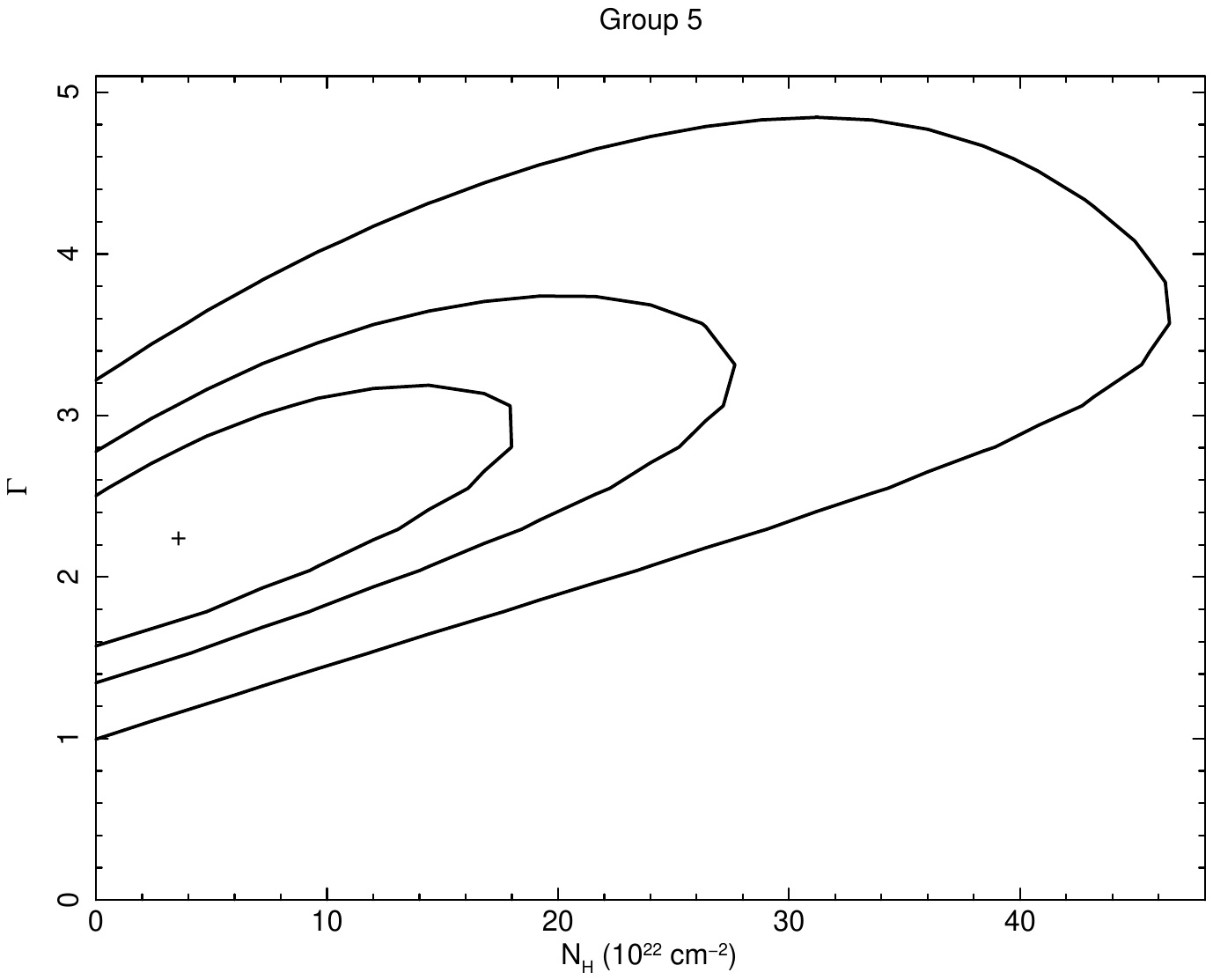}{0.45\textwidth}{}}
          \vspace*{-1.5cm}
\gridline{\fig{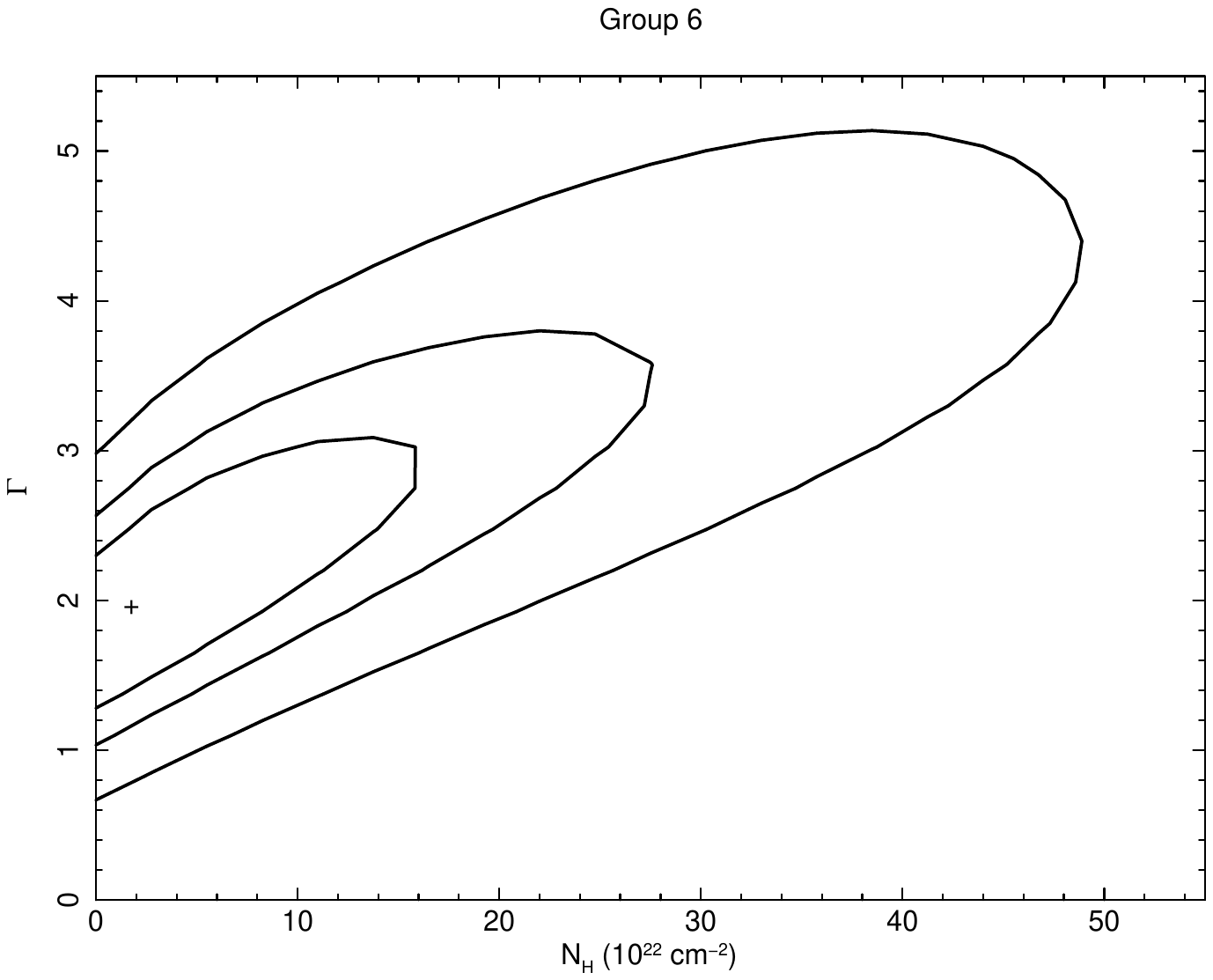}{0.45\textwidth}{}
          \hspace*{-1.5cm}\fig{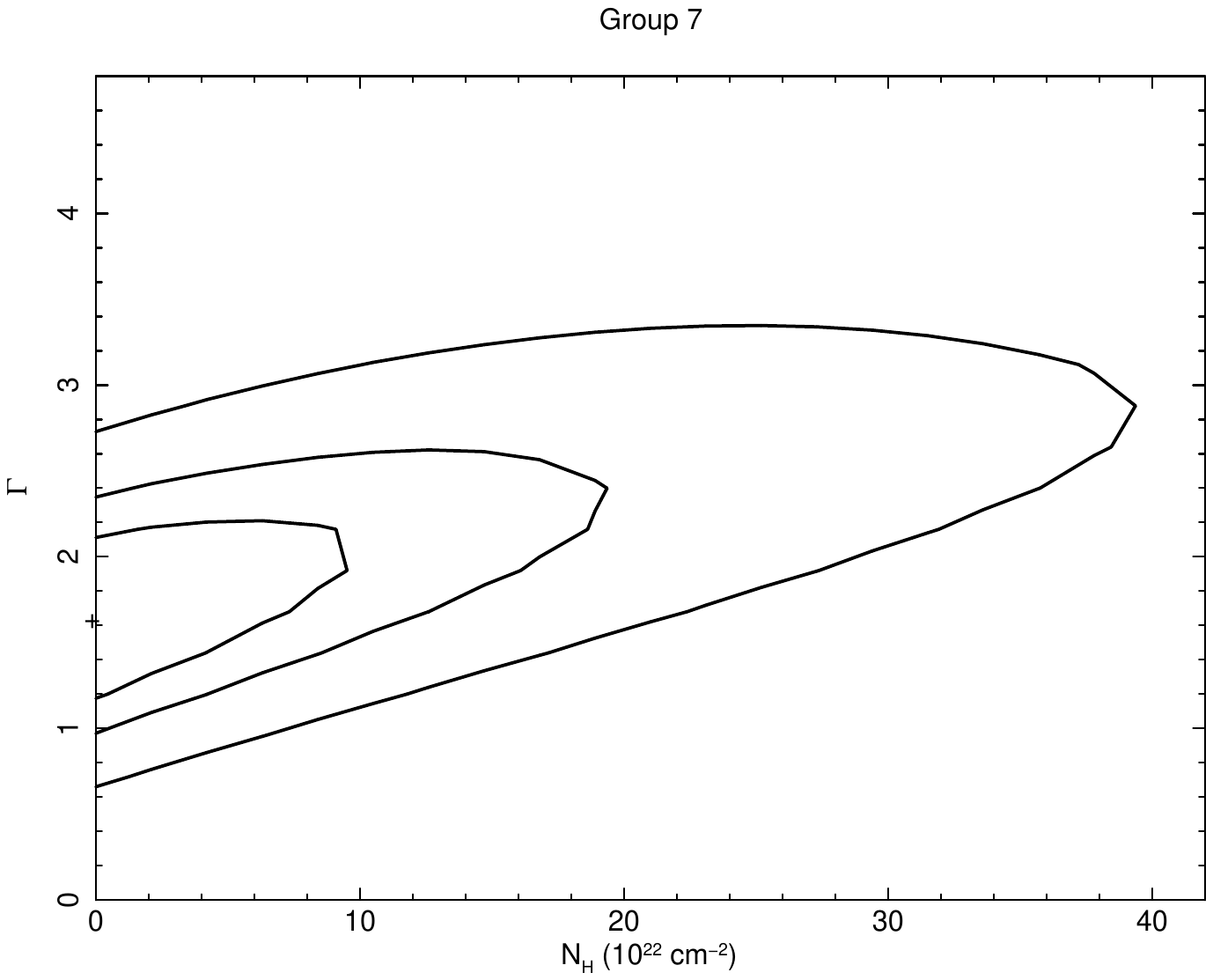}{0.45\textwidth}{}}
\caption{Same as Figure~\ref{fig:cont} but for the \lledd\ bins given in Table~\ref{tab:stack_lledd}.}
\label{fig:lledd_cont}
\end{figure*}

In the second and third runs, we followed the procedures of the first run and joint-fitted groups of sources in order of increasing \luv\ and \lledd, respectively. The grouping was performed such that each group has $\sim100$ counts. All fits were performed at $>2$ keV in the rest frame using the average redshift of each group. Tables~\ref{tab:stack_luv} and \ref{tab:stack_lledd} present the results of these joint fits. Figures~\ref{fig:luv_cont} and \ref{fig:lledd_cont} show the respective 68\%, 90\%, and 99\% confidence regions for the photon index versus intrinsic neutral absorption column density of each group. We do not find any trends between $\Gamma$ and either \luv\ or \lledd\ in these joint fits. As in the first run, the joint fits resulted in upper-limits on the neutral absorption column density (see Tables~\ref{tab:stack_luv} and \ref{tab:stack_lledd}).

Given the large uncertainties on both $\Gamma$ and \nh\ values in all of these joint fits, deeper \xray\ observations of all of our sources are required to obtain accurate, individual $\Gamma$ values that would allow us to search for trends among the optical-UV Eddington luminosity ratio diagnostics and this parameter, similar to the analysis we performed for \aox.

\begin{deluxetable}{lccccc}
\tabletypesize{\footnotesize}
\tablecolumns{6}
\tablecaption{Joint Fitting\label{tab:stack_z}}
\tablehead{
\colhead{} &
\colhead{} &
\colhead{} &
\colhead{} &
\colhead{} &
\colhead{$N_{\rm H}$} \\
\colhead{Sample} &
\colhead{$N$} &
\colhead{Counts\tablenotemark{a}} &
\colhead{$\left<z\right>$} &
\colhead{$\Gamma$} &
\colhead{($10^{22}$~cm$^{-2}$)} 
}
\startdata
GNIRS-DQS all\tablenotemark{b} & 55 & 781 & 2.106 & $1.81\pm 0.14$ & $< 1.64$ \\
GNIRS-DQS at $z\sim1.5$\tablenotemark{c} & 17 & 231 & 1.596 & $1.80\pm 0.27$ & $< 1.64$ \\
GNIRS-DQS at $z\sim2$\tablenotemark{d} & 36 & 524 & 2.282 & $1.82^{+0.23}_{-0.17}$ & $< 2.03$ \\
GNIRS-DQS targeted\tablenotemark{e} & 50 & 650 & 2.067 & $1.84\pm 0.16$ & $< 1.64$
\enddata
\tablenotetext{a}{Number of total counts above 2 keV in the rest frame, using the average redshift of each subsample.}
\tablenotetext{b}{Excludes four sources with $> 100$ counts and four sources with insufficient counts, SDSS~J081019.47$+$095040.9, SDSS J094637.83$+$012411.5, SDSS J095823.07$+$371218.3, SDSS J110810.87$+$014140.7 (see Table~\ref{tab:chandra_uv-opt}).}
\tablenotetext{c}{Includes only targeted sources (see Table~\ref{tab:chandra_uv-opt}).}
\tablenotetext{d}{Includes both targeted and M22 sources; excludes four sources with $> 100$ counts and four sources with insufficient counts from Table~\ref{tab:chandra_uv-opt}.}
\tablenotetext{e}{Includes all targeted sources; excludes four sources with insufficient counts from Table~\ref{tab:chandra_uv-opt}.}
\end{deluxetable}

\begin{deluxetable}{lccccc}
\tabletypesize{\footnotesize}
\tablecolumns{6}
\tablecaption{Joint Fitting by \luv\label{tab:stack_luv}}
\tablehead{
\colhead{} &
\colhead{} &
\colhead{} &
\colhead{} &
\colhead{} &
\colhead{$N_{\rm H}$} \\
\colhead{Sample} &
\colhead{$N$} &
\colhead{Counts\tablenotemark{a}} &
\colhead{$\left<\log L_{2500\,\text{\AA}}\right>$} &
\colhead{$\Gamma$} &
\colhead{($10^{22}$~cm$^{-2}$)} 
}
\startdata
Group 1 & 6 & 102 & $31.25\pm 0.06$ & $2.16^{+1.77}_{-1.25}$ & $<29.95$ \\
Group 2 & 8 & 94 & $31.38\pm 0.03$ & $2.20^{+1.74}_{-1.09}$ & $<13.27$ \\
Group 3 & 5 & 98 & $31.46\pm 0.02$ & $2.16^{+1.27}_{-0.73}$ & $<25.26$ \\
Group 4 & 9 & 103 & $31.56\pm 0.04$ & $1.95^{+1.12}_{-0.62}$ & $<10.64$ \\
Group 5 & 10 & 105 & $31.67\pm 0.03$ & $2.02^{+1.21}_{-0.61}$ & $<25.49$ \\
Group 6 & 4 & 108 & $31.71\pm 0.01$ & $2.01^{+0.99}_{-0.53}$ & $<7.90$ \\
Group 7 & 8 & 96 & $31.80\pm 0.07$ & $2.53^{+1.52}_{-1.21}$ & $<62.18$ \\
Group 8 & 5 & 75 & $31.96\pm 0.03$ & $1.53^{+0.69}_{-0.61}$ & $<6.42$
\enddata
\tablenotetext{a}{Number of total counts above 2 keV in the rest frame, using the average redshift of each group.}
\end{deluxetable}

\begin{deluxetable}{lccccc}
\tabletypesize{\footnotesize}
\tablecolumns{6}
\tablecaption{Joint Fitting by \lledd\label{tab:stack_lledd}}
\tablehead{
\colhead{} &
\colhead{} &
\colhead{} &
\colhead{} &
\colhead{} &
\colhead{$N_{\rm H}$} \\
\colhead{Sample} &
\colhead{$N$} &
\colhead{Counts\tablenotemark{a}} &
\colhead{$\left<L/L_{\rm Edd}\right>$} &
\colhead{$\Gamma$} &
\colhead{($10^{22}$~cm$^{-2}$)} 
}
\startdata
Group 1 & 6 & 93 & $0.23\pm{0.06}$ & $2.03^{+1.26}_{-1.00}$ & $<24.05$ \\
Group 2 & 6 & 107 & $0.39\pm{0.08}$ & $2.32^{+0.70}_{-0.67}$ & $<13.16$ \\
Group 3 & 7 & 129 & $0.55\pm{0.03}$ & $1.76^{+0.52}_{-0.47}$ & $<3.29$ \\
Group 4 & 8 & 106 & $0.78\pm{0.08}$ & $1.98^{+1.15}_{-0.75}$ & $<15.93$ \\
Group 5 & 8 & 103 & $1.08\pm{0.08}$ & $2.24^{+1.06}_{-0.71}$ & $<20.07$ \\
Group 6 & 6 & 110 & $1.32\pm{0.07}$ & $1.96^{+1.27}_{-0.66}$ & $<18.41$ \\
Group 7 & 14 & 133 & $4.09\pm{4.36}$ & $1.63^{+0.53}_{-0.50}$ & $<6.58$
\enddata
\tablenotetext{a}{Number of total counts above 2 keV in the rest frame, using the average redshift of each group.}
\end{deluxetable}

\newpage

\newpage

\section{Summary}
\label{sec:summary}
We present \chandra\ observations of 63 GNIRS-DQS sources, 54 of which are targeted, snapshot observations obtained in Cycle~24. We utilize these data to investigate if \xray s can contribute to providing a robust Eddington luminosity ratio estimate in quasars. This is performed by searching for correlations between \aox, \daox, \luv, \feii-corrected \hb-based \lledd, and the \civdist\ parameters. Our results confirm previous findings identifying the \civdist\ parameter as a robust \lledd\ indicator among optical-UV diagnostics of this fundamental quasar property up to, at least, $z\sim3.5$. Our results also suggest that \aox\ does not contribute any significant improvements to \lledd\ estimates indicated by the \civdist\ parameter.

We derive hard-\xray\ photon index ($\Gamma$) values for a small subset of our GNIRS-DQS sources for which a sufficient number of \xray\ photons are available. The $\Gamma$ values of these sources are consistent with their \lledd\ values derived from optical-UV diagnostics. We also obtain average $\Gamma$ values for our sources through joint-fitting their \xray\ spectra and separating them by redshift, optical luminosity, and Eddington luminosity ratio. Overall, the results of these joint-fit exercises indicate that the average $\Gamma$ is consistent with the relatively high average \hb-based \lledd\ value of these sources. Deeper \xray\ observations of our \xray-detected GNIRS-DQS sources are required to test whether $\Gamma$ can serve as a robust, un-biased Eddington luminosity ratio indicator in quasars.

\acknowledgements
The scientific results presented in this paper are based on observations made by the \chandra\ \xray\ Observatory and on data obtained from the \chandra\ Data Archive. Support for this work was provided by the National Aeronautics and Space Administration (NASA) through \chandra\ award No. GO3-24089X (A.M., O.S.) issued by the \chandra\ \xray\ Observatory Center (CXC), which is operated by the Smithsonian Astrophysical Observatory for and on behalf of NASA under contract NAS8-03060. Support for this work was also provided by NASA under award No. 80NSSC24K1468 (O.S., G.T.R.). W.N.B. acknowledges support from the Penn State Eberly Endowment. We thank an anonymous referee for providing valuable feedback that helped to improve this manuscript. This research has made use of the NASA/IPAC Extragalactic Database (NED) which is operated by the Jet Propulsion Laboratory, California Institute of Technology, under contract with the National Aeronautics and Space Administration, as well as NASA’s Astrophysics Data System Bibliographic Services.

This paper employs a list of Chandra datasets, obtained by the Chandra \xray\ Observatory, contained in the Chandra Data Collection (CDC) 342~\dataset[doi:10.25574/cdc.342]{https://doi.org/10.25574/cdc.342}

\clearpage
\appendix
\section{Source Overlap}\label{sec:A}
\begin{figure*}
\epsscale{1.1}
\plotone{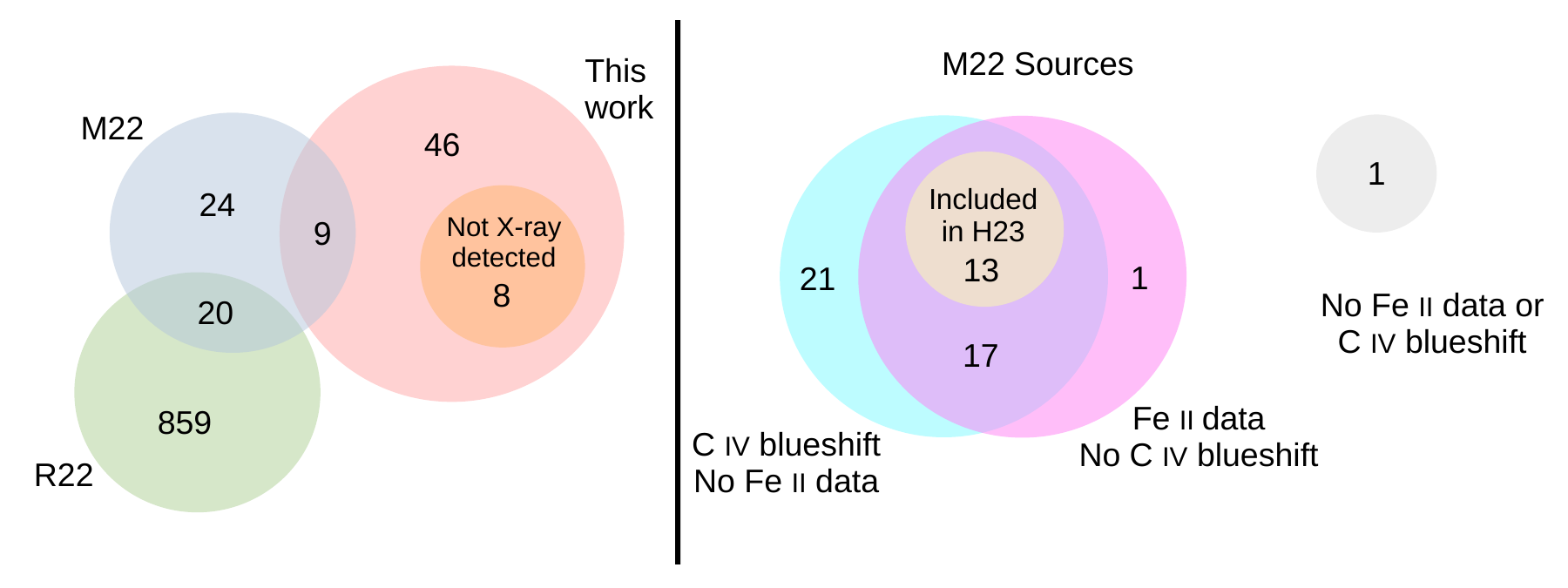}
\caption{Left: Venn-diagram of the overlap in sources between this work, M22, and R22. Right: The breakdown of all 53 M22 sources in regards to which sources have \civ\ data, \feii\ data, both, or neither (see also Table~\ref{tab:2022}).} 
\label{fig:venn_diagram}
\end{figure*}

In this work, we employ the use of three parameters to search for a robust accretion-rate indicator for quasars: \aox\ from \xray\ observations, \civdist\ from the \civ\ parameter space, and the \feii-corrected \hb-based \lledd. R22 introduced the concept of \civdist, and compared to \aox, but lacked the \feii\ correction to \lledd. H23 introduced this \feii\ correction, and compared with \civdist, but lacked \xray\ information. This work is the first to exploit all three parameters. For our analysis, we used multiple combinations of the aforementioned studies, along with subsets of the M22 sample for which \civdist\ or \feii\ information was available (see Figure~\ref{fig:venn_diagram} and Appendix~\ref{sec:C}). The red circle on the left side of Figure~\ref{fig:venn_diagram} shows the 63 GNIRS sources: eight of which were not detected in \xray s, and, therefore, not included in any correlation involving \aox; nine come from archival observations that were presented in M22. The green circle on the left is the R22 ``Good" sample of 879 sources, 20 of which overlap with M22 (see Appendix~\ref{sec:B}). The blue circle on the left shows the 53 sources from M22 - the right side of Figure~\ref{fig:venn_diagram} shows the breakdown of how many sources have \civ\ vs. \feii\ information, with one source having neither; of the 30 sources that have both parameters, 13 are included in the 248 H23 sample.

\section{Cross Matching Sources Between R22 and M22}\label{sec:B}
Table~\ref{tab:R22_overlap} presents the 20 sources that are crossed matched between R22 and M22, and, therefore, are only counted once for the purpose of the correlations in Section~\ref{sec:aox_luv}. The \aox\ and \luv\ values are taken from M22, and except for two sources, \civdist\ values are taken from R22.
\begin{deluxetable}{lcccc}
\tablecolumns{5}
\tablecaption{Cross Matched Sources Between M22 and R22\label{tab:R22_overlap}}
\tablehead{
\colhead{Quasar} &
\colhead{\aox} &
\colhead{\daox} &
\colhead{log \luv} &
\colhead{\civdist} 
}
\startdata
\object{SDSS~J002019.22$-$110609.2} & $-1.60$ & $-0.17$ & 30.17 & 0.462 \\
\object{SDSS~J030341.04$-$002321.9} & $-1.82$ & $-0.01$ & 32.08 & 0.688 \\
\object{SDSS~J082024.21$+$233450.4} & $-1.48$ & $-0.02$ & 30.32 & 0.506 \\
\object{SDSS~J082658.85$+$061142.6} & $-1.56$ & $-0.11$ & 30.25 & 0.498 \\
\object{SDSS~J083332.92$+$164411.0} & $-1.51$ & $-0.06$ & 30.26 & 0.608 \\
\object{SDSS~J083510.36$+$035901.1} & $-1.62$ & $-0.17$ & 30.24 & 0.543 \\
\object{SDSS~J091451.42$+$421957.0} & $-1.71$ & $-0.25$ & 30.31 & 0.481 \\
\object{SDSS~J094202.04$+$042244.5} & $-1.71$ & $+0.11$ & 32.14 & 0.594 \\
\object{SDSS~J100054.96$+$262242.4} & $-1.64$ & $-0.19$ & 30.26 & 0.611 \\
\object{SDSS~J111138.66$+$575030.0} & $-1.64$ & $-0.19$ & 30.27 & 0.545 \\
\object{SDSS~J112614.93$+$310146.6} & $-1.44$ & $-0.01$ & 30.15 & 0.393 \\
\object{SDSS~J113327.78$+$032719.1} & $-1.48$ & $-0.02$ & 30.31 & 0.252 \\
\object{SDSS~J123734.47$+$444731.7} & $-1.54$ & $-0.08$ & 30.32 & 0.629 \\
\object{SDSS~J125415.55$+$480850.6} & $-1.47$ & $0.00$ & 30.36 & 0.420 \\
\object{SDSS~J134701.54$+$215401.1} & $-1.48$ & $-0.05$ & 30.16 & 0.494 \\
\object{SDSS~J135023.68$+$265243.1} & $-1.71$ & $-0.02$ & 31.45 & 0.583 \\
\object{SDSS~J141141.96$+$140233.9} & $-1.42$ & $+0.16$ & 30.93 & 1.065\tablenotemark{a} \\
\object{SDSS~J141730.92$+$073320.7} & $-1.56$ & $+0.05$ & 31.08 & 1.271\tablenotemark{a} \\
\object{SDSS~J152654.61$+$565512.3} & $-1.59$ & $-0.15$ & 30.21 & 0.551 \\
\object{SDSS~J212329.46$-$005052.9} & $-1.75$ & $+0.09$ & 32.24 & 0.935
\enddata
\tablenotetext{a}{Value taken from H23.}
\end{deluxetable}

\section{M22 Sources}\label{sec:C}
Table~\ref{tab:2022} presents the complete M22 sample of 53 sources with \civdist\ and \feii-corrected \hb-based \lledd\ values derived for the purpose of this work. Of these sources, 22 lack \feii\ data, and were not included in the correlations performed in Section~\ref{sec:aox_lledd}. Two sources lack \civ\ data, and were not included in the correlations performed in Section~\ref{sec:aox_civ}. Additionally, 30 sources have both the \feii\ and \civ\ data necessary for our analyses. Of these, nine are the archival GNIRS-DQS sources included in the sample of this work, and an additional four are included in H23, leaving 17 sources that were added to the correlations performed in Section~\ref{sec:civ_lledd}.
\startlongtable
\begin{deluxetable*}{lccccc}
\tablecolumns{6}
\tablecaption{All M22 Sources \label{tab:2022}}
\tablehead{ 
\nocolhead{} &
\nocolhead{} &
\nocolhead{} &
\nocolhead{} &
\nocolhead{} &
\nocolhead{} \\
\colhead{Quasar} &
\colhead{\civdist} &
\colhead{Ref.} &
\colhead{$R_{\rm Fe~{\sc II}}$} &
\colhead{Ref.} &
\colhead{\lledd} \\
\colhead{(1)} &
\colhead{(2)} &
\colhead{(3)} &
\colhead{(4)} &
\colhead{(5)} &
\colhead{(6)}
}
\startdata
\object{SDSS~J002019.22$-$110609.2} & 0.462 & 1 & 0.38 & 2 & 0.30 \\
\object{SDSS~J005709.94$+$144610.1} & 0.299 & 3 & 0.16 & 2 & 0.02 \\
\object{SDSS~J014812.83$+$000322.9} & 0.687 & 3 & \nodata & \nodata & \nodata \\
\object{SDSS~J015950.23$+$002340.9} & 0.366 & 3 & 0.51 & 2 & 0.55 \\
\object{SDSS~J030341.04$-$002321.9} & 0.688 & 1 & \nodata & \nodata & \nodata \\
\object{SDSS~J032349.53$-$002949.8} & 0.511 & 3 & \nodata & \nodata & \nodata \\
\object{SDSS~J080117.79$+$521034.5} & 0.959 & 4 & 0.64 & 4 & 1.63 \\
\object{SDSS~J082024.21$+$233450.4} & 0.506 & 1 & \nodata & \nodata & \nodata \\
\object{SDSS~J082658.85$+$061142.6} & 0.498 & 1 & 0.54 & 2 & 0.97 \\
\object{SDSS~J083332.92$+$164411.0} & 0.608 & 1 & \nodata & \nodata & \nodata \\
\object{SDSS~J083510.36$+$035901.1} & 0.543 & 1 & 0.44 & 2 & 0.28 \\
\object{SDSS~J084846.11$+$611234.6} & 0.563 & 4 & 0.52 & 4 & 1.48 \\
\object{SDSS~J085116.14$+$424328.8} & 0.634 & 3 & 0.52 & 2 & 0.18 \\
\object{SDSS~J090033.50$+$421547.0} & 0.571 & 3 & \nodata & \nodata & \nodata \\
\object{SDSS~J091451.42$+$421957.0} & 0.481 & 1 & 0.58 & 2 & 0.44 \\
\object{SDSS~J093502.52$+$433110.6} & 0.548 & 3 & 0.47 & 2 & 0.10 \\
\object{SDSS~J094202.04$+$042244.5} & 0.594 & 1 & \nodata & \nodata & \nodata \\
\object{SDSS~J094602.31$+$274407.0} & 1.359 & 4 & 1.65 & 4 & 2.89 \\
\object{SDSS~J094646.94$+$392719.0} & 0.989 & 4 & 1.10 & 4 & 1.08 \\
\object{SDSS~J095852.19$+$120245.0} & 0.715 & 4 & 0.25 & 4 & 1.07 \\
\object{SDSS~J100054.96$+$262242.4} & 0.611 & 1 & \nodata & \nodata & \nodata \\
\object{SDSS~J102907.09$+$651024.6} & 0.760 & 4 & 0.48 & 4 & 0.90 \\
\object{SDSS~J103320.65$+$274024.2} & 0.496 & 3 & \nodata & \nodata & \nodata \\
\object{SDSS~J111119.10$+$133603.8} & 0.718 & 4 & 0.33 & 4 & 0.70 \\
\object{SDSS~J111138.66$+$575030.0} & 0.545 & 1 & 0.51 & 2 & 1.28 \\
\object{SDSS~J111830.28$+$402554.0} & 0.458 & 3 & 0.52 & 2 & 0.16 \\
\object{SDSS~J111908.67$+$211918.0} & 0.408 & 3 & \nodata & \nodata & \nodata \\
\object{SDSS~J111941.12$+$595108.7} & 0.661 & 3 & 0.76 & 2 & 1.73 \\
\object{SDSS~J112224.15$+$031802.6} & 0.663 & 3 & 0.84 & 2 & 0.45 \\
\object{SDSS~J112614.93$+$310146.6} & 0.393 & 1 & \nodata & \nodata & \nodata \\
\object{SDSS~J113327.78$+$032719.1} & 0.252 & 1 & 0.40 & 2 & 0.20 \\
\object{SDSS~J115954.33$+$201921.1} & 0.760 & 3 & \nodata & \nodata & \nodata \\
\object{SDSS~J123734.47$+$444731.7} & 0.629 & 1 & 0.64 & 2 & 0.22 \\
\object{SDSS~J125415.55$+$480850.6} & 0.420 & 1 & 0.22 & 2 & 0.25 \\
\object{SDSS~J131627.84$+$315825.7} & 0.517 & 3 & \nodata & \nodata & \nodata \\
\object{SDSS~J134701.54$+$215401.1} & 0.494 & 1 & \nodata & \nodata & \nodata \\
\object{SDSS~J135023.68$+$265243.1} & 0.583 & 1 & \nodata & \nodata & \nodata \\
\object{SDSS~J140331.29$+$462804.8} & 0.376 & 3 & 0.47 & 2 & 0.51 \\
\object{SDSS~J140621.89$+$222346.5} & \nodata & \nodata & \nodata & \nodata & \nodata \\
\object{SDSS~J141028.14$+$135950.2} & 0.637 & 4 & 0.82 & 4 & 0.87 \\
\object{SDSS~J141141.96$+$140233.9} & 1.065 & 4 & 1.41 & 4 & 1.06 \\
\object{SDSS~J141730.92$+$073320.7} & 1.271 & 4 & 1.65 & 4 & 2.95 \\
\object{SDSS~J141949.39$+$060654.0} & 0.801 & 3 & \nodata & \nodata & \nodata \\
\object{SDSS~J141951.84$+$470901.3} & 0.909 & 4 & 0.70 & 4 & 1.19 \\
\object{SDSS~J144741.76$-$020339.1} & 0.865 & 4 & 1.60 & 4 & 5.27 \\
\object{SDSS~J145334.13$+$311401.4} & 0.475 & 3 & \nodata & \nodata & \nodata \\
\object{SDSS~J152156.48$+$520238.5} & 1.214 & 4 & 1.64 & 4 & 3.27 \\
\object{SDSS~J152654.61$+$565512.3} & 0.551 & 1 & \nodata & \nodata & \nodata \\
\object{SDSS~J155837.77$+$081345.8} & 0.536 & 3 & \nodata & \nodata & \nodata \\
\object{SDSS~J212329.46$-$005052.9} & 0.935 & 1 & \nodata & \nodata & \nodata \\
\object{SDSS~J230301.45$-$093930.7} & 0.600 & 3 & \nodata & \nodata & \nodata \\
\object{SDSS~J234145.51$-$004640.5} & 0.422 & 3 & 0.42 & 2 & 0.09 \\
\object{SDSS~J235321.62$-$002840.6} & \nodata & \nodata & 0.16 & 2 & 0.24
\enddata
\tablereferences{(1) R22, (2) derived using EW(\feii) and EW(\hb) values from Dong \et (2011), (3) computed for this work, (4) H23.}
\end{deluxetable*}



\clearpage
\begin{thebibliography}{}
%
\bibitem[Ahmed et al.(2024)]{2024ApJ...968...77A} Ahmed, H., Shemmer, O., Matthews, B., et al.\ 2024, \apj, 968, 77. doi:10.3847/1538-4357/ad3e69
%
\bibitem[Arnaud(1996)]{1996ASPC..101...17A} Arnaud, K.~A.\ 1996, Astronomical Data Analysis Software and Systems V, 101, 17
%
\bibitem[Bahk et al.(2019)]{2019ApJ...875...50B} Bahk, H., Woo, J.-H., \& Park, D.\ 2019, \apj, 875, 50. doi:10.3847/1538-4357/ab100d
%
\bibitem[Ba{\~n}ados et al.(2016)]{2016ApJS..227...11B} Ba{\~n}ados, E., Venemans, B.~P., Decarli, R., et al.\ 2016, \apjs, 227, 11. doi:10.3847/0067-0049/227/1/11
%
\bibitem[Baskin \& Laor(2004)]{2004MNRAS.350L..31B} Baskin, A. \& Laor, A.\ 2004, \mnras, 350, L31. doi:10.1111/j.1365-2966.2004.07833.x
%
\bibitem[Brightman et al.(2013)]{2013MNRAS.433.2485B} Brightman, M., Silverman, J.~D., Mainieri, V., et al.\ 2013, \mnras, 433, 2485. doi:10.1093/mnras/stt920
%
\bibitem[Cash(1979)]{1979ApJ...228..939C} Cash, W.\ 1979, \apj, 228, 939. doi:10.1086/156922
%
\bibitem[Coatman et al.(2016)]{2016MNRAS.461..647C} Coatman, L., Hewett, P.~C., Banerji, M., et al.\ 2016, \mnras, 461, 647. doi:10.1093/mnras/stw1360
%
\bibitem[Constantin et al.(2009)]{2009ApJ...705.1336C} Constantin, A., Green, P., Aldcroft, T., et al.\ 2009, \apj, 705, 1336. doi:10.1088/0004-637X/705/2/1336
%
\bibitem[Dalla Bont{\`a} et al.(2020)]{2020ApJ...903..112D} Dalla Bont{\`a}, E., Peterson, B.~M., Bentz, M.~C., et al.\ 2020, \apj, 903, 112. doi:10.3847/1538-4357/abbc1c
%
\bibitem[Dickey \& Lockman(1990)]{1990ARA&A..28..215D} Dickey, J.~M. \& Lockman, F.~J.\ 1990, \araa, 28, 215. doi:10.1146/annurev.aa.28.090190.001243
%
\bibitem[Dong et al.(2011)]{2011ApJ...736...86D} Dong, X.-B., Wang, J.-G., Ho, L.~C., et al.\ 2011, \apj, 736, 86. doi:10.1088/0004-637X/736/2/86
%
\bibitem[Du \& Wang(2019)]{2019ApJ...886...42D} Du, P. \& Wang, J.-M.\ 2019, \apj, 886, 42. doi:10.3847/1538-4357/ab4908
%
\bibitem[Du et al.(2018)]{2018ApJ...856....6D} Du, P., Zhang, Z.-X., Wang, K., et al.\ 2018, \apj, 856, 6. doi:10.3847/1538-4357/aaae6b
%
\bibitem[Fanali et al.(2013)]{2013MNRAS.433..648F} Fanali, R., Caccianiga, A., Severgnini, P., et al.\ 2013, \mnras, 433, 648. doi:10.1093/mnras/stt757
%
\bibitem[Freeman et al.(2002)]{2002ApJS..138..185F} Freeman, P.~E., Kashyap, V., Rosner, R., et al.\ 2002, \apjs, 138, 185. doi:10.1086/324017
%
\bibitem[Gallagher et al.(2006)]{2006ApJ...644..709G} Gallagher, S.~C., Brandt, W.~N., Chartas, G., et al.\ 2006, \apj, 644, 709. doi:10.1086/503762
%
\bibitem[Garmire et al.(2003)]{2003SPIE.4851...28G} Garmire, G.~P., Bautz, M.~W., Ford, P.~G., et al.\ 2003, \procspie, 4851, 28. doi:10.1117/12.461599
%
\bibitem[Gehrels(1986)]{1986ApJ...303..336G} Gehrels, N.\ 1986, \apj, 303, 336. doi:10.1086/164079
%
\bibitem[Gibson et al.(2008)]{2008ApJ...685..773G} Gibson, R.~R., Brandt, W.~N., \& Schneider, D.~P.\ 2008, \apj, 685, 773. doi:10.1086/590403
%
\bibitem[Giustini \& Proga(2019)]{2019A&A...630A..94G} Giustini, M. \& Proga, D.\ 2019, \aap, 630, A94. doi:10.1051/0004-6361/201833810
%
\bibitem[Grier et al.(2017)]{2017ApJ...851...21G} Grier, C.~J., Trump, J.~R., Shen, Y., et al.\ 2017, \apj, 851, 21. doi:10.3847/1538-4357/aa98dc
%
\bibitem[Grupe et al.(2010)]{2010ApJS..187...64G} Grupe, D., Komossa, S., Leighly, K.~M., et al.\ 2010, \apjs, 187, 64. doi:10.1088/0067-0049/187/1/64
%
\bibitem[Ha et al.(2023)]{2023ApJ...950...97H} Ha, T., Dix, C., Matthews, B.~M., et al.\ 2023, \apj, 950, 97. doi:10.3847/1538-4357/acd04d
%
\bibitem[Haardt \& Maraschi(1991)]{1991ApJ...380L..51H} Haardt, F. \& Maraschi, L.\ 1991, \apjl, 380, L51. doi:10.1086/186171
%
\bibitem[Jin et al.(2012)]{2012MNRAS.425..907J} Jin, C., Ward, M., \& Done, C.\ 2012, \mnras, 425, 907. doi:10.1111/j.1365-2966.2012.21272.x
%
\bibitem[Just et al.(2007)]{2007ApJ...665.1004J} Just, D.~W., Brandt, W.~N., Shemmer, O., et al.\ 2007, \apj, 665, 1004. doi:10.1086/519990
%
\bibitem[Kellermann et al.(1989)]{1989AJ.....98.1195K} Kellermann, K.~I., Sramek, R., Schmidt, M., et al.\ 1989, \aj, 98, 1195. doi:10.1086/115207
%
\bibitem[Kraft et al.(1991)]{1991ApJ...374..344K} Kraft, R.~P., Burrows, D.~N., \& Nousek, J.~A.\ 1991, \apj, 374, 344. doi:10.1086/170124
%
\bibitem[Kubota \& Done(2018)]{2018MNRAS.480.1247K} Kubota, A. \& Done, C.\ 2018, \mnras, 480, 1247. doi:10.1093/mnras/sty1890
%
\bibitem[Laor(1998)]{1998ApJ...505L..83L} Laor, A.\ 1998, \apjl, 505, L83. doi:10.1086/311619
%
\bibitem[Liu et al.(2021)]{2021ApJ...910..103L} Liu, H., Luo, B., Brandt, W.~N., et al.\ 2021, \apj, 910, 103. doi:10.3847/1538-4357/abe37f
%
\bibitem[Luo et al.(2015)]{2015ApJ...805..122L} Luo, B., Brandt, W.~N., Hall, P.~B., et al.\ 2015, \apj, 805, 122. doi:10.1088/0004-637X/805/2/122
%
\bibitem[Lusso et al.(2010)]{2010A&A...512A..34L} Lusso, E., Comastri, A., Vignali, C., et al.\ 2010, \aap, 512, A34. doi:10.1051/0004-6361/200913298
%
\bibitem[Lusso et al.(2020)]{2020A&A...642A.150L} Lusso, E., Risaliti, G., Nardini, E., et al.\ 2020, \aap, 642, A150. doi:10.1051/0004-6361/202038899
%
\bibitem[Maithil et al.(2022)]{2022MNRAS.515..491M} Maithil, J., Brotherton, M.~S., Shemmer, O., et al.\ 2022, \mnras, 515, 491. doi:10.1093/mnras/stac1748
%
\bibitem[Maithil et al.(2024)]{2024MNRAS.528.1542M} Maithil, J., Brotherton, M.~S., Shemmer, O., et al.\ 2024, \mnras, 528, 2, 1542. doi:10.1093/mnras/stae115
%
\bibitem[Marlar et al.(2022)]{2022ApJ...931...41M} Marlar, A., Shemmer, O., Brotherton, M.~S., et al.\ 2022, \apj, 931, 41. doi:10.3847/1538-4357/ac5f58
%
\bibitem[Mart{\'\i}nez-Aldama et al.(2018)]{2018A&A...618A.179M} Mart{\'\i}nez-Aldama, M.~L., del Olmo, A., Marziani, P., et al.\ 2018, \aap, 618, A179. doi:10.1051/0004-6361/201833541
%
\bibitem[Matthews et al.(2023)]{2023ApJ...950...95M} Matthews, B.~M., Dix, C., Shemmer, O., et al.\ 2023, \apj, 950, 95. doi:10.3847/1538-4357/acd04c
%
\bibitem[Matthews et al.(2021)]{2021ApJS..252...15M} Matthews, B.~M., Shemmer, O., Dix, C., et al.\ 2021, \apjs, 252, 15. doi:10.3847/1538-4365/abc705
%
\bibitem[Mej{\'\i}a-Restrepo et al.(2016)]{2016MNRAS.460..187M} Mej{\'\i}a-Restrepo, J.~E., Trakhtenbrot, B., Lira, P., et al.\ 2016, \mnras, 460, 187. doi:10.1093/mnras/stw568
%
\bibitem[Merloni et al.(2012)]{2012arXiv1209.3114M} Merloni, A., Predehl, P., Becker, W., et al.\ 2012, arXiv:1209.3114. doi:10.48550/arXiv.1209.3114
%
\bibitem[Miller et al.(2011)]{2011ApJ...726...20M} Miller, B.~P., Brandt, W.~N., Schneider, D.~P., et al.\ 2011, \apj, 726, 20. doi:10.1088/0004-637X/726/1/20
%
\bibitem[Ni et al.(2018)]{2018MNRAS.480.5184N} Ni, Q., Brandt, W.~N., Luo, B., et al.\ 2018, \mnras, 480, 5184. doi:10.1093/mnras/sty1989
%
\bibitem[Onoue et al.(2019)]{2019ApJ...880...77O} Onoue, M., Kashikawa, N., Matsuoka, Y., et al.\ 2019, \apj, 880, 77. doi:10.3847/1538-4357/ab29e9
%
\bibitem[Page et al.(2005)]{2005MNRAS.364..195P} Page, K.~L., Reeves, J.~N., O'Brien, P.~T., et al.\ 2005, \mnras, 364, 195. doi:10.1111/j.1365-2966.2005.09550.x
%
\bibitem[Park et al.(2006)]{2006ApJ...652..610P} Park, T., Kashyap, V.~L., Siemiginowska, A., et al.\ 2006, \apj, 652, 610. doi:10.1086/507406
%
\bibitem[Pu et al.(2020)]{2020ApJ...900..141P} Pu, X., Luo, B., Brandt, W.~N., et al.\ 2020, \apj, 900, 141. doi:10.3847/1538-4357/abacc5
%
\bibitem[Rankine et al.(2020)]{2020MNRAS.492.4553R} Rankine, A.~L., Hewett, P.~C., Banerji, M., et al.\ 2020, \mnras, 492, 4553. doi:10.1093/mnras/staa130
%
\bibitem[Reed et al.(2019)]{2019MNRAS.487.1874R} Reed, S.~L., Banerji, M., Becker, G.~D., et al.\ 2019, \mnras, 487, 1874. doi:10.1093/mnras/stz1341
%
\bibitem[Richards et al.(2011)]{2011AJ....141..167R} Richards, G.~T., Kruczek, N.~E., Gallagher, S.~C., et al.\ 2011, \aj, 141, 167. doi:10.1088/0004-6256/141/5/167
%
\bibitem[Risaliti et al.(2009)]{2009ApJ...700L...6R} Risaliti, G., Young, M., \& Elvis, M.\ 2009, \apjl, 700, L6. doi:10.1088/0004-637X/700/1/L6
%
\bibitem[Rivera et al.(2020)]{2020ApJ...899...96R} Rivera, A.~B., Richards, G.~T., Hewett, P.~C., et al.\ 2020, \apj, 899, 96. doi:10.3847/1538-4357/aba62c
%
\bibitem[Rivera et al.(2022)]{2022ApJ...931..154R} Rivera, A.~B., Richards, G.~T., Gallagher, S.~C., et al.\ 2022, \apj, 931, 154. doi:10.3847/1538-4357/ac6a5d
%
\bibitem[Shemmer et al.(2006)]{2006ApJ...646L..29S} Shemmer, O., Brandt, W.~N., Netzer, H., et al.\ 2006, \apjl, 646, L29. doi:10.1086/506911
%
\bibitem[Shemmer et al.(2008)]{2008ApJ...682...81S} Shemmer, O., Brandt, W.~N., Netzer, H., et al.\ 2008, \apj, 682, 81. doi:10.1086/588776
%
\bibitem[Shemmer \& Lieber(2015)]{2015ApJ...805..124S} Shemmer, O. \& Lieber, S.\ 2015, \apj, 805, 124. doi:10.1088/0004-637X/805/2/124
%
\bibitem[Shen \& Liu(2012)]{2012ApJ...753..125S} Shen, Y. \& Liu, X.\ 2012, \apj, 753, 125. doi:10.1088/0004-637X/753/2/125
%
\bibitem[Skrutskie et al.(2006)]{2006AJ....131.1163S} Skrutskie, M.~F., Cutri, R.~M., Stiening, R., et al.\ 2006, \aj, 131, 1163. doi:10.1086/498708
%
\bibitem[Spergel et al.(2007)]{2007ApJS..170..377S} Spergel, D.~N., Bean, R., Dor{\'e}, O., et al.\ 2007, \apjs, 170, 377. doi:10.1086/513700
%
\bibitem[Steffen et al.(2006)]{2006AJ....131.2826S} Steffen, A.~T., Strateva, I., Brandt, W.~N., et al.\ 2006, \aj, 131, 2826. doi:10.1086/503627
%
\bibitem[Tee et al.(2023)]{2023ApJ...956...52T} Tee, W.~L., Fan, X., Wang, F., et al.\ 2023, \apj, 956, 52. doi:10.3847/1538-4357/acf12d
%
\bibitem[Temple et al.(2023)]{2023MNRAS.523..646T} Temple, M.~J., Matthews, J.~H., Hewett, P.~C., et al.\ 2023, \mnras, 523, 646. doi:10.1093/mnras/stad1448
%
\bibitem[Thomas et al.(2020)]{2020SPIE11445E..0IT} Thomas, S.~J., Barr, J., Callahan, S., et al.\ 2020, \procspie, 11445, 114450I. doi:10.1117/12.2561581
%
\bibitem[Timlin et al.(2020)]{2020MNRAS.492..719T} Timlin, J.~D., Brandt, W.~N., Ni, Q., et al.\ 2020, \mnras, 492, 719. doi:10.1093/mnras/stz3433
%
\bibitem[Vasudevan \& Fabian(2007)]{2007MNRAS.381.1235V} Vasudevan, R.~V. \& Fabian, A.~C.\ 2007, \mnras, 381, 1235. doi:10.1111/j.1365-2966.2007.12328.x
%
\bibitem[Vestergaard \& Peterson(2006)]{2006ApJ...641..689V} Vestergaard, M. \& Peterson, B.~M.\ 2006, \apj, 641, 689. doi:10.1086/500572
%
\bibitem[Vignali et al.(2003)]{2003AJ....125..433V} Vignali, C., Brandt, W.~N., \& Schneider, D.~P.\ 2003, \aj, 125, 2, 433. doi:10.1086/345973
%
\bibitem[Wang et al.(2021)]{2021ApJ...907L...1W} Wang, F., Yang, J., Fan, X., et al.\ 2021, \apjl, 907, L1. doi:10.3847/2041-8213/abd8c6
%
\bibitem[Weisskopf et al.(2000)]{2000SPIE.4012....2W} Weisskopf, M.~C., Tananbaum, H.~D., Van Speybroeck, L.~P., et al.\ 2000, \procspie, 4012, 2. doi:10.1117/12.391545
%
\bibitem[Wu et al.(2012)]{2012ApJ...747...10W} Wu, J., Brandt, W.~N., Anderson, S.~F., et al.\ 2012, \apj, 747, 10. doi:10.1088/0004-637X/747/1/10
%
\bibitem[Wu et al.(2012)]{2012ApJS..201...10W} Wu, J., Vanden Berk, D., Grupe, D., et al.\ 2012, \apjs, 201, 10. doi:10.1088/0067-0049/201/2/10
%
\bibitem[York et al.(2000)]{2000AJ....120.1579Y} York, D.~G., Adelman, J., Anderson, J.~E., et al.\ 2000, \aj, 120, 1579. doi:10.1086/301513
%
\bibitem[Zhu et al.(2020)]{2020MNRAS.496..245Z} Zhu, S.~F., Brandt, W.~N., Luo, B., et al.\ 2020, \mnras, 496, 245. doi:10.1093/mnras/staa1411
%
\bibitem[Zhu et al.(2021)]{2021MNRAS.505.1954Z} Zhu, S.~F., Timlin, J.~D., \& Brandt, W.~N.\ 2021, \mnras, 505, 1954. doi:10.1093/mnras/stab1406
%
\end{thebibliography}
\end{document}